\documentclass[12pt]{article}
\usepackage[authoryear,longnamesfirst]{natbib}
\usepackage{amsmath}
\usepackage{caption}
\usepackage{subcaption}
\usepackage{graphicx}
\usepackage{multirow}
\usepackage{booktabs}
\usepackage{microtype}
\usepackage[headheight=0.7in,margin=0.7in]{geometry}
\usepackage{fancyhdr}
\usepackage{hyperref}
\usepackage{cleveref}
\usepackage{amsfonts}
\usepackage[table]{xcolor}
\usepackage{colortbl}
\usepackage{xstring}
\usepackage{tikz}

\usetikzlibrary{shapes.geometric, arrows}

\pgfdeclarelayer{background}
\pgfdeclarelayer{foreground}
\pgfsetlayers{background,main,foreground}

\tikzstyle{startstop} = [rectangle, rounded corners, minimum width=3cm, minimum height=1cm,text centered, draw=black, fill=white!30]
\tikzstyle{digitalSignatures} = [rectangle, minimum width=3cm, minimum height=1cm, text centered, draw=black, fill=orange!30]
\tikzstyle{pois} = [rectangle, minimum width=3cm, minimum height=1cm, text centered, draw=black, fill=green!30]
\tikzstyle{clustering} = [rectangle, minimum width=3cm, minimum height=1cm, text centered, draw=black, fill=blue!30]
\tikzstyle{modelling} = [rectangle, minimum width=3cm, minimum height=1cm, text centered, draw=black, fill=purple!30]
\tikzstyle{arrow} = [thick,->,>=stealth]

\hypersetup{
    colorlinks=true,
    linkcolor=blue,
    filecolor=magenta,
    urlcolor=red,
    citecolor=red}
\begin{document}
 \lhead{ArXiv}
\begin{center}
{\large \textbf{Unveiling social vibrancy in urban spaces with app usage} \\}
\bigskip
{
    Thomas Collins$^{*,1}$,
    Diogo Pacheco$^1$,
    Riccardo Di Clemente$^{3,4}$,
    and Federico Botta$^{1,2}$\\
}
\bigskip
{
    {\small\textit{
    $^1$ Department of Computer Science, University of Exeter, Exeter, EX4 4QF, United Kingdom.\\
    $^2$ The Alan Turing Institute, London, NW1 2DB, United Kingdom.\\
    $^3$ Complex Connections Lab, Network Science Institute, Northeastern University London, London, E1W 1LP, United Kingdom.\\
    $^4$ ISI Foundation Via Chisola 5, 10126 Turin, Italy\\
}
}
}
$^*$ trc207@exeter.ac.uk
\end{center}

\begin{abstract}
Urban vibrancy is an important measure of the energetic nature of a city that is related to why and how people use urban spaces, and it is inherently connected with our social behaviour. Increasingly, people use a wide range of mobile phone apps in their daily lives to connect socially, search for information, make decisions, and arrange travel, amongst many other reasons. However, the relationship between online app usage and urban vibrancy remains unclear, particularly regarding how sociospatial behaviours interact with urban features. Here, we use app-usage data as a digital signature to investigate this question. To do this, we use a high-resolution data source of mobile service-level traffic volumes across eighteen cities in France. We investigate the social component of cities using socially relevant urban features constructed from \texttt{OpenStreetMap} `Points of Interest'. We developed a methodology for identifying and classifying multidimensional app usage time series based on similarity. We used these in predictive models to interpret the results for each city and across France. Across cities, there were spatial behavioural archetypes, characterised by multidimensional properties. We found patterns between the week and the weekend, and across cities, and the country. These archetypes correspond to changes in socially relevant urban features that impact urban vibrancy. Our results add further evidence for the importance of using computational approaches to understand urban environments, the use of sociological concepts in computational science, and urban vibrancy in cities.
\end{abstract}
\section{Introduction}\label{sec:usv-introduction}
Cities are among the largest and most complex human-made systems~\citep{battySizeScaleShape2008}. As they continue to grow and evolve, they face increasing pressure to accommodate larger populations and create more opportunities. This rapid urbanisation exacerbates existing challenges such as overcrowding~\citep{tangriUrbanGrowthHousing1968,kitilaDriversProspectsUrbanization2023b}, traffic congestion~\citep{wangUrbanMorphologyTraffic2021}, air pollution~\citep{undesaWorldUrbanizationProspects2019}, crime~\citep{malikUrbanizationCrimeRelational2016}, and uneven development~\citep{loganUnevenDevelopmentNature1987}. When population growth outpaces economic development, over-urbanisation emerges, intensifying these issues. Addressing these challenges requires robust development policies to achieve a sustainable balance between growth and equity~\citep{kitilaDriversProspectsUrbanization2023b}. Such policies, in turn, depend on the ability to measure and understand urban dynamics effectively.

Researchers and urban planners have traditionally relied on low-frequency data gathering to measure various aspects of urban environments, such as population movement, land use, and infrastructure development~\citep{frenchHowShouldUrban2017}. However, the recent explosion of digital data~\citep{zhengUrbanComputingConcepts2014}, combined with advancements in computational methods~\citep{battySmartCitiesFuture2012}, has enabled unprecedented high-frequency urban measurement~\citep{kandtSmartCitiesBig2021}. This transformation has also allowed researchers to explore and define the concept of \emph{urban vibrancy} (or \emph{urban vitality}), a term first introduced by Jane Jacobs in her seminal work, \emph{The Death and Life of Great American Cities}~\citep{jacobsDeathLifeGreat1961}. Urban vibrancy refers to the liveliness and energy of places. It is characterised by a high concentration of people, driven primarily by pedestrian activity rather than residential presence. It is a multi-faceted concept linked to the social, economic, and cultural vitality of urban spaces~\citep{sulisUsingMobilityData2018, bottaModellingUrbanVibrancy2021,wangMeasuringUrbanVibrancy2021}.

Urban vibrancy is increasingly recognised as a critical metric for understanding the dynamics of urbanisation. It offers insights into how cities evolve as they grow and densify, helping to uncover patterns in population movement, land use, and the interactions that shape urban spaces. Beyond these structural aspects, urban vibrancy also relates to many social dimensions of city life, providing a lens through which to understand how and why people engage with urban environments. Recent advancements in technology have made it common to use mobile phone data as a proxy for measuring urban vibrancy~\citep{tangExploringInfluenceUrban2018}, leveraging the widespread adoption of mobile devices to effectively capture the location and movement of people within urban settings~\citep{bottaQuantifyingCrowdSize2015b,bottaMeasuringSizeCrowd2020}.

Many characteristics of urban environments influence activity levels and patterns, each contributing to the overall vibrancy of city life~\citep{bottaModellingUrbanVibrancy2021,yueMeasurementsPOIbasedMixed2017,wuUrbanFormBreeds2018,collinsSpatiotemporalGenderDifferences2023a}. For example, buildings serve multiple purposes, functioning as homes, workplaces, or venues for social interactions, and thus play a key role in shaping urban activity. These spaces often overlap with what are known as `\emph{Points of Interest}' (`POIs')—locations that attract people due to their opportunities, such as access to resources, employment, or social activities. It is theorised that both the density and diversity of such urban features are particularly influential in fostering vibrancy, as they draw a larger variety of people to these locations~\citep{kumakoshiDiversityDensityUrban2021,jacobsDeathLifeGreat1961}. The density and diversity of POIs, in particular, are frequently found to correlate positively with urban vibrancy~\citep{tuPortrayingSpatialDynamics2020b}. This relationship underscores the significance of the built environment in fostering dynamic, lively urban spaces.

While researchers often use mobile phone data to examine how urban vibrancy changes with urban features~\citep{tangExploringInfluenceUrban2018}, the sociospatial behaviours that drive these patterns remain less understood. In particular, there is limited knowledge about how such behaviours vary during periods of leisure versus work. A more granular understanding of these temporal differences could offer valuable insights into the dynamics of urban life and how people interact with their environments.

\emph{Social vibrancy}, a crucial component of urban vibrancy, shapes how people interact and behave within city spaces~\citep{gaoExploringRelationshipUrban2024}.  While urban vibrancy encompasses all dynamic activities within city environments, social vibrancy focuses on the patterns of human interaction and social behaviour that animate these spaces. It is often measured by large sources of behavioural data~\citep{huangAnalyticsLocationbasedBig2021}. This could come from people's social interactions on social media platforms but could also come from location-based usage and services. Often people rely on single social media platforms but, with the release of multi-app datasets, it is becoming possible to understand a more detailed world of online behaviours across space and time.

Here, we explore the relationship between app usage across a range of app categories and the urban environments in which apps are used. We focus on eighteen metropolitan regions of France: Bordeaux, Dijon, Grenoble, Lille, Lyon, Mans, Marseille, Metz, Montpellier, Nancy, Nantes, Nice, Orleans, Paris, Rennes, Strasbourg, Toulouse, and Tours. These areas are home to more than one-third of the total population in the country~\citep{martinez-duriveNetMob23DatasetHighresolution2023}. We analyse how mobile app usage data varies across different parts of each city and how this relates to socially relevant urban features. Additionally, we investigate whether this relationship changes between weekdays, typically reserved for work and study, and weekends, when social and leisure activities are more common. This analysis is conducted at two spatial levels: (1) a `local level,' examining each city as a single unit, and (2) a `global level,' comparing cities to one another.

Our study contributes to the understanding of urban and social vibrancy in cities, particularly the role of \emph{third places}~\citep{bottaModellingUrbanVibrancy2021, collinsSpatiotemporalGenderDifferences2023a}—social spaces known to foster urban vitality. Using the \texttt{NetMob23} dataset, a high-resolution spatiotemporal source of \emph{mobile service-level traffic volumes} data provided by \emph{Orange}~\citep{martinez-duriveNetMob23DatasetHighresolution2023}, we demonstrate how urban vibrancy is shaped by spatial and temporal factors. Specifically, we reveal how app usage patterns correspond to the vibrancy of socially relevant urban features, offering insights into the behavioural dynamics of urban environments. Our findings highlight the ways in which weekdays and weekends influence urban vibrancy, as well as the importance of distinguishing between local and global spatial perspectives.

\section {Data}\label{sec:usv-data}

Our analyses rely on two data sources: (1) the \texttt{NetMob23} data set~\citep{martinez-duriveNetMob23DatasetHighresolution2023} which contains the usage of mobile applications over time and across eighteen cities of France (see~\cref{tab:usv-table-city-order-count-summary} for summary statistics), and (2) \texttt{OpenStreetMap} data (`OSM')~\citep{osmOpenStreetMapContributors2017}--a crowdsourced data set containing detailed information about the characteristics of urban environments. For each data type, we gather data for only the `metropolitan areas' of each city. These regions encompass the most densely populated areas of cities and consist of urban, suburban, and some rural areas. They are characterised by a connection to a city's shared culture, or economy, and share a close geographic location. \Cref{tab:usv-table-city-order-count-summary} gives the total cell count and total area per region.
\begin{table}[h]
\centering
\tiny
\caption[Geographical statistics for each city]{Geographical statistics for each city in the analysis. These values are based on the metropolitan areas of each city in the analysis.}

\NewDocumentCommand{\rot}{O{90} O{1em} m}{\makebox[#2][l]{\rotatebox{#1}{#3}}}
\NewDocumentCommand{\rotf}{O{45} O{1em} m}{\makebox[#2][c]{\rotatebox{#1}{#3}}}

\newcolumntype{g}{>{\columncolor{gray!20}}c}

\setlength{\tabcolsep}{8pt}
\renewcommand{\arraystretch}{4}

\begin{tabular}{l|g c g c g c g c g c g c g c g c g c}
\toprule
\textbf{Statistic} & \rot{Paris} & \rot{Rennes} & \rot{Lille} & \rot{Bordeaux} & \rot{Grenoble} & \rot{Lyon} & \rot{Nantes} & \rot{Toulouse} & \rot{Montpellier} & \rot{Tours} & \rot{Strasbourg} & \rot{Orleans} & \rot{Metz} & \rot{Mans} & \rot{Dijon} & \rot{Marseille} & \rot{Nice} & \rot{Nancy} \\
\midrule
\textbf{Number of Cells} & \rotf{81731} & \rotf{71597} & \rotf{67750} & \rotf{57846} & \rotf{54412} & \rotf{54013} & \rotf{53370} & \rotf{46269} & \rotf{44148} & \rotf{39131} & \rotf{34023} & \rotf{33758} & \rotf{30744} & \rotf{26859} & \rotf{23682} & \rotf{23663} & \rotf{16349} & \rotf{14353} \\
\textbf{Total Area ($\text{km}^2$)} & \rotf{816.26} & \rotf{714.96} & \rotf{676.85} & \rotf{577.32} & \rotf{543.08} & \rotf{539.15} & \rotf{532.86} & \rotf{461.69} & \rotf{440.52} & \rotf{390.71} & \rotf{339.78} & \rotf{337.09} & \rotf{307.06} & \rotf{268.20} & \rotf{236.45} & \rotf{236.10} & \rotf{163.14} & \rotf{143.34} \\
\bottomrule
\end{tabular}
\label{tab:usv-table-city-order-count-summary}
\end{table}

\subsection{\texttt{NetMob23} mobile network
traffic}\label{subsec:usv-netmob23-mobile-network-traffic}

We retrieved mobile network traffic volume data for each study area. This Data was made available by Orange within the framework of the Netmob 2023 Challenge~\citep{martinez-duriveNetMob23DatasetHighresolution2023}. The data consists of 77 days from the 16\textsuperscript{th} of March 2019, to the 31\textsuperscript{st} of May 2019. The data are time series for the downlink and uplink of the volume of demand for 68 mobile services (see apps in \cref{fig:app-sankey}). For instance, these services include social media-type apps like \emph{Facebook} and messaging-type apps like \emph{WhatsApp}. The data are available at a temporal granularity of fifteen-minute intervals. The values have undergone a normalisation that conceals sensitive information whilst preserving the signal in the data. This means that there are no units of measurement. At download, coverage is represented as probabilities of association and is partitioned into tiles measuring $100 \times 100 \, \text{m}^2$ each. For further details on the generation of the data set, see~\citet{martinez-duriveNetMob23DatasetHighresolution2023}.

\subsection{\texttt{OpenStreetMap}}\label{subsec:usv-openstreetmap-data}

We use \texttt{OpenStreetMap}~\citep{osmOpenStreetMapContributors2017} to construct urban features for our analysis (downloaded on the 13\textsuperscript{th} of February 2024). \texttt{OpenStreetMap} is a data repository built and maintained by crowdsourcing and collaboration. Volunteer users collect geospatial data and upload it to the open-source repository. These digitised representations of urban environments contain a range of urban attributes such as transportation networks or POIs~\citep{boeingOSMnxNewMethods2017}. We download data for all cells in the analysis. From this data, we calculate urban feature metrics.

\section{Methods}\label{sec:usv-methods}

\subsection{Digital signatures}\label{subsec:usv-digital-signatures}

We construct \emph{digital signatures} from the \texttt{NetMob23}~\citep{martinez-duriveNetMob23DatasetHighresolution2023} data to characterise each urban area according to its app usage. First, we aggregate the downlink and uplink data volume to calculate the total volume of data traffic for each app. Next, the data is aggregated by day of the week into two temporal categories: `weekdays,' which consists of all data from Monday to Thursday, and `weekends,' which includes all data from Friday to Sunday. Friday is included as part of the weekend due to the known leisure patterns associated with this day~\citep{suRhythmStreetsStreet2022}, which aligns with our previous methodology~\citep{collinsSpatiotemporalGenderDifferences2023a}. Aggregating in this way enables the study of behavioural differences between weekdays, typically associated with work and study, and weekends, when social and leisure activities are more prominent.

Next, we group each of the 68 mobile services in the \texttt{NetMob23} data set according to the \emph{Apple Store} app categorisation. The \emph{Apple Store} is used solely as a way to categorise the apps based on their names. However, this does not imply that users in our data set are necessarily \emph{Apple} users. The same grouping could have been achieved using \emph{Google Play}, yielding identical results. As a result of this grouping, we identify thirty app categories (see app categories in \cref{fig:app-sankey}). For ease of presentation, we refer to each app category simply as `apps' in the following text to streamline the language. It should be understood, however, that this term always refers to the aggregation of all apps within a specific category (e.g., the `Messaging' category includes the apps \emph{Apple iMessage}, \emph{Facebook Messenger}, \emph{Skype}, \emph{Telegram}, and \emph{WhatsApp}; see \cref{fig:app-sankey}).

Next, we aggregate the original data into 2-hour intervals over a twenty-four-hour period, resulting in 12 categories. Each category represents a two-hour range spanning the entire day, from 00:00 to 23:59.\footnote{The specific ranges are: [00:00–01:59, 02:00–03:59, ..., 22:00–23:59].} For each interval, a thirty-dimensional vector is constructed, referred to as the digital signature of a cell. This aggregation produces a time series of digital signatures for each $100 \times 100 \text{m}^2$ cell.

To standardise the data, normalisation is applied using the \emph{relative risk}. This approach ensures that each row (i.e., location) accounts for variations across 2-hour ranges within each temporal category (weekdays and weekends). The relative risk is calculated by dividing the value in the current location by the mean of all other locations within the same temporal group, excluding the current location. As a result, a ratio is obtained that highlights deviations relative to the overall dataset:

\begin{equation}\label{eq:usv-relative-risk}
\text{Relative Risk}_{ij} = \frac{X_{ij}}{\frac{\sum_{k \neq i} X_{kj}}{n-1}}
\end{equation}

\noindent where $X_{ij}$ is the value in the $i$th row (location) and $j$th column (app category), $\sum_{k \neq i} X_{kj}$ is the sum of values in the $j$th column excluding the $i$th row, and $n$ is the total number of rows.

Our analysis focuses on the relative distribution of app usage rather than absolute data volumes. By normalising the data, we enable meaningful comparisons between apps that inherently transfer varying amounts of data, ranging from low-data text-based apps to high-data video-based apps.

To provide both detailed and broad insights, the data is analysed at two spatial scales: a `local level,' which examines patterns within individual cities, and a `global level,' which considers the entire dataset. This dual-scale approach uncovers both city-specific patterns and country-wide trends.

\subsection{Multidimensional time series $k$-means clustering}\label{subsec:multidimensional-time-series-$k$-means-clustering}

At each of the temporal categories, `weekdays' and `weekends,' and for both the local and global levels, we implement a time series $k$-means clustering algorithm using the \texttt{tslearn} python package~\citep{tavenardTslearnMachineLearning2020}.

The methodology clusters spatial data based on temporal app usage patterns. Each spatial unit is represented by 30 time series (one per app category), creating a multidimensional dataset structured as (locations $\times$ time series $\times$ app categories). To determine the optimal number of clusters, values ranging from 3 to 10 were tested, and the silhouette method~\citep{rousseeuwSilhouettesGraphicalAid1987} was used to evaluate the quality of cluster separation. Clustering is conducted at both local and global scales, providing insights into cross-scale patterns through metric evaluation.

A metric comparison approach is applied to compare clusters at the local and global levels, offering insights into how scale and granularity influence our understanding.

\subsection{Social and geographical features}\label{subsec:usv-independent-variables}

For each city in the analysis, we retrieve POI data from the crowdsourced geospatial repository \texttt{OpenStreetMap}~\citep{osmOpenStreetMapContributors2017}. Our focus is on four categories of OSM features relevant to urban activity: `amenity,' `leisure,' `shop,' and `sport.' These categories include a diverse range of locations, such as essential services, leisure destinations, retail spaces, and sports facilities.

To enhance the signal and concentrate the analysis on the most relevant POIs, we refined the dataset by removing duplicates that occurred fewer than 10 times. The urban feature data is then aggregated to the same spatial cells as the \texttt{NetMob23} data, with each cell measuring $100 \times 100 , \text{m}^2$. Using a socially-focused classification system for `third places,' originally introduced by~\citet{oldenburgThirdPlace1982} and further developed by~\citet{jeffresImpactThirdPlaces2009}, we manually label the urban feature data.

Third places, which are neither homes nor workplaces, are spaces where people gather and socialise. They play a critical role in enhancing urban vibrancy~\citep{jeffresImpactThirdPlaces2009}, facilitating spontaneous interactions in urban settings, and contributing positively to communities~\citep{bottaModellingUrbanVibrancy2021}. Given the substantial amount of leisure time people spend in these places, they have become a significant focus of computational social science research~\citep{bottaModellingUrbanVibrancy2021, collinsSpatiotemporalGenderDifferences2023a, moriTimespaceDynamicsIncome2024}. We categorise the POIs into five third-place categories: (1) \emph{commercial services}, (2) \emph{commercial venues}, (3) \emph{eating and drinking}, (4) \emph{outdoor}, and (5) \emph{organised activities}. This categorisation aligns with the work of~\citet{jeffresImpactThirdPlaces2009} and expands upon it as described in the work of~\citet{collinsSpatiotemporalGenderDifferences2023a}.

After classification, a total of 497,296 urban features were identified as third places, encompassing 338 unique labelled urban features. Using this classification system, we calculate the count and diversity of third-place POIs. Specifically, we calculate:

\begin{itemize}
\item[\textbf{(1)}] Total third place count: The total count of third-place labels assigned to the urban features within each cell. This measure represents the total number of POIs categorised as third places. A higher count of third places is believed to enhance urban vibrancy by providing more opportunities for interactions~\citep{jacobsDeathLifeGreat1961,bottaModellingUrbanVibrancy2021}.

\item [\textbf{(2)}] Total third place diversity: The variety of third-place labels assigned to the urban features within each cell. This measure captures the diversity of POIs categorised as third places. Greater diversity in third-place types has been shown to enhance urban vibrancy by encouraging more interactions and attracting a wider variety of people~\citep{jacobsDeathLifeGreat1961,bottaModellingUrbanVibrancy2021}.

\item[\textbf{(3--7)}] Within-category count for each cell: The number of urban features (e.g., `swimming-pool') within each third-place category (e.g., the count of POIs within the commercial venues category).

\item[\textbf{(8--12)}] Within-category diversity for each cell: The diversity of urban features within each third-place category. This measure includes the variety of different POIs, such as restaurants, bars, and cafes, within the commercial venue category.

\end{itemize}

The Shannon-Wiener Index (H)~\citep{shannonMathematicalTheoryCommunication1948} is calculated using the following formula:

\begin{equation}\label{eq:usv-diverstity}
H = -\sum_{i=1}^{M} P_i \log_2 P_i
\end{equation}

\noindent where $H$ is the Shannon-Wiener Index, $M$ is the number of different categories, $P_i$ is the proportion of POIs in category $i$, and $\log_2$ is the base-2 logarithm.

\subsection{Statistical approach}\label{subsec:usv-statistical-approach}

Our approach to understanding the relationship between urban vibrancy and urban features uses the clusters projected onto the grid of $100 \times 100 \, \text{m}^2$ cells. We use those geometries as a reference with our urban features data. Since we are interested in the relationship between urban features and urban vibrancy, we included urban features as independent variables in the models (see~\cref{sec:usv-methods}) (for an overview of the methodology, see \cref{fig:flowchart}). We aim to model the covariates: total third-place count, total third-place diversity, and the count and diversity of the POIs within each subcategory of third place: commercial services, commercial venues, eating and drinking, outdoor, or organised activities. The dependent variable in our models is the cluster label. Since it is categorical with more than two classes, we use regularised multinomial logistic regression models predicting the cluster label generated by the digital signatures--we regularise the models using L2 regularisation (Ridge), which prevents overfitting by using a penalty term within the cost function and controls model complexity to make the models stable and interpretable.

Fitting this model gives a series of regression coefficients describing the strength of association between the app category complex and a socially relevant urban feature variable and membership of a particular cluster. We receive one coefficient per cluster and predictor. We ran models at different levels for comparisons. To compare work- versus leisure-type behaviours, we make models for each temporal category: weekdays and weekends. To understand the contents of each city, we perform the analysis at the city level, whereas, to understand country-level trends, we perform the analysis on the aggregation of all cities together.

We use performance statistics to measure each model's overall performance: we consider the \emph{accuracy} of each model which is the ratio of the number of correct predictions to the total number of predictions made by each model:

\begin{equation}\label{eq:usv-accuracy}
    \text{Accuracy} = \frac{\sum_{i=1}^{N} \mathbf{1}\{y_i = \hat{y}_i\}}{N}
\end{equation}

\noindent where $N$ is the total number of instances, $y_i$ is the true label for instance $i$, $\hat{y}_i$ is the predicted label for instance $i$, and $\mathbf{1}\{\cdot\}$ is the indicator function, which is 1 if the argument is true, and 0 otherwise.\\

We consider the \emph{macro F1 score}, which is the unweighted mean of the F1 scores for each class. This assesses the performance of each class independently and is important in class imbalance:

\begin{equation}\label{eq:usv-f1_i}
    F1_i = 2 \times \frac{\text{Precision}_i \times \text{Recall}_i}{\text{Precision}_i + \text{Recall}_i}
\end{equation}
\begin{equation}\label{eq:usv-macro-f1-score}
    \text{Macro F1 Score} = \frac{1}{C} \sum_{i=1}^{C} F1_i
\end{equation}
\noindent where $C$ is the number of classes, $F1_i$ is the F1 score for class $i$, $\text{Precision}_i$ is the $\frac{\text{TP}_i}{\text{TP}_i + \text{FP}_i}$ precision for class $i$, $\text{Recall}_i$ is the $\frac{\text{TP}_i}{\text{TP}_i + \text{FN}_i}$ recall for class $i$, $\text{TP}_i$ is the number of true positives for class $i$, $\text{FP}_i$ is the number of false positives for class $i$, and $\text{FN}_i$ is the number of false negatives for class $i$.\\

Finally, to account for class imbalance, we consider the \emph{weighted-f1-score}, which is similar to the macro F1 score, but instead of treating all classes equally, it weights the F1 score of each class by the number of true instances for that class (i.e., the support):

\begin{equation}\label{eq:usv-weighted-f1-score}
    \text{Weighted F1 Score} = \frac{1}{N} \sum_{i=1}^{C} \left( \text{Support}_i \times F1_i \right)
\end{equation}

\noindent where $N$ is the total number of instances across all classes, $C$ is the number of classes, $\text{Support}_i$ is the number of true instances for class $i$ (i.e., the number of times class $i$ appears in the dataset), and $F1_i$ is the F1 score for class $i$.\\

\section{Results}\label{sec:usv-results}

\subsection{Cluster number}\label{subsec:usv-results-cluster-number}

We assess the optimal number of clusters using the silhouette method~\citep{rousseeuwSilhouettesGraphicalAid1987} and find that three clusters were preferred most often for both the `weekday' and `weekend' data subsets (\cref{fig:usv-local_paris_week,fig:usv-local_paris_weekend} and~\cref{fig:usv-all_plot_all_cluster_maps_week_local,fig:usv-all_plot_all_cluster_maps_weekend_local,fig:usv-all_plot_all_cluster_maps_week_global,fig:usv-all_plot_all_cluster_maps_weekend_global}).

\subsection{Spatial arrangement}\label{subsec:usv-results-spatial-arrangement}

At both local and global scales, clusters consistently exhibited either monocentric patterns (a single core) or polycentric zones (multiple cores), typically concentrated around city centers (\cref{fig:usv-local_paris_week,fig:usv-local_paris_weekend}, \cref{fig:usv-all_plot_all_cluster_maps_week_local,fig:usv-all_plot_all_cluster_maps_weekend_local,fig:usv-all_plot_all_cluster_maps_week_global,fig:usv-all_plot_all_cluster_maps_weekend_global}). This aligns with established patterns of urban activity, where the main commercial or administrative core attracts the highest levels of activity~\citep{battinoAnalyzingCentralBusiness2012,doorleyRevUrbUnderstandingUrban2019}. These patterns were observed consistently on both weekdays and weekends.

Cities were typically divided into an outskirts area (always labelled as Cluster 1) and two central zones (labelled as Clusters 2 and 3). This consistency between weekdays and weekends underscores the enduring importance of central zones as activity hubs. Cluster 3 was most often located in the city centre, although exceptions existed where this cluster was not centrally positioned. At the local scale, clusters exhibited greater detail and definition, reflecting the expected outcome that finer spatial resolution reveals clearer urban activity patterns.

During weekdays, all cities had at least three clusters, with exceptions such as Tours (five clusters), and Strasbourg, Mans, Marseille, and Nancy (four clusters) (\cref{fig:usv-all_plot_all_cluster_maps_week_local}). On weekends, Mans was the only exception with four clusters (\cref{fig:usv-all_plot_all_cluster_maps_weekend_local}). Notably, the fourth cluster in Mans highlighted the Bellevue-Carnac district, a predominantly social housing area. This could indicate a diverse demographic within the district, though further investigation is needed to confirm this. The increased number of clusters on weekdays may reflect the complexity of work-related movement patterns, contrasting with the leisure-oriented activities more prevalent on weekends.

At the global scale, a different pattern emerged. On weekdays, Cluster 1 often dominated as the sole cluster~(\cref{fig:usv-all_plot_all_cluster_maps_week_global,fig:usv-all_plot_all_cluster_maps_weekend_global}). This suggests that global-scale data is more homogenous, highlighting central zones of activity but revealing little variation across cities. On weekends, a small section of the city centre was defined as Cluster 2, with larger cities also displaying a re-emergence of Cluster 3. This may indicate a more dispersed activity pattern across urban areas. However, Paris deviated significantly from this pattern. On weekdays, Cluster 2 in Paris was relatively large, growing even larger on weekends.

At the local level, weekday clusters in Paris closely matched the global-scale clusters. However, this alignment did not persist on weekends. On a global scale, weekend clusters in Paris were larger and less defined. In contrast, local-scale clusters revealed reduced outskirts and an additional interim cluster (Cluster 2) between the outskirts and the central area (Cluster 3).

Paris, as a major global city, behaves as an outlier. Weekend activity in Paris extends beyond the Boulevard Périphérique, possibly driven by an influx of tourists exploring the city's many vibrant areas. The mismatch between local and global clusters on weekends may reflect widespread activity patterns recognised at the global scale, while local clusters capture concentrated activity in specific areas.

\subsection{Usage intensity and variability}\label{subsec:usv-results-usage-intensity-and-variability}

\begin{figure}
    \captionsetup{font=large, labelfont=sc}
    \centering
    \begin{subfigure}[t]{0.44\linewidth}
        \centering
        \includegraphics[width=\linewidth]{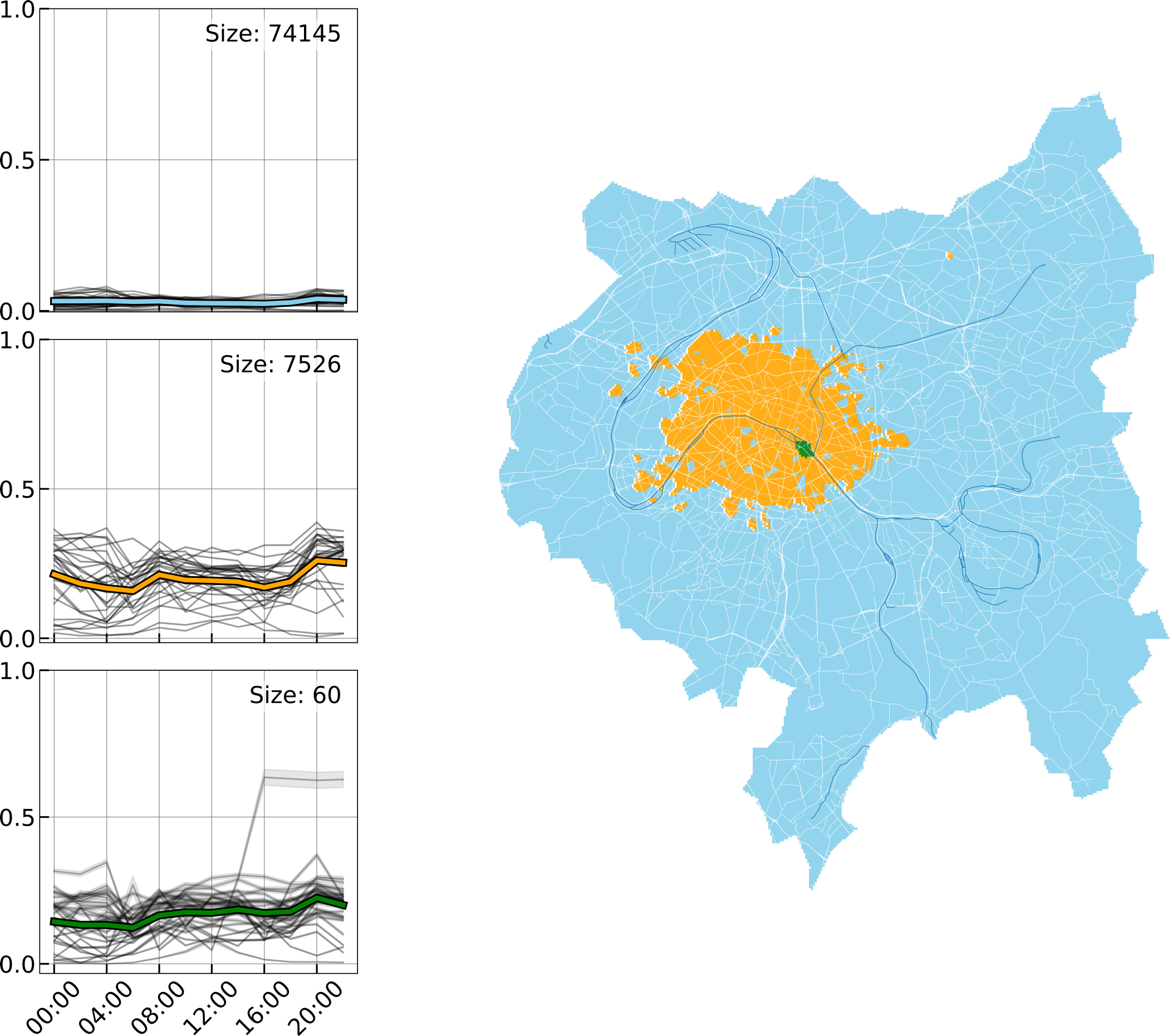}
        \caption{Week}
        \label{fig:usv-local_paris_week}
    \end{subfigure}
    \hfill
    \begin{subfigure}[t]{0.44\linewidth}
        \centering
        \includegraphics[width=\linewidth]{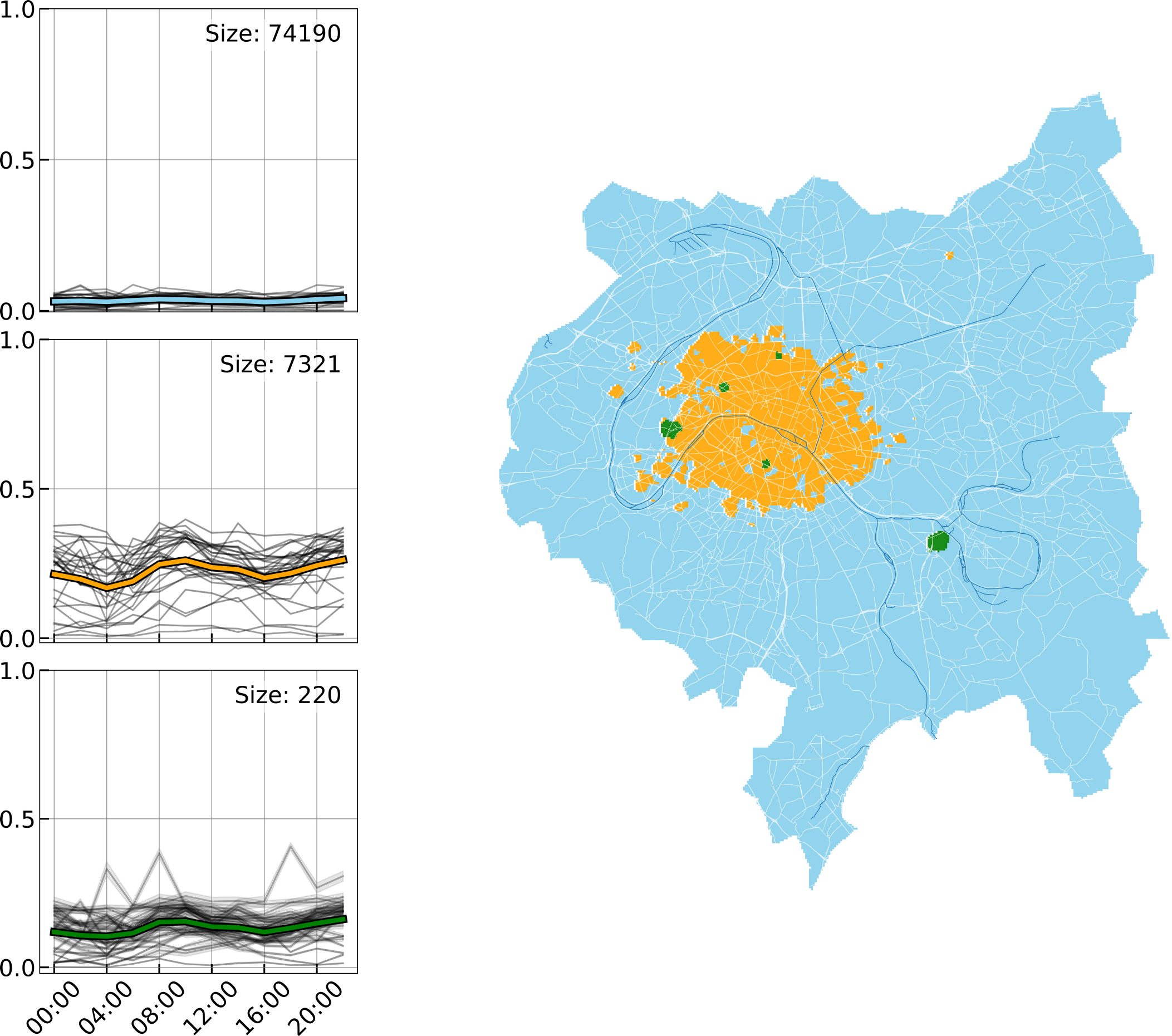}
        \caption{Weekend}
        \label{fig:usv-local_paris_weekend}
    \end{subfigure}
    \hfill
    \begin{subfigure}[t]{0.10\linewidth}
        \centering
        \includegraphics[width=\linewidth]{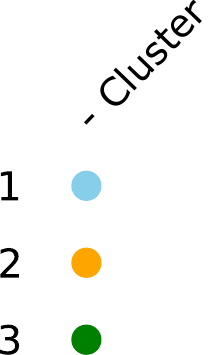}
        \label{fig:legend}
    \end{subfigure}

    \captionsetup{font=small, labelfont=normal}
    \caption[Clustered app usage in space and time for Paris]{
        Clustered app usage in space and time for Paris. Time series and maps for each cluster for weekdays (Monday--Thursday) (\subref{fig:usv-local_paris_week}) and weekends (Friday--Sunday) (\subref{fig:usv-local_paris_weekend}) for Paris at the local scale. The time series displays the average app usage for all cells in Paris. To facilitate comparison across app categories, the data were scaled using the min-max scaling technique, but this scaling was applied only for visualization purposes. The averaged app categories are shown as black lines, with each cluster centre visualized at the forefront of these. The size of each cluster is indicated in the top corner of each figure. Clusters are presented in size order, with the largest cluster assigned the smallest number. The maps depict the spatial arrangement of the clusters, and the legend shows the colour coding for the clusters in both the time series and the maps.
    }
\end{figure}

At the local scale, the largest cluster in each city (Cluster 1) consistently exhibited low relative usage intensity and variability compared to other clusters. This pattern was evident across both temporal scales—weekdays and weekends—and is illustrated in the cluster time series plots (\cref{fig:usv-local_paris_week,fig:usv-local_paris_weekend}, \cref{fig:usv-all_plot_local_timeseries_week_second_half,fig:usv-usv-all_plot_local_timeseries_weekend_first_half,fig:all_plot_local_timeseries_weekend_second_half,fig:usv-all_plot_global_timeseries_week,fig:usv-all_plot_global_timeseries_weekend}). These stable patterns suggest that Cluster 1 represents predominantly residential areas, where activity levels remain relatively constant.

Clusters closer to city centres tended to exhibit higher relative usage intensity and variability. Smaller clusters, such as Clusters 2 and 3, often corresponded to more diverse urban areas. Cluster 2 typically showed mid-level intensity and variability, while Cluster 3 consistently had the highest levels. These clusters likely represent commercial zones or entertainment districts, which are dynamic areas characterised by shifting activity peaks, such as during lunch hours or social gatherings.

In cases where cities had more than three clusters, the additional clusters often reflected areas with similar usage intensity but differing variability. For example, in Tours, Cluster 3 displayed high intensity during the morning, while Clusters 4 and 5 had high intensity in the afternoon. This highlights within-city temporal complexity, where the most dynamic zones differ in their temporal activity patterns or `fingerprints.'

\begin{figure*}[htp]
	\includegraphics[width=\textwidth]{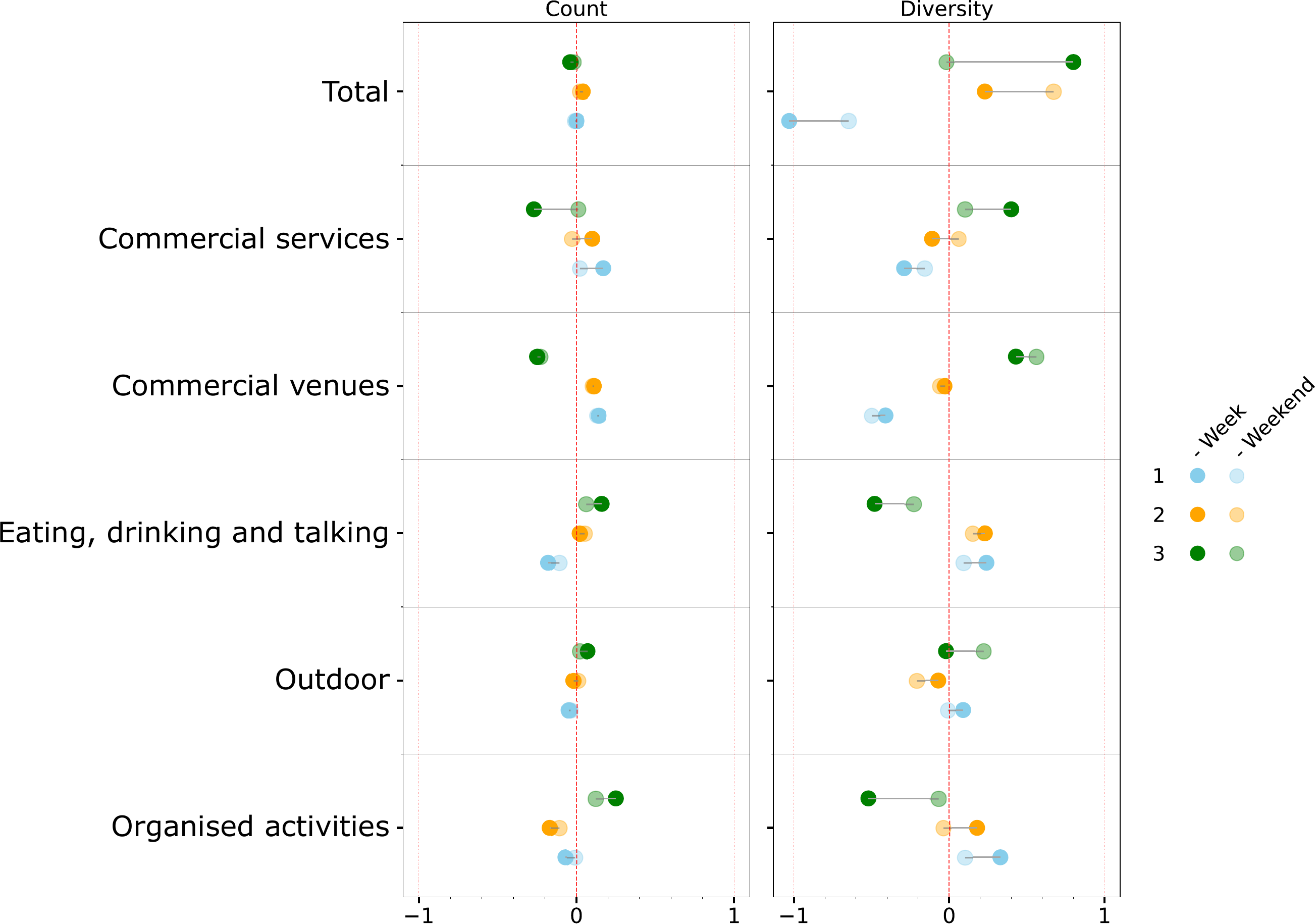}
	\caption[Local-level multinomial logistic regression predictive model estimates]{
	Local-level multinomial logistic regression predictive model estimates for the weekdays and weekends in Paris. The figure contains estimates for the count and diversity of all \emph{third place} features as well as each third place category. The y-axis displays each coefficient. Coefficients for the week (darker shade) and weekend (lighter shade). Data preprocessing was applied to the \texttt{NetMob23} data set~\citep{martinez-duriveNetMob23DatasetHighresolution2023}. The legend shows the colour coding for the clusters.
}
	\label{fig:usv-local_paris_coeff}
\end{figure*}

\begin{figure*}[htp]
	\includegraphics[width=\textwidth]{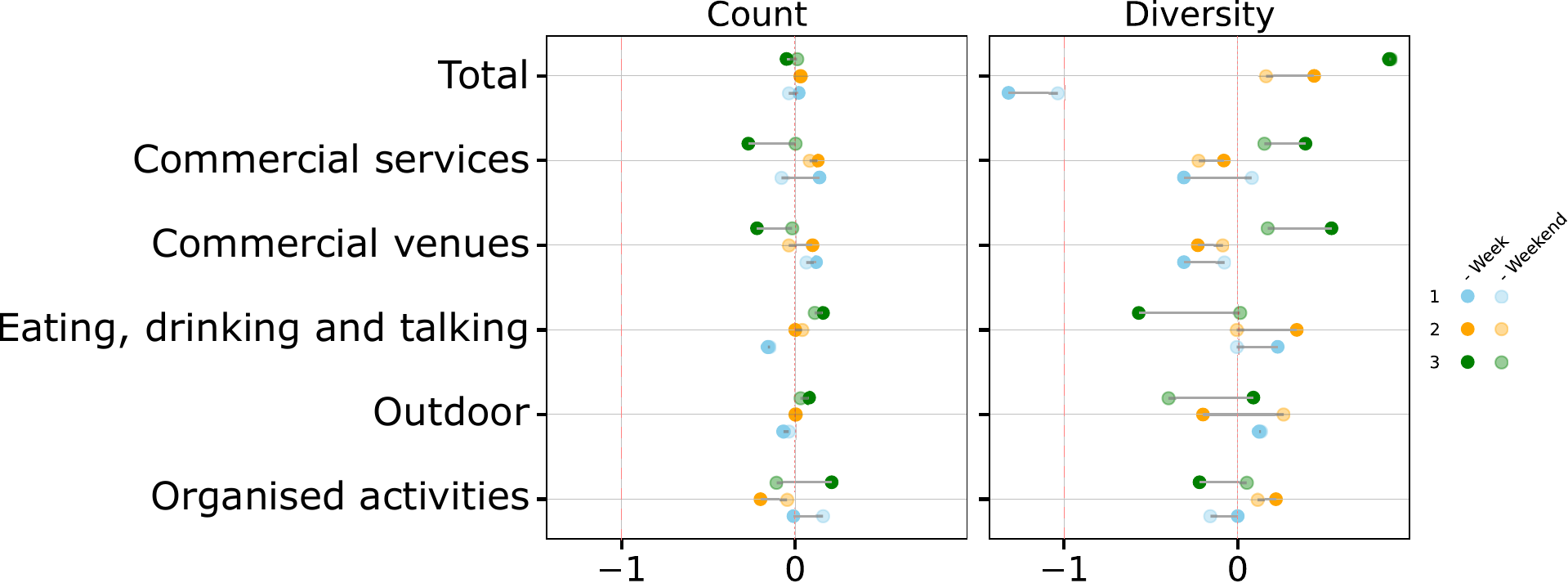}
	\caption[Global-level multinomial logistic regression predictive model estimates]{
	Global-level multinomial logistic regression predictive model estimates for the weekdays and weekends. The figure contains estimates for the count and diversity of all \emph{third place} features as well as each third place category. The y-axis displays each coefficient. Coefficients for the week (darker shade) and weekend (lighter shade). Data preprocessing was applied to the \texttt{NetMob23} data set~\citep{martinez-duriveNetMob23DatasetHighresolution2023}. The legend shows the colour coding for the clusters.
}
	\label{fig:usv-plot_all_dot_coefficients_global_plot_all_dots}
\end{figure*}

\subsection{Multinomial logistic regression}\label{subsec:usv-results-multinomial-logistic-regression}

\subsubsection{Diversity in third places}

Across both temporal categories—weekdays and weekends—and at both local and global scales, diversity coefficient estimates consistently exceeded count estimates (\cref{fig:usv-local_paris_coeff,fig:usv-plot_all_dot_coefficients_global_plot_all_dots}, \cref{tab:total-count-coefficients,tab:commercial-service-count-coefficients,tab:commercial-venue-count-coefficients,tab:eating-drinking-talking-count-coefficients,tab:organised-activity-count-coefficients,tab:outdoor-count-coefficients,tab:total-diversity-coefficients,tab:commercial-service-diversity-coefficients,tab:commercial-venue-diversity-coefficients,tab:eating-drinking-talking-diversity-coefficients,tab:organised-activity-diversity-coefficients,tab:outdoor-diversity-coefficients}). This indicates that diversity in urban features is often a stronger predictor in the models than simple counts.

Regarding total diversity coefficients (\cref{tab:total-diversity-coefficients}), significant variation was observed across cities. In Paris, Cluster 1 was negatively associated with diversity on both weekdays and weekends, while Cluster 2 showed stronger positive associations on weekends. Cluster 3, however, had coefficients close to zero on weekdays, suggesting a potential switching relationship between Clusters 1 and 2 regarding third-place diversity. Similar trends were observed in Nantes, Tours, and Strasbourg. For example, Strasbourg’s Cluster 4 showed the strongest positive association with diversity on weekdays, indicating an intensive level of third-place diversity in this area. A general trend across cities was the negative relationship between Cluster 1 and diversity, while Cluster 3 often exhibited strong positive associations. These findings suggest that Cluster 3 areas are vibrant, mixed-use zones with a variety of services, activities, and venues, whereas Cluster 1 areas are more homogenous or residential in nature.

For commercial services (\cref{tab:commercial-service-diversity-coefficients}), city-level differences were prominent. In Paris, Cluster 1 and Cluster 2 showed negative associations with commercial service diversity on weekdays, with coefficients of -0.29 and -0.11, respectively. Conversely, Cluster 3 exhibited a strong positive association (0.40), suggesting higher diversity in commercial services in central areas. In most cities, however, Cluster 1 showed stronger positive associations with diversity, while Cluster 3 was negatively associated, indicating that commercial diversity was often concentrated in Cluster 1 areas, typically hubs of commercial activity. Notably, exceptions like Nancy displayed unique trends; for example, Cluster 4 in Nancy had a strong negative association with diversity, but only on weekdays.

Commercial venue diversity (\cref{tab:commercial-venue-diversity-coefficients}) was generally a weaker predictor. Most cities showed negative associations for Cluster 1 and positive associations for Cluster 3, with Nantes as a notable exception where Cluster 3 exhibited strong positive associations.

For eating and drinking diversity (\cref{tab:eating-drinking-talking-diversity-coefficients}), cities like Paris, Rennes, and Lille demonstrated positive associations with Cluster 1. Rennes had particularly strong coefficients (1.23 on weekdays and 1.14 on weekends), suggesting high eating and drinking diversity in these areas. In Paris, the strength of associations decreased between weekdays and weekends for all clusters. For example, Cluster 1 dropped from 0.24 to 0.09, while Cluster 3 decreased from -0.48 to -0.23. In Tours, diversity patterns shifted noticeably: Cluster 1 had a positive association on weekdays (0.44) but dropped sharply on weekends (0.05). Meanwhile, Cluster 2 flipped from a negative association (-0.24) on weekdays to a positive one (0.26) on weekends. Overall, many cities exhibited positive associations with Cluster 1 for eating and drinking diversity, contrary to expectations, suggesting these areas may be highly diverse.

For organised activity diversity (\cref{tab:organised-activity-diversity-coefficients}) and outdoor diversity (\cref{tab:outdoor-diversity-coefficients}), the results were less pronounced. Organised activity diversity showed positive associations with Cluster 3 across cities, indicating that these areas may be focal points for structured events and activities.

\subsubsection{Count in third places}

Fewer models were significant for count variables compared to diversity variables (\cref{tab:total-count-coefficients,tab:commercial-service-count-coefficients,tab:commercial-venue-count-coefficients,tab:eating-drinking-talking-count-coefficients,tab:organised-activity-count-coefficients,tab:outdoor-count-coefficients}). However, there were notable variations across cities in the coefficients for the total count of third places (\cref{tab:total-count-coefficients}). For example, Paris displayed coefficients close to zero for both weekdays and weekends, indicating a weak relationship between the covariates and the cluster structure. In contrast, cities like Lille and Nice showed stronger coefficients. Lille, for instance, had a large negative coefficient (-0.34) for Cluster 3 on weekends, with relatively small differences between weekdays and weekends.

In the commercial service third-place category (\cref{tab:commercial-service-count-coefficients}), patterns varied between weekdays and weekends. In Paris, weekday coefficients were positive for Cluster 1 (0.17) and Cluster 2 (0.10), while Cluster 3 had a negative coefficient (-0.27), suggesting that Cluster 1 and Cluster 2 were more strongly associated with commercial services during weekdays. On weekends, the associations were weaker across all clusters. In Lille, both weekdays and weekends showed strong positive coefficients for Cluster 3 (0.68 on weekdays, 0.76 on weekends), indicating that Cluster 3 was significantly associated with commercial services. Across most cities, Cluster 3 consistently demonstrated the strongest positive association with commercial services, whereas Clusters 1 and 2 were often negatively associated, particularly in Lille and Nice.

For the commercial venue third-place category (\cref{tab:commercial-venue-count-coefficients}), city-level differences between weekdays and weekends were minimal. However, general trends emerged: Cluster 1 tended to show positive associations with commercial venues across most cities, suggesting dominance in commercial areas regardless of time. Cluster 2 exhibited mixed associations, while Cluster 3 was often negatively associated in many cities.

The eating, drinking, and talking third-place category (\cref{tab:eating-drinking-talking-count-coefficients}) followed a similar pattern. Most cities showed little variation between weekdays and weekends, with exceptions such as Nice. In Nice, Cluster 1 and Cluster 2 had positive associations on weekends, while Cluster 3 was negatively associated; weekday coefficients in all clusters were closer to zero. In general, this category displayed an inverse pattern compared to commercial venues: Cluster 1 tended to have negative coefficients, while Cluster 3 was mostly positive. Cluster 2 continued to show mixed results across cities and time periods.

\subsection{Model performance}\label{subsec:usv-results-model-performance}

The models demonstrated consistent performance across weekdays and weekends. During weekdays, the local models achieved an average accuracy of 0.88, while the global models had a higher accuracy of 0.99. For weekend data, the local models performed slightly better, with an accuracy of 0.89, compared to 0.95 for the global models. This suggests that local models are slightly more accurate on weekend data, whereas global models perform better on weekday data.

The macro F1 score, which measures the balance of precision and recall across all classes, averaged 0.41 for the local models and 0.38 for the global models during weekdays. On weekends, these scores improved slightly to 0.44 for the local models and 0.40 for the global models. These relatively low macro F1 scores highlight difficulties in achieving balanced performance across all classes, particularly for less frequent or harder-to-predict categories.

To address class imbalance, we also calculated the weighted F1 score, which accounts for the relative frequency of each class. During weekdays, the weighted F1 score averaged 0.84 for the local models and 0.98 for the global models. On weekends, the scores were 0.86 for the local models and 0.93 for the global models. These higher weighted F1 scores indicate that the models perform robustly overall, even across different time periods, despite some challenges in handling class imbalances.

\section{Discussion}\label{sec:usv-discussion}

Our analysis examines social vibrancy patterns across eighteen French cities using the \texttt{NetMob23} dataset ~\citep{martinez-duriveNetMob23DatasetHighresolution2023}. By clustering app usage patterns, we investigate how digital behavioural signatures correspond to distinct urban spaces. Our approach suggests that app usage patterns create unique spatial fingerprints that may reveal the presence and characteristics of third places—social spaces that mediate between home and work environments delineating specific third place urban features. We tested this via two computational approaches: (1) we use a time series $k$-means clustering~\citep{tavenardTslearnMachineLearning2020} approach to maintain the structure of the apps whilst reducing the dimensionality, and (2) we use multinomial logistic regression models to test the cluster labels to understand how app usage interacts with neighbouring urban features. We have done this for each city and a global data aggregate. Our main aim has been to explore the relationship between social vibrancy behaviours and their connection with socially focused urban features. To do this, we utilised a collection of relevant variables associated with urban features known to contribute to urban vibrancy.

Our analysis has yielded some intriguing insights. We have demonstrated that it is possible to leverage highly detailed and high-frequency app usage data alongside geospatial data to (1) observe patterns and variations throughout the week, and (2) identify differences and commonalities among the largest cities in France. We do this with high model performance values that indicate generally that our models for the most part have high predictive power. Our multidimensional time series clustering methodology found distinct spatial clusters using the app usage data (\cref{fig:usv-local_paris_coeff}); these were most detailed at the local level.

We found Cluster 1 always had low relative usage intensity and variability compared to the other clusters showing that the largest areas were stable in terms of their activity suggesting areas of residence. Cluster 2 often had mid-level usage whereas Cluster 1 had the highest. This suggests that most often cities contain polycentric centres with a core of intense usage in terms of social vibrancy. In these places, social activities occur most and with the most diversity. This makes sense in the centres of cities as we find that these places contain high building density where they are more diverse in terms of POIs and so more diverse opportunities for social activities occur. Importantly, this was not the case in Paris. Instead, Paris contained mid-level pockets within and around a large core area of the most intense and variable relative usage. This could be because Paris contains multiple vibrant areas used for widespread activities. Contained within this central zone are pockets where the usage is less variable and less intense.

The overall pattern indicates that POI diversity is tied to specific classes of areas, with distinct spatial patterns that vary by city and time of week. Where many count coefficients were clustered around zero, the diversity estimates were more spread out. A common trend was that diversity in third places was positively associated with Cluster 3 whereas there was a negative association for Cluster 1. This is confirming evidence that Cluster 1 locations lack diverse social places which makes sense because they mainly consist of residential places. There were negative coefficients across cities. This indicates that as the value of these predictors increases, the probability of the observation being classified into certain classes decreases. Negative estimates occurred in different cities, predictors, and temporal categories with large variations. Conversely, some coefficients were positive, suggesting that higher values of these predictors increase the probability of classification into certain classes. Again, there was variation across cities, predictors, and temporal categories. Total diversity had a recurring shape across cities: Cluster 1 showed a negative association, meaning that there is a lower probability of the observation being classified into that class; Cluster 2 showed an interim level of association. Cluster 3 showed a strong association. At the global level, the results were similar to the coefficients of Paris perhaps due to the greater size of Paris.

We also wish to highlight and discuss some of the limitations of our analysis. We exploit our data set to uncover behavioural patterns in cities through app usage. However, the data set likely contains demographic biases which can affect our results. We also note here that our global-scale model seemed to be influenced heavily by Paris, due to its size. The data set only covers a specific time period, which is not necessarily representative of the general behaviour of people. This may be particularly true since the time period is before the COVID-19 pandemic, which is likely to have changed many of the existing patterns. A further potential problem is that the app usage data and the \texttt{OpenStreetMap} data are from differing time periods; the app usage data are from 2019, whereas the \texttt{OpenStreetMap} data are from 2024. This may introduce certain biases into the analysis; however, we anticipate that these biases are minimal and unlikely to significantly impact the overall results.

In this study, we have considered the modelling of \emph{social vibrancy} for numerous cities and across temporal categories. We found a large variation in the results: we found that the diversity of third places was associated with the centres of most cities whereas the opposite was true in the outskirts. This differed from Paris in that the centre of Paris was much larger with pockets of reduced social vibrancy. This gives more evidence for the importance of understanding cities through the lens of third places.\\

To conclude, our results add further evidence for the importance of using computational approaches for understanding urban environments, the use of sociological concepts in computational science and for understanding urban vibrancy in cities.
\section*{Acknowledgements}\label{sec:usv-acknowledgements}
For the purpose of open access, the author has applied a Creative Commons Attribution (CC BY) licence to any Author Accepted Manuscript version arising from this submission.
 \section*{Funding}\label{sec:usv-cues-funding}
The authors received no financial support for the research, authorship, and/or publication of this article. This research did not receive any specific grant from funding agencies in the public, commercial, or not-for-profit sectors.

\bibliographystyle{abbrvnat}
\bibliography{jr_lib}

\appendix
\renewcommand{\thesection}{Supplementary Information \arabic{section}}
\renewcommand{\thefigure}{SI~\arabic{figure}}
\setcounter{figure}{0}
\renewcommand{\thetable}{SI~\arabic{table}}
\setcounter{table}{0}
 \clearpage

\section*{Supplemental material}\label{sec:usv-supplemental-materiall}

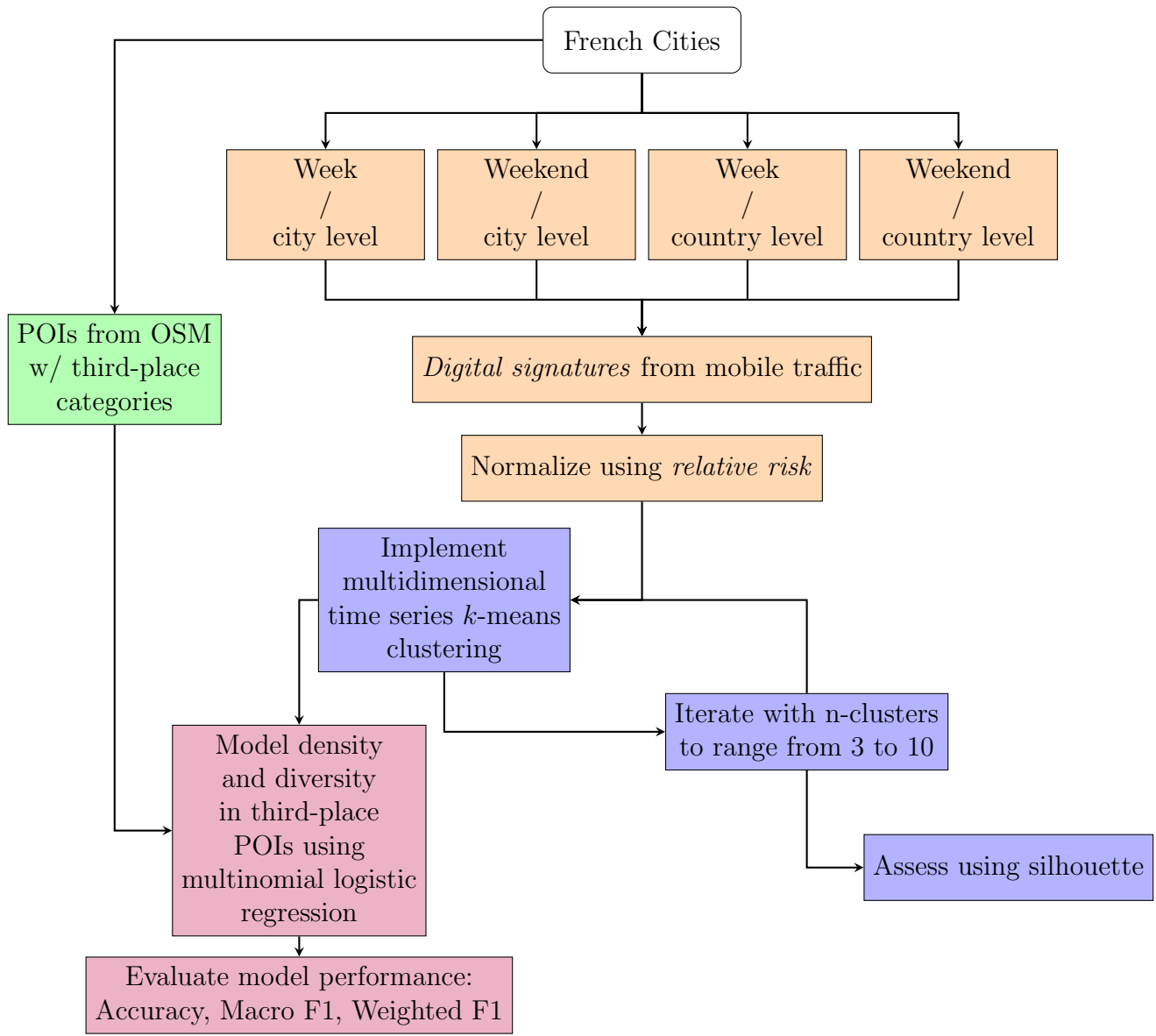
\begin{figure}[!h]
    \centering
    \begin{flushleft}
    \begin{tikzpicture}[node distance=1.5cm]

    \node (start) [startstop] at (2, 0) {French Cities};

    \node (weekCityLevel) [digitalSignatures, below of=start, yshift=-1cm, xshift=-4.8cm, align=center] {Week\\ / \\city level};
    \node (weekendCityLevel) [digitalSignatures, below of=start, yshift=-1cm, xshift=-1.6cm, align=center] {Weekend\\ / \\city level};
    \node (weekCountryLevel) [digitalSignatures, below of=start, yshift=-1cm, xshift=1.6cm, align=center] {Week\\ / \\country level};
    \node (weekendCountryLevel) [digitalSignatures, below of=start, yshift=-1cm, xshift=4.8cm, align=center] {Weekend\\ / \\country level};

    \node (digitalSignatures) [digitalSignatures, below of=start, yshift=-3.5cm] {\emph{Digital signatures} from mobile traffic};

    \node (poiData) [pois, below of=start, yshift=-3.5cm, xshift=-8cm, align=center] {POIs from OSM \\w/ third-place \\ categories};

    \node (normalize) [digitalSignatures, below of=digitalSignatures, align=center] {Normalize using \emph{relative risk}};
    \node (clustering) [clustering, below of=normalize, yshift=-.5cm, xshift=-3cm, align=center] {Implement\\multidimensional\\time series $k$-means\\clustering};
    \node (iterate) [clustering, right of=clustering, xshift=4cm, yshift=-2cm, align=center] {Iterate with n-clusters \\ to range from 3 to 10};
    \node (silhouette) [clustering, below right of=iterate, xshift=2cm, yshift=-1cm] {Assess using silhouette};

    \node (regressionModel) [modelling, below of=clustering, yshift=-2cm, xshift=-2.2cm, align=center] {Model density \\ and diversity \\ in third-place \\ POIs using \\ multinomial logistic \\ regression};
    \node (performance) [modelling, below of=regressionModel, yshift=-1cm, align=center] {Evaluate model performance: \\ Accuracy, Macro F1, Weighted F1};

    \begin{pgfonlayer}{background}
        \draw [arrow] (start.south) -- ++(0,-.6) -| (weekCityLevel);
        \draw [arrow] (start.south) -- ++(0,-.6) -| (weekendCityLevel);
        \draw [arrow] (start.south) -- ++(0,-.6) -| (weekCountryLevel);
        \draw [arrow] (start.south) -- ++(0,-.6) -| (weekendCountryLevel);

        \draw [arrow] (weekCityLevel.south) -- ++(0,-.6) -| (digitalSignatures);
        \draw [arrow] (weekendCityLevel.south) -- ++(0,-.6) -| (digitalSignatures);
        \draw [arrow] (weekCountryLevel.south) -- ++(0,-.6) -| (digitalSignatures);
        \draw [arrow] (weekendCountryLevel.south) -- ++(0,-.6) -| (digitalSignatures);

        \draw [arrow] (digitalSignatures) -- (normalize);

        \draw [arrow] (normalize) |- (clustering);

        \draw [arrow] (clustering) |- (iterate);

        \draw [arrow] (iterate) |- (clustering);
        \draw [arrow] (iterate) |- (silhouette);

        \draw [arrow] (clustering.west) -| (regressionModel);

        \draw [arrow] (start) -| (poiData);
        \draw [arrow] (poiData) |- (regressionModel);
        \draw [arrow] (regressionModel) -- (performance);
    \end{pgfonlayer}
    \end{tikzpicture}
    \end{flushleft}
    \caption[Flowchart showing digital signatures, POIs, clustering, and modelling process in 18 French cities]{Flowchart showing digital signatures, POIs, clustering, and modelling process in 18 French cities.}
    \label{fig:flowchart}
\end{figure}

\begin{figure*}[htp]
	\includegraphics[width=\linewidth]{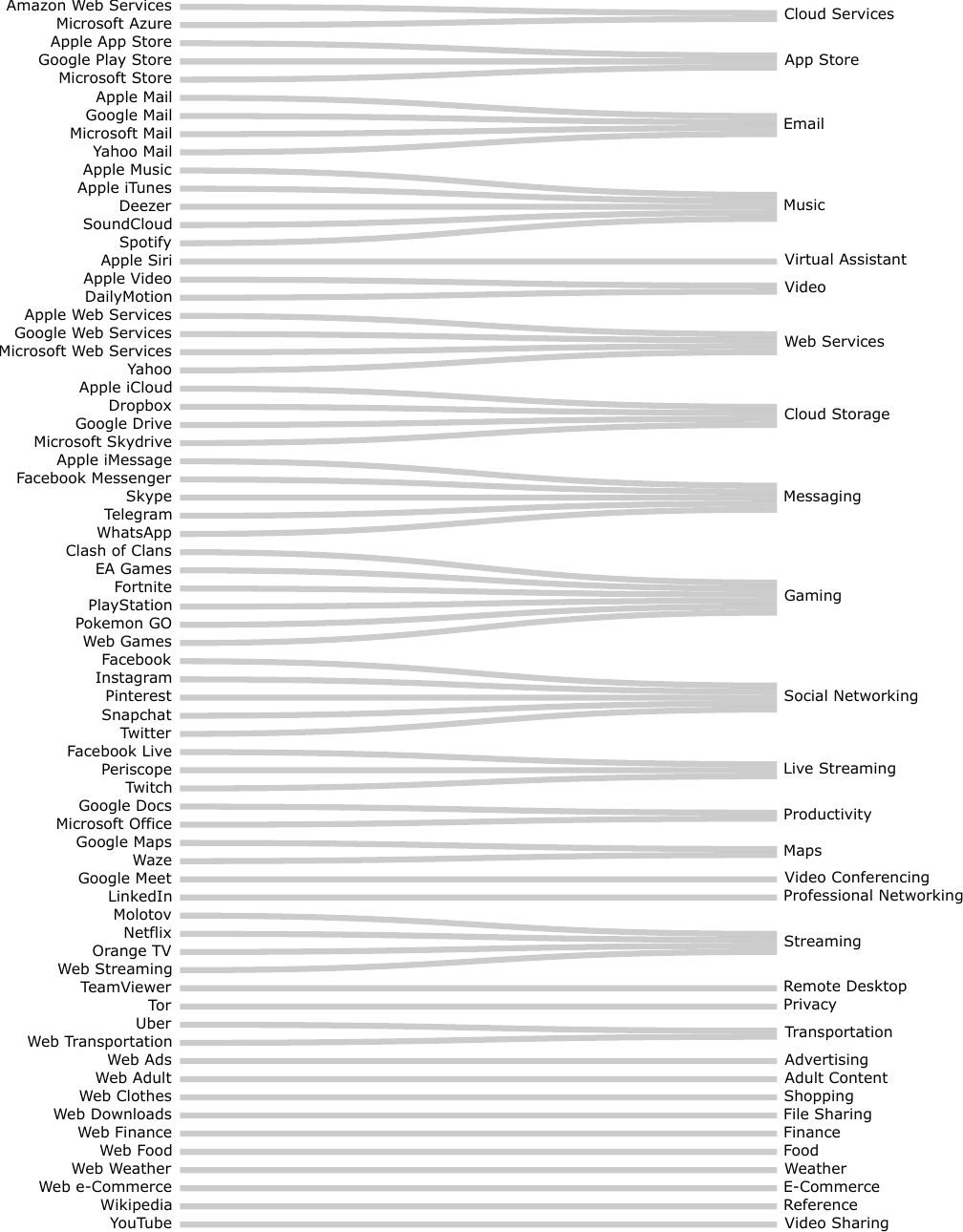}
	\caption[Apps and \emph{Apple Store} categories]{
        Mobile apps and their assigned \emph{Apple Store} categories. The apps are given in the left-hand
        column and the app categories are given in the right-hand column.
    }
	\label{fig:app-sankey}
\end{figure*}

\begin{figure*}[htp]
	\centering
	\includegraphics[height=0.8\textheight]{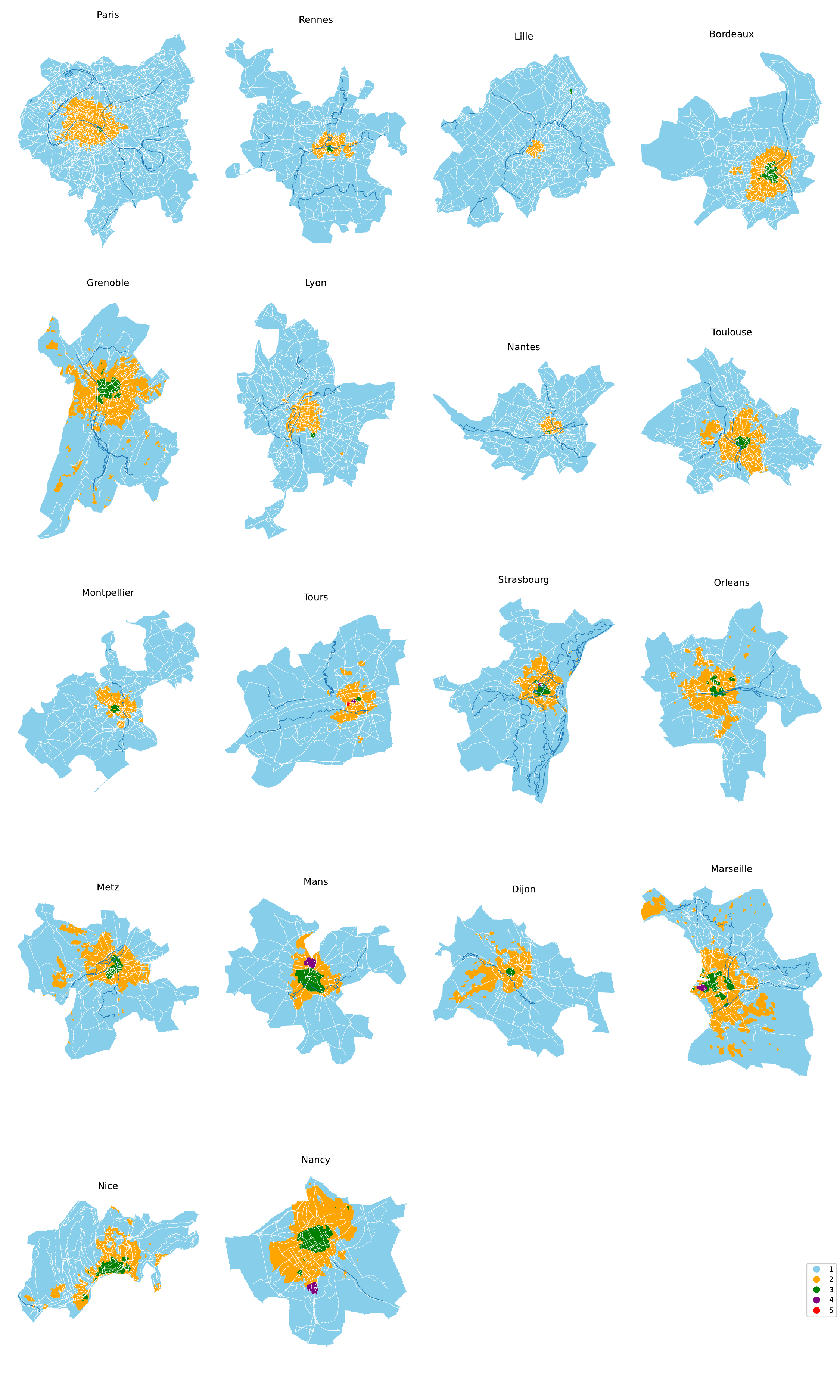}
	\caption[Cluster maps for the week at local level]{
	Cluster maps for the week (Monday--Thursday) at the local level: Paris--Nancy. The maps depict the spatial arrangement of the clusters. Clusters are given in size order where the largest is given the smallest number.
}
	\label{fig:usv-all_plot_all_cluster_maps_week_local}
\end{figure*}

\begin{figure*}[htp]
	\centering
	\includegraphics[height=0.8\textheight]{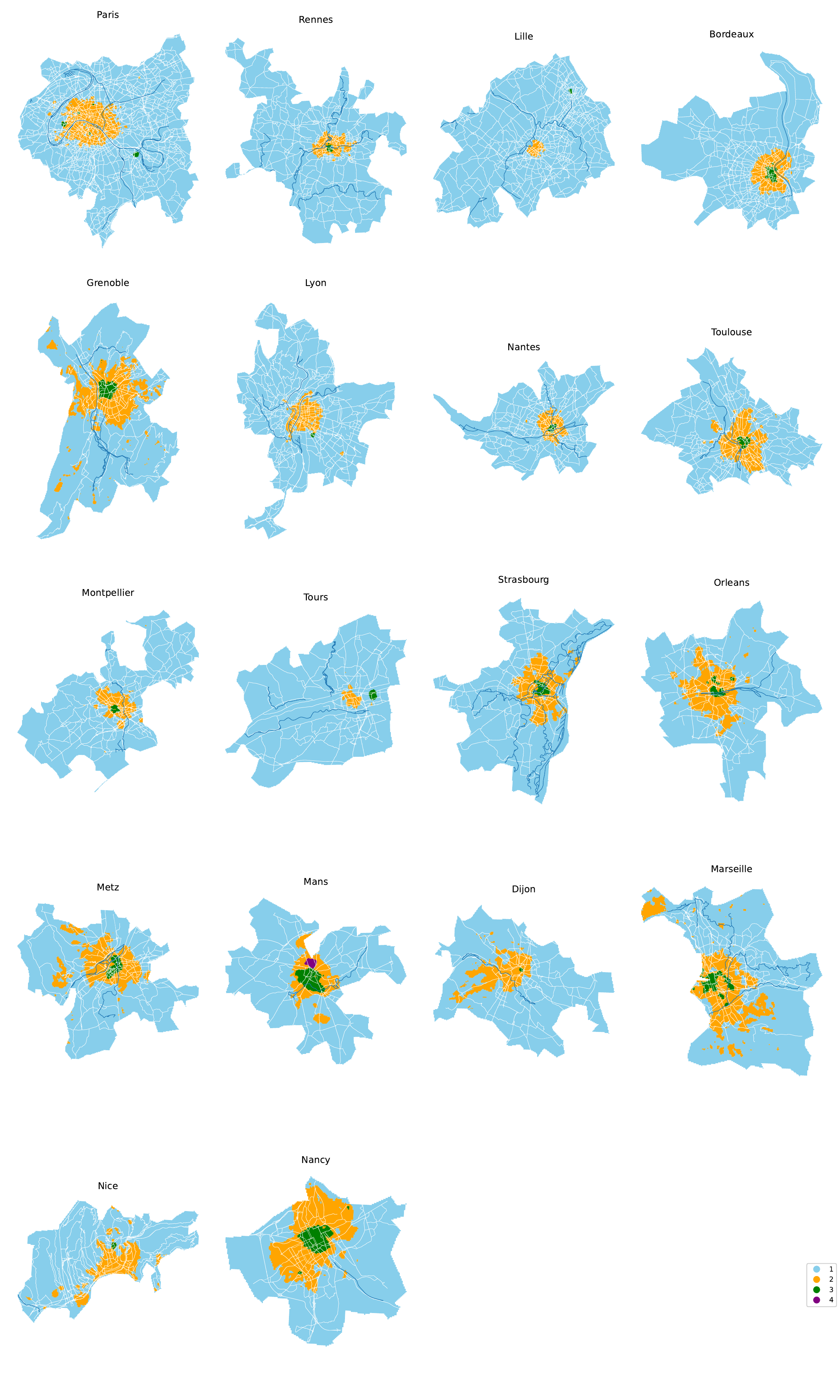}
	\caption[Cluster maps for the weekend at local level]{
	Cluster maps for the weekend (Friday--Sunday) at the local level: Paris--Nancy. The maps depict the spatial arrangement of the clusters. Clusters are given in size order where the largest is given the smallest number.
}
	\label{fig:usv-all_plot_all_cluster_maps_weekend_local}
\end{figure*}

\begin{figure*}[htp]
	\centering
	\includegraphics[height=0.8\textheight]{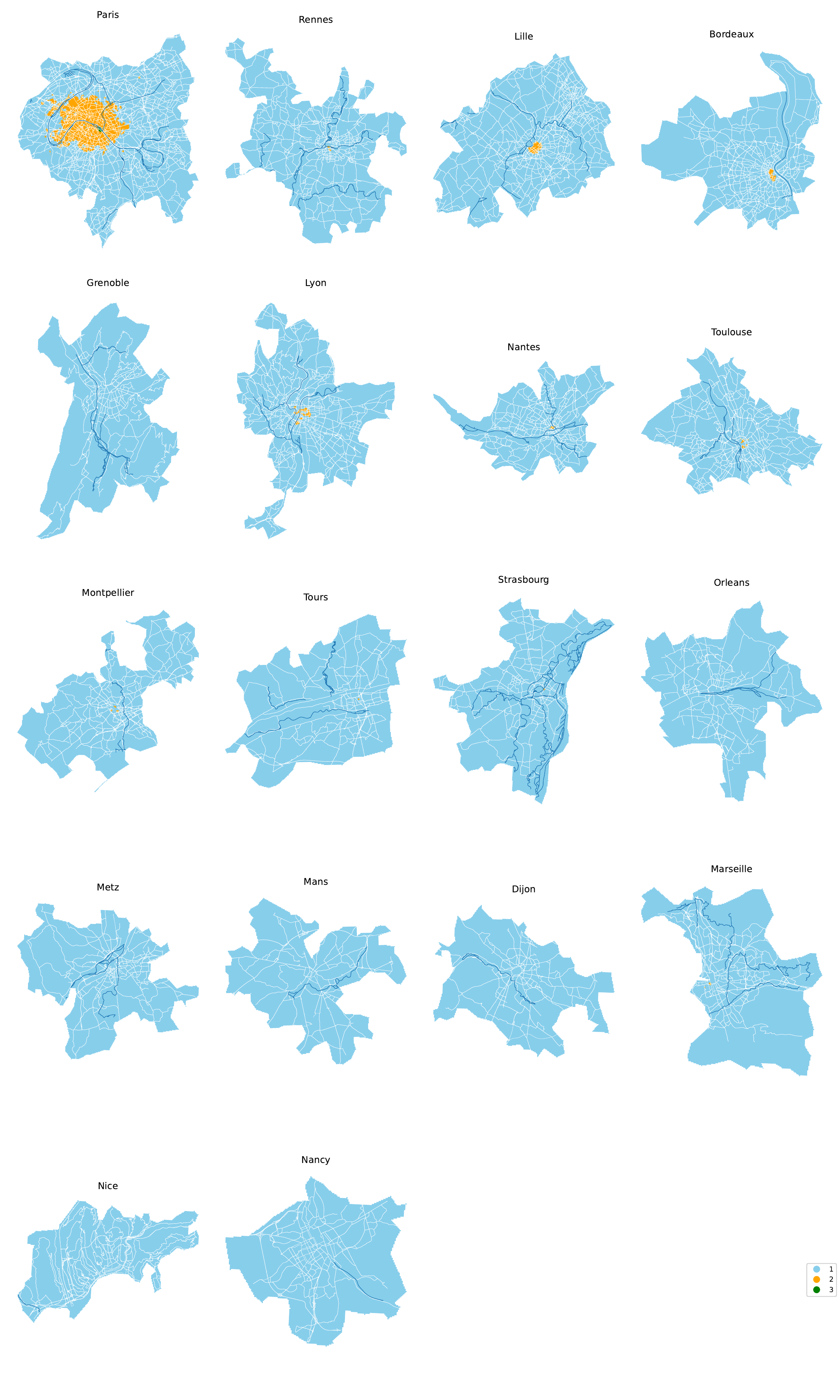}
	\caption[Cluster maps for the week at global level]{
	Cluster maps for the week (Monday--Thursday) at the global level: Paris--Nancy. The maps depict the spatial arrangement of the clusters. Clusters are given in size order where the largest is given the smallest number.
}
	\label{fig:usv-all_plot_all_cluster_maps_week_global}
\end{figure*}

\begin{figure*}[htp]
	\centering
	\includegraphics[height=0.8\textheight]{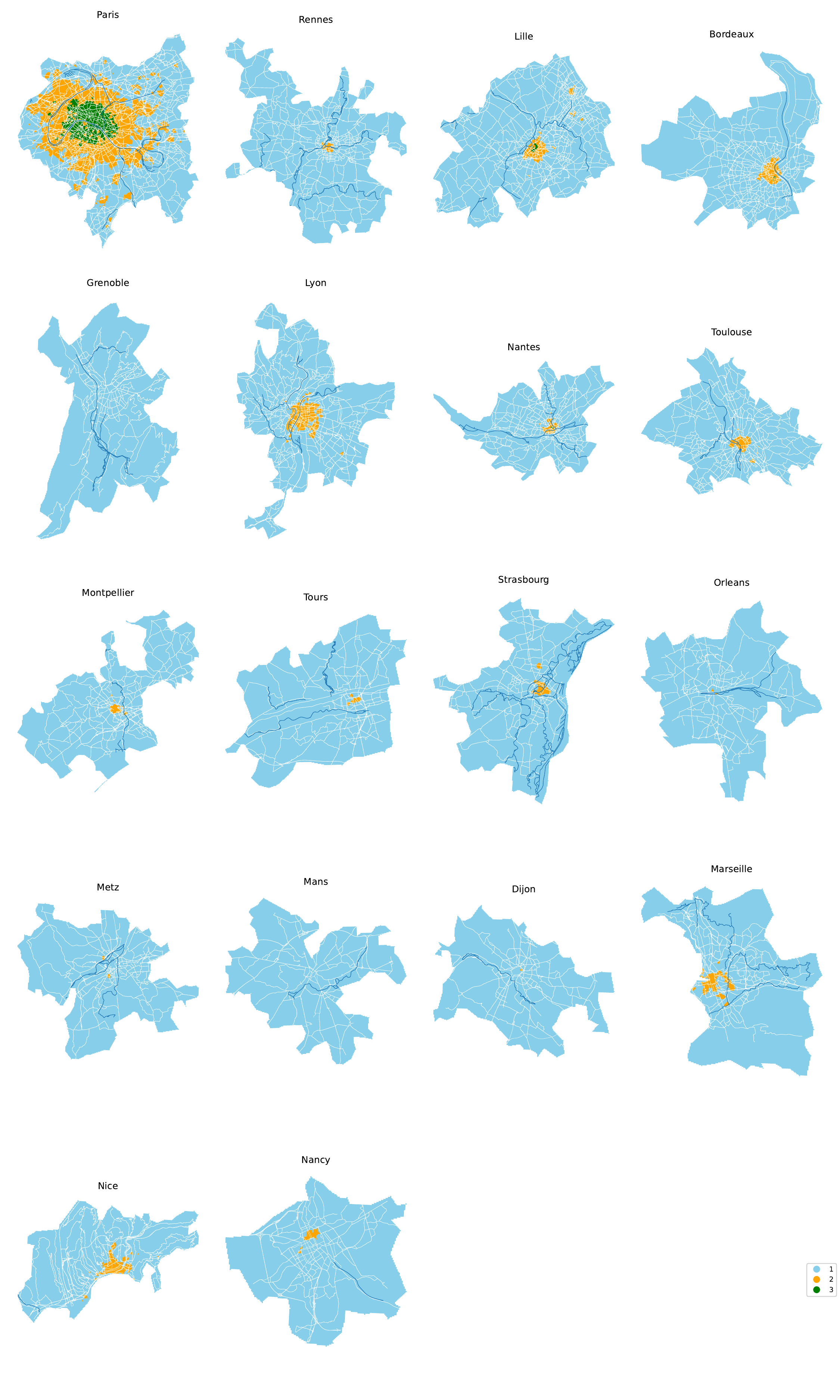}
	\caption[Cluster maps for the weekend at global level]{
	Cluster maps for the weekend (Friday--Sunday) at the global level: Paris--Nancy. The maps depict the spatial arrangement of the clusters. Clusters are given in size order where the largest is given the smallest number.
}
	\label{fig:usv-all_plot_all_cluster_maps_weekend_global}
\end{figure*}

\begin{figure*}[htp]
    \centering
	\includegraphics[height=0.8\textheight]{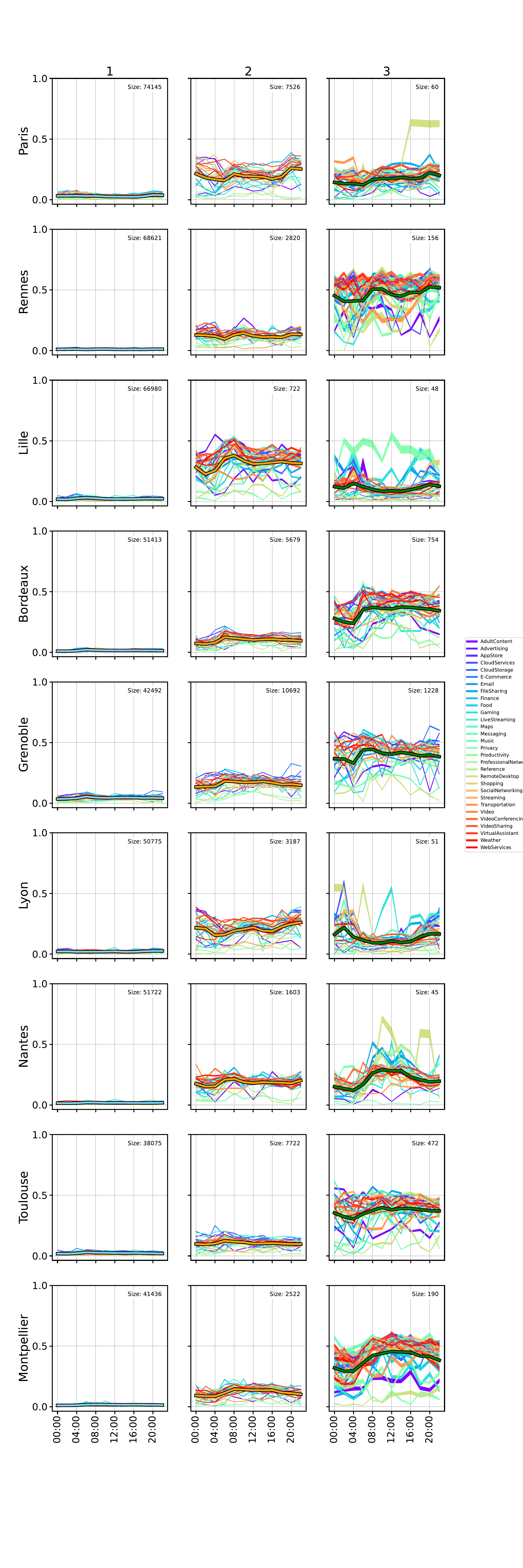}
	\caption[Time series for the week at local level: Paris--Montpellier]{
        Time series for the week (Monday--Thursday) at the local level: Paris--Montpellier. The time series contains the scaled app usage for each category. Each cluster centre is visualised at the forefront of these. The size of the cluster is shown in the top corner of each figure. Clusters are given in size order where the largest is given the smallest number.
}
	\label{fig:usv-all_plot_local_timeseries_week_first_half}
\end{figure*}

\begin{figure*}[htp]
    \centering
	\includegraphics[height=0.8\textheight]{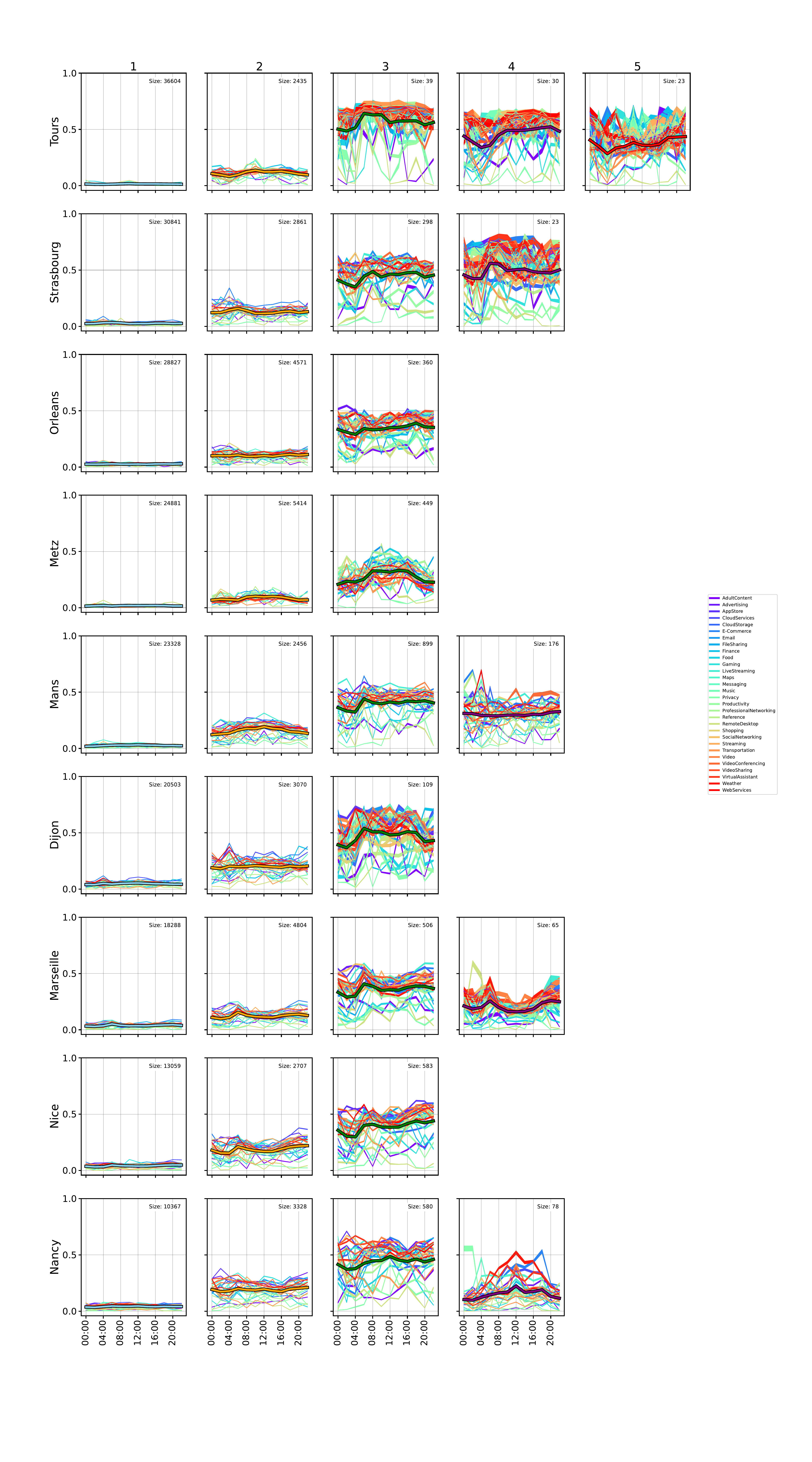}
	\caption[Time series for the week at local level: Tours--Nancy]{
        Time series for the week (Monday--Thursday) at the local level: Tours--Nancy. The time series contains the scaled app usage for each category. Each cluster centre is visualised at the forefront of these. The size of the cluster is shown in the top corner of each figure. Clusters are given in size order where the largest is given the smallest number.
}
	\label{fig:usv-all_plot_local_timeseries_week_second_half}
\end{figure*}

\begin{figure*}[htp]
    \centering
	\includegraphics[height=0.8\textheight]{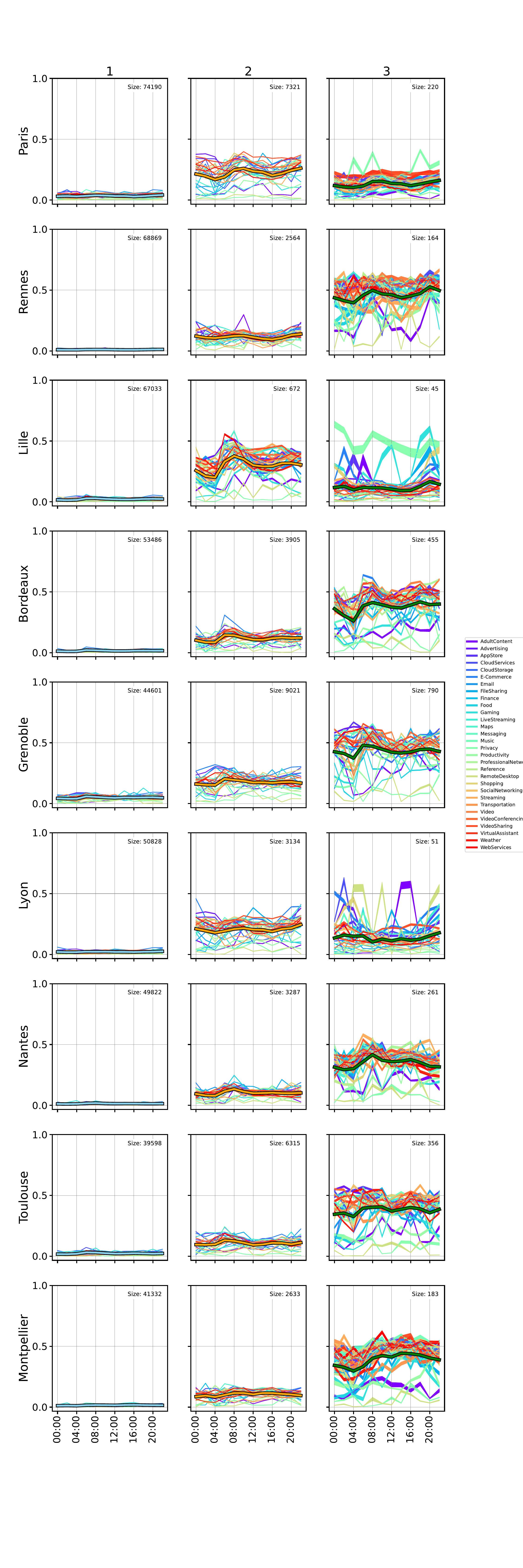}
	\caption[Time series for the weekend at local level: Paris--Montpellier]{
        Time series for the weekend (Friday--Sunday) at the local level: Paris--Montpellier. The time series contains the scaled app usage for each category. Each cluster centre is visualised at the forefront of these. The size of the cluster is shown in the top corner of each figure. Clusters are given in size order where the largest is given the smallest number.
    }
	\label{fig:usv-usv-all_plot_local_timeseries_weekend_first_half}
\end{figure*}

\begin{figure*}[htp]
    \centering
	\includegraphics[height=0.8\textheight]{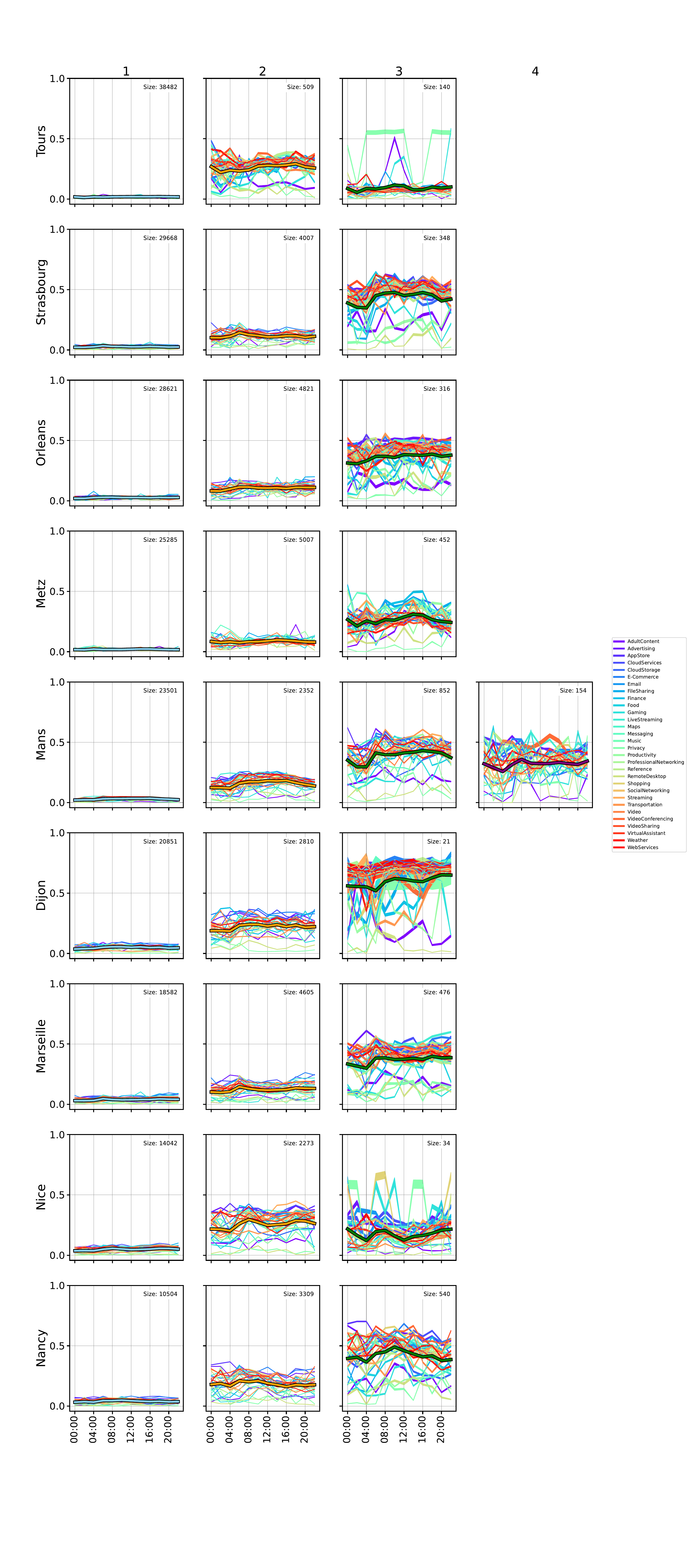}
	\caption[Time series for the weekend at local level: Tours--Nancy]{
        Time series for the weekend (Friday--Sunday) at the local level: Tours--Nancy. The time series contains the scaled app usage for each category. Each cluster centre is visualised at the forefront of these. The size of the cluster is shown in the top corner of each figure. Clusters are given in size order where the largest is given the smallest number.
}
	\label{fig:all_plot_local_timeseries_weekend_second_half}
\end{figure*}

\begin{figure}[htbp]
    \centering
    \begin{subfigure}[b]{0.9\textwidth}
        \centering
        \includegraphics[width=\textwidth]{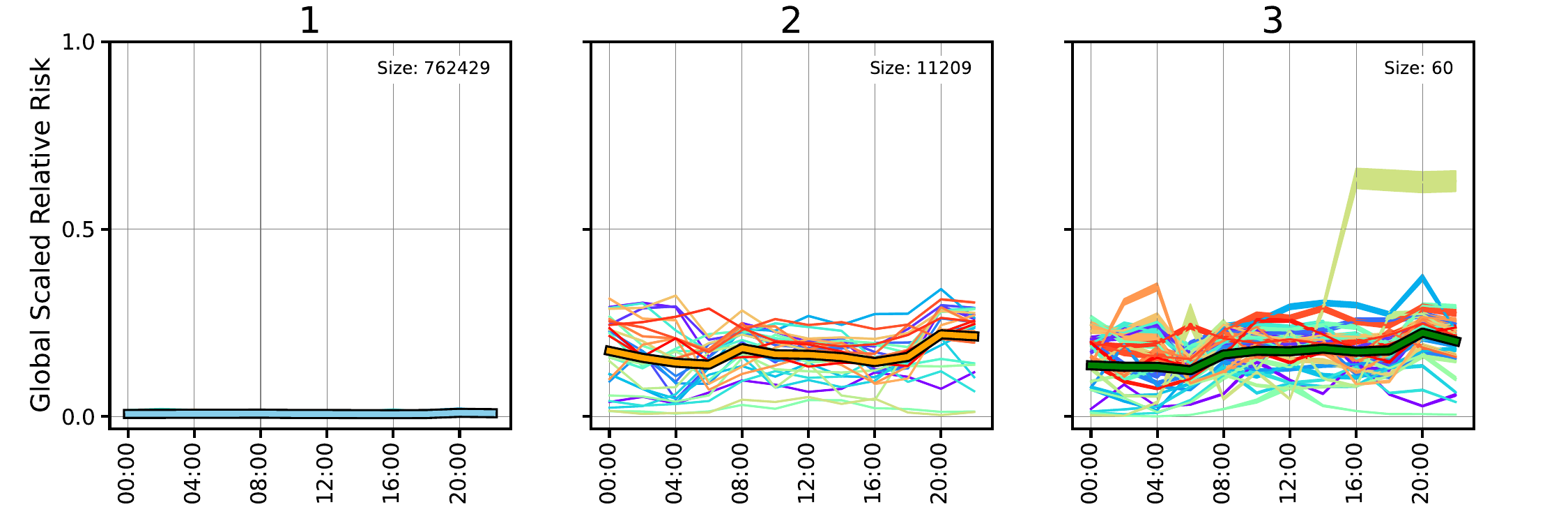}
        \caption[Time series for the week at global level: Paris--Nancy]{
            Week (Monday--Thursday)
        }
        \label{fig:usv-all_plot_global_timeseries_week}
    \end{subfigure}

    \vspace{1cm}

    \begin{subfigure}[b]{0.9\textwidth}
        \centering
        \includegraphics[width=\textwidth]{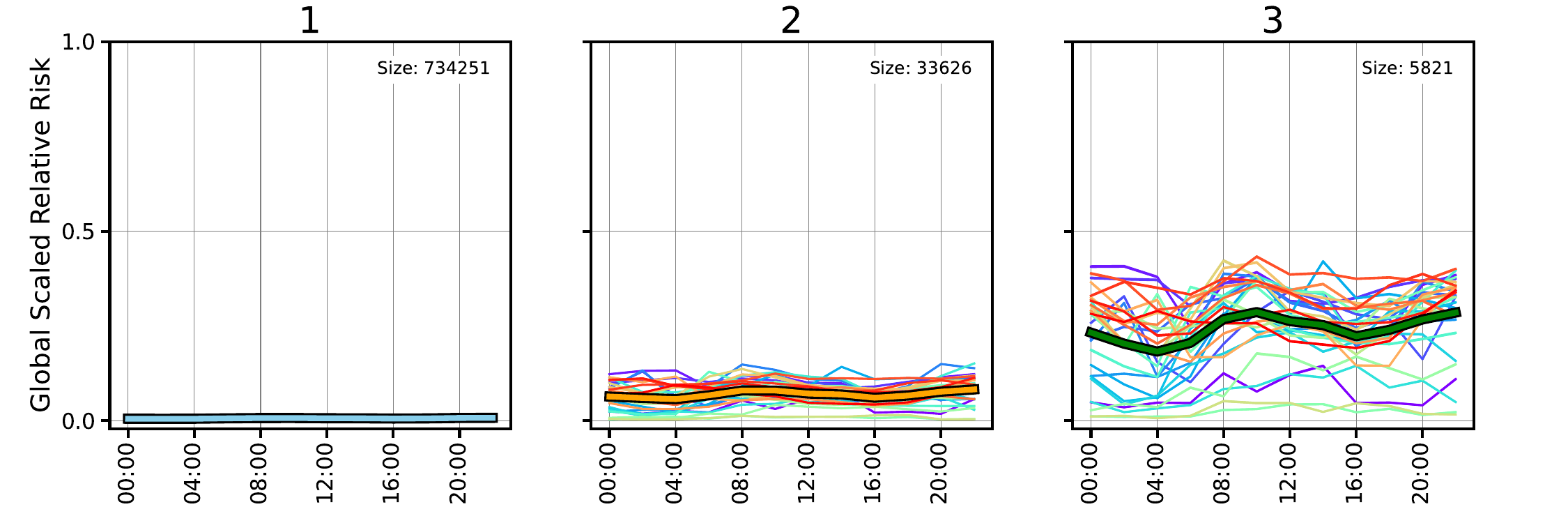}
        \caption[Time series for the weekend at global level: Paris--Nancy]{
            Weekend (Friday--Sunday)
        }
        \label{fig:usv-all_plot_global_timeseries_weekend}
    \end{subfigure}
    \caption[Time series for the week and weekend at global level: Paris--Nancy]{
        Time series for the week- (\subref{fig:usv-all_plot_global_timeseries_week}) (Monday--Thursday) and weekend-scale (\subref{fig:usv-all_plot_global_timeseries_weekend}) (Friday--Sunday) at the global level: Paris--Nancy. The time series contains the scaled app usage for each category. Each cluster centre is visualised at the forefront of these. The size of the cluster is shown in the top corner of each figure. Clusters are given in size order where the largest is given the smallest number.
    }
    \label{fig:usv-all_plot_global_timeseries}
\end{figure}

\begin{table}
\centering
\caption[Total diversity coefficient estimates for local-level multinomial logistic regression predictive model]{Total diversity coefficient estimates for local-level multinomial logistic regression predictive model for the week and weekend and by cluster and by city (Paris--Nancy shown in size order). The figure contains the strength of the association between each of the covariates. Coefficients are coloured according to the strength of the relationship. Data preprocessing was applied to the \texttt{NetMob23} dataset~\citep{martinez-duriveNetMob23DatasetHighresolution2023}. City names and Week/Weekend labels are abbreviated for brevity.}
\label{tab:total-diversity-coefficients}
\begin{tabular}{llrrrrr}
\toprule
 & Class & 1 & 2 & 3 & 4 & 5 \\
City & WK/WKND &  &  &  &  &  \\
\midrule
\multirow[c]{2}{*}{Pari} & WK & {\cellcolor[HTML]{779AF7}} \color[HTML]{F1F1F1} \color{black} -1.03 & {\cellcolor[HTML]{F0CDBB}} \color[HTML]{000000} \color{black} 0.23 & {\cellcolor[HTML]{F29274}} \color[HTML]{F1F1F1} \color{black} 0.80 & {\cellcolor[HTML]{000000}} \color[HTML]{F1F1F1} {\cellcolor[HTML]{FFFFFF}} \color{black} --- & {\cellcolor[HTML]{000000}} \color[HTML]{F1F1F1} {\cellcolor[HTML]{FFFFFF}} \color{black} --- \\
 & WKND & {\cellcolor[HTML]{A2C1FF}} \color[HTML]{000000} \color{black} -0.65 & {\cellcolor[HTML]{F6A283}} \color[HTML]{000000} \color{black} 0.67 & {\cellcolor[HTML]{DFDBD9}} \color[HTML]{000000} \color{black} -0.02 & {\cellcolor[HTML]{000000}} \color[HTML]{F1F1F1} {\cellcolor[HTML]{FFFFFF}} \color{black} --- & {\cellcolor[HTML]{000000}} \color[HTML]{F1F1F1} {\cellcolor[HTML]{FFFFFF}} \color{black} --- \\
\cline{1-7}
\multirow[c]{2}{*}{Renn} & WK & {\cellcolor[HTML]{4961D2}} \color[HTML]{F1F1F1} \color{black} -1.47 & {\cellcolor[HTML]{F0CDBB}} \color[HTML]{000000} \color{black} 0.22 & {\cellcolor[HTML]{D24B40}} \color[HTML]{F1F1F1} \color{black} 1.25 & {\cellcolor[HTML]{000000}} \color[HTML]{F1F1F1} {\cellcolor[HTML]{FFFFFF}} \color{black} --- & {\cellcolor[HTML]{000000}} \color[HTML]{F1F1F1} {\cellcolor[HTML]{FFFFFF}} \color{black} --- \\
 & WKND & {\cellcolor[HTML]{465ECF}} \color[HTML]{F1F1F1} \color{black} -1.49 & {\cellcolor[HTML]{F0CDBB}} \color[HTML]{000000} \color{black} 0.22 & {\cellcolor[HTML]{D0473D}} \color[HTML]{F1F1F1} \color{black} 1.27 & {\cellcolor[HTML]{000000}} \color[HTML]{F1F1F1} {\cellcolor[HTML]{FFFFFF}} \color{black} --- & {\cellcolor[HTML]{000000}} \color[HTML]{F1F1F1} {\cellcolor[HTML]{FFFFFF}} \color{black} --- \\
\cline{1-7}
\multirow[c]{2}{*}{Lill} & WK & {\cellcolor[HTML]{94B6FF}} \color[HTML]{000000} \color{black} -0.77 & {\cellcolor[HTML]{E8765C}} \color[HTML]{F1F1F1} \color{black} 0.99 & {\cellcolor[HTML]{CDD9EC}} \color[HTML]{000000} \color{black} -0.23 & {\cellcolor[HTML]{000000}} \color[HTML]{F1F1F1} {\cellcolor[HTML]{FFFFFF}} \color{black} --- & {\cellcolor[HTML]{000000}} \color[HTML]{F1F1F1} {\cellcolor[HTML]{FFFFFF}} \color{black} --- \\
 & WKND & {\cellcolor[HTML]{93B5FE}} \color[HTML]{000000} \color{black} -0.78 & {\cellcolor[HTML]{EC8165}} \color[HTML]{F1F1F1} \color{black} 0.92 & {\cellcolor[HTML]{D6DCE4}} \color[HTML]{000000} \color{black} -0.13 & {\cellcolor[HTML]{000000}} \color[HTML]{F1F1F1} {\cellcolor[HTML]{FFFFFF}} \color{black} --- & {\cellcolor[HTML]{000000}} \color[HTML]{F1F1F1} {\cellcolor[HTML]{FFFFFF}} \color{black} --- \\
\cline{1-7}
\multirow[c]{2}{*}{Bord} & WK & {\cellcolor[HTML]{8FB1FE}} \color[HTML]{000000} \color{black} -0.81 & {\cellcolor[HTML]{F1CDBA}} \color[HTML]{000000} \color{black} 0.23 & {\cellcolor[HTML]{F7AC8E}} \color[HTML]{000000} \color{black} 0.58 & {\cellcolor[HTML]{000000}} \color[HTML]{F1F1F1} {\cellcolor[HTML]{FFFFFF}} \color{black} --- & {\cellcolor[HTML]{000000}} \color[HTML]{F1F1F1} {\cellcolor[HTML]{FFFFFF}} \color{black} --- \\
 & WKND & {\cellcolor[HTML]{8BADFD}} \color[HTML]{000000} \color{black} -0.86 & {\cellcolor[HTML]{EAD4C8}} \color[HTML]{000000} \color{black} 0.13 & {\cellcolor[HTML]{F49A7B}} \color[HTML]{000000} \color{black} 0.73 & {\cellcolor[HTML]{000000}} \color[HTML]{F1F1F1} {\cellcolor[HTML]{FFFFFF}} \color{black} --- & {\cellcolor[HTML]{000000}} \color[HTML]{F1F1F1} {\cellcolor[HTML]{FFFFFF}} \color{black} --- \\
\cline{1-7}
\multirow[c]{2}{*}{Gren} & WK & {\cellcolor[HTML]{779AF7}} \color[HTML]{F1F1F1} \color{black} -1.03 & {\cellcolor[HTML]{E4D9D2}} \color[HTML]{000000} \color{black} 0.04 & {\cellcolor[HTML]{E7745B}} \color[HTML]{F1F1F1} \color{black} 1.00 & {\cellcolor[HTML]{000000}} \color[HTML]{F1F1F1} {\cellcolor[HTML]{FFFFFF}} \color{black} --- & {\cellcolor[HTML]{000000}} \color[HTML]{F1F1F1} {\cellcolor[HTML]{FFFFFF}} \color{black} --- \\
 & WKND & {\cellcolor[HTML]{6F92F3}} \color[HTML]{F1F1F1} \color{black} -1.11 & {\cellcolor[HTML]{E1DAD6}} \color[HTML]{000000} \color{black} -0.00 & {\cellcolor[HTML]{DF634E}} \color[HTML]{F1F1F1} \color{black} 1.11 & {\cellcolor[HTML]{000000}} \color[HTML]{F1F1F1} {\cellcolor[HTML]{FFFFFF}} \color{black} --- & {\cellcolor[HTML]{000000}} \color[HTML]{F1F1F1} {\cellcolor[HTML]{FFFFFF}} \color{black} --- \\
\cline{1-7}
\multirow[c]{2}{*}{Lyon} & WK & {\cellcolor[HTML]{7DA0F9}} \color[HTML]{F1F1F1} \color{black} -0.97 & {\cellcolor[HTML]{F5C2AA}} \color[HTML]{000000} \color{black} 0.36 & {\cellcolor[HTML]{F7A889}} \color[HTML]{000000} \color{black} 0.61 & {\cellcolor[HTML]{000000}} \color[HTML]{F1F1F1} {\cellcolor[HTML]{FFFFFF}} \color{black} --- & {\cellcolor[HTML]{000000}} \color[HTML]{F1F1F1} {\cellcolor[HTML]{FFFFFF}} \color{black} --- \\
 & WKND & {\cellcolor[HTML]{7DA0F9}} \color[HTML]{F1F1F1} \color{black} -0.98 & {\cellcolor[HTML]{F5C1A9}} \color[HTML]{000000} \color{black} 0.37 & {\cellcolor[HTML]{F7A98B}} \color[HTML]{000000} \color{black} 0.61 & {\cellcolor[HTML]{000000}} \color[HTML]{F1F1F1} {\cellcolor[HTML]{FFFFFF}} \color{black} --- & {\cellcolor[HTML]{000000}} \color[HTML]{F1F1F1} {\cellcolor[HTML]{FFFFFF}} \color{black} --- \\
\cline{1-7}
\multirow[c]{2}{*}{Nant} & WK & {\cellcolor[HTML]{8CAFFE}} \color[HTML]{000000} \color{black} -0.84 & {\cellcolor[HTML]{F7B89C}} \color[HTML]{000000} \color{black} 0.47 & {\cellcolor[HTML]{F5C2AA}} \color[HTML]{000000} \color{black} 0.36 & {\cellcolor[HTML]{000000}} \color[HTML]{F1F1F1} {\cellcolor[HTML]{FFFFFF}} \color{black} --- & {\cellcolor[HTML]{000000}} \color[HTML]{F1F1F1} {\cellcolor[HTML]{FFFFFF}} \color{black} --- \\
 & WKND & {\cellcolor[HTML]{779AF7}} \color[HTML]{F1F1F1} \color{black} -1.03 & {\cellcolor[HTML]{E6D7CF}} \color[HTML]{000000} \color{black} 0.07 & {\cellcolor[HTML]{E97A5F}} \color[HTML]{F1F1F1} \color{black} 0.96 & {\cellcolor[HTML]{000000}} \color[HTML]{F1F1F1} {\cellcolor[HTML]{FFFFFF}} \color{black} --- & {\cellcolor[HTML]{000000}} \color[HTML]{F1F1F1} {\cellcolor[HTML]{FFFFFF}} \color{black} --- \\
\cline{1-7}
\multirow[c]{2}{*}{Toul} & WK & {\cellcolor[HTML]{89ACFD}} \color[HTML]{000000} \color{black} -0.87 & {\cellcolor[HTML]{DDDCDC}} \color[HTML]{000000} \color{black} -0.04 & {\cellcolor[HTML]{EC8165}} \color[HTML]{F1F1F1} \color{black} 0.92 & {\cellcolor[HTML]{000000}} \color[HTML]{F1F1F1} {\cellcolor[HTML]{FFFFFF}} \color{black} --- & {\cellcolor[HTML]{000000}} \color[HTML]{F1F1F1} {\cellcolor[HTML]{FFFFFF}} \color{black} --- \\
 & WKND & {\cellcolor[HTML]{84A7FC}} \color[HTML]{F1F1F1} \color{black} -0.92 & {\cellcolor[HTML]{DEDCDB}} \color[HTML]{000000} \color{black} -0.03 & {\cellcolor[HTML]{EB7D62}} \color[HTML]{F1F1F1} \color{black} 0.94 & {\cellcolor[HTML]{000000}} \color[HTML]{F1F1F1} {\cellcolor[HTML]{FFFFFF}} \color{black} --- & {\cellcolor[HTML]{000000}} \color[HTML]{F1F1F1} {\cellcolor[HTML]{FFFFFF}} \color{black} --- \\
\cline{1-7}
\multirow[c]{2}{*}{Mont} & WK & {\cellcolor[HTML]{5A78E4}} \color[HTML]{F1F1F1} \color{black} -1.29 & {\cellcolor[HTML]{EBD3C6}} \color[HTML]{000000} \color{black} 0.14 & {\cellcolor[HTML]{DA5A49}} \color[HTML]{F1F1F1} \color{black} 1.16 & {\cellcolor[HTML]{000000}} \color[HTML]{F1F1F1} {\cellcolor[HTML]{FFFFFF}} \color{black} --- & {\cellcolor[HTML]{000000}} \color[HTML]{F1F1F1} {\cellcolor[HTML]{FFFFFF}} \color{black} --- \\
 & WKND & {\cellcolor[HTML]{5875E1}} \color[HTML]{F1F1F1} \color{black} -1.32 & {\cellcolor[HTML]{EAD5C9}} \color[HTML]{000000} \color{black} 0.12 & {\cellcolor[HTML]{D75445}} \color[HTML]{F1F1F1} \color{black} 1.20 & {\cellcolor[HTML]{000000}} \color[HTML]{F1F1F1} {\cellcolor[HTML]{FFFFFF}} \color{black} --- & {\cellcolor[HTML]{000000}} \color[HTML]{F1F1F1} {\cellcolor[HTML]{FFFFFF}} \color{black} --- \\
\cline{1-7}
\multirow[c]{2}{*}{Tour} & WK & {\cellcolor[HTML]{5A78E4}} \color[HTML]{F1F1F1} \color{black} -1.29 & {\cellcolor[HTML]{D9DCE1}} \color[HTML]{000000} \color{black} -0.09 & {\cellcolor[HTML]{F18D6F}} \color[HTML]{F1F1F1} \color{black} 0.83 & {\cellcolor[HTML]{F7AA8C}} \color[HTML]{000000} \color{black} 0.60 & {\cellcolor[HTML]{DDDCDC}} \color[HTML]{000000} \color{black} -0.05 \\
 & WKND & {\cellcolor[HTML]{84A7FC}} \color[HTML]{F1F1F1} \color{black} -0.92 & {\cellcolor[HTML]{F59D7E}} \color[HTML]{000000} \color{black} 0.71 & {\cellcolor[HTML]{EFCEBD}} \color[HTML]{000000} \color{black} 0.21 & {\cellcolor[HTML]{000000}} \color[HTML]{F1F1F1} {\cellcolor[HTML]{FFFFFF}} \color{black} --- & {\cellcolor[HTML]{000000}} \color[HTML]{F1F1F1} {\cellcolor[HTML]{FFFFFF}} \color{black} --- \\
\cline{1-7}
\multirow[c]{2}{*}{Stra} & WK & {\cellcolor[HTML]{3B4CC0}} \color[HTML]{F1F1F1} \color{black} -1.62 & {\cellcolor[HTML]{AEC9FC}} \color[HTML]{000000} \color{black} -0.54 & {\cellcolor[HTML]{F6A586}} \color[HTML]{000000} \color{black} 0.64 & {\cellcolor[HTML]{B40426}} \color[HTML]{F1F1F1} \color{black} 1.52 & {\cellcolor[HTML]{000000}} \color[HTML]{F1F1F1} {\cellcolor[HTML]{FFFFFF}} \color{black} --- \\
 & WKND & {\cellcolor[HTML]{6F92F3}} \color[HTML]{F1F1F1} \color{black} -1.10 & {\cellcolor[HTML]{D8DCE2}} \color[HTML]{000000} \color{black} -0.10 & {\cellcolor[HTML]{D75445}} \color[HTML]{F1F1F1} \color{black} 1.20 & {\cellcolor[HTML]{000000}} \color[HTML]{F1F1F1} {\cellcolor[HTML]{FFFFFF}} \color{black} --- & {\cellcolor[HTML]{000000}} \color[HTML]{F1F1F1} {\cellcolor[HTML]{FFFFFF}} \color{black} --- \\
\cline{1-7}
\multirow[c]{2}{*}{Orle} & WK & {\cellcolor[HTML]{8BADFD}} \color[HTML]{000000} \color{black} -0.85 & {\cellcolor[HTML]{DFDBD9}} \color[HTML]{000000} \color{black} -0.02 & {\cellcolor[HTML]{EF886B}} \color[HTML]{F1F1F1} \color{black} 0.87 & {\cellcolor[HTML]{000000}} \color[HTML]{F1F1F1} {\cellcolor[HTML]{FFFFFF}} \color{black} --- & {\cellcolor[HTML]{000000}} \color[HTML]{F1F1F1} {\cellcolor[HTML]{FFFFFF}} \color{black} --- \\
 & WKND & {\cellcolor[HTML]{92B4FE}} \color[HTML]{000000} \color{black} -0.79 & {\cellcolor[HTML]{E2DAD5}} \color[HTML]{000000} \color{black} 0.02 & {\cellcolor[HTML]{F39475}} \color[HTML]{000000} \color{black} 0.78 & {\cellcolor[HTML]{000000}} \color[HTML]{F1F1F1} {\cellcolor[HTML]{FFFFFF}} \color{black} --- & {\cellcolor[HTML]{000000}} \color[HTML]{F1F1F1} {\cellcolor[HTML]{FFFFFF}} \color{black} --- \\
\cline{1-7}
\multirow[c]{2}{*}{Metz} & WK & {\cellcolor[HTML]{6788EE}} \color[HTML]{F1F1F1} \color{black} -1.18 & {\cellcolor[HTML]{DCDDDD}} \color[HTML]{000000} \color{black} -0.06 & {\cellcolor[HTML]{D44E41}} \color[HTML]{F1F1F1} \color{black} 1.23 & {\cellcolor[HTML]{000000}} \color[HTML]{F1F1F1} {\cellcolor[HTML]{FFFFFF}} \color{black} --- & {\cellcolor[HTML]{000000}} \color[HTML]{F1F1F1} {\cellcolor[HTML]{FFFFFF}} \color{black} --- \\
 & WKND & {\cellcolor[HTML]{6788EE}} \color[HTML]{F1F1F1} \color{black} -1.18 & {\cellcolor[HTML]{DFDBD9}} \color[HTML]{000000} \color{black} -0.02 & {\cellcolor[HTML]{D75445}} \color[HTML]{F1F1F1} \color{black} 1.20 & {\cellcolor[HTML]{000000}} \color[HTML]{F1F1F1} {\cellcolor[HTML]{FFFFFF}} \color{black} --- & {\cellcolor[HTML]{000000}} \color[HTML]{F1F1F1} {\cellcolor[HTML]{FFFFFF}} \color{black} --- \\
\cline{1-7}
\multirow[c]{2}{*}{Mans} & WK & {\cellcolor[HTML]{86A9FC}} \color[HTML]{F1F1F1} \color{black} -0.89 & {\cellcolor[HTML]{EDD1C2}} \color[HTML]{000000} \color{black} 0.17 & {\cellcolor[HTML]{F7B396}} \color[HTML]{000000} \color{black} 0.52 & {\cellcolor[HTML]{EFCFBF}} \color[HTML]{000000} \color{black} 0.20 & {\cellcolor[HTML]{000000}} \color[HTML]{F1F1F1} {\cellcolor[HTML]{FFFFFF}} \color{black} --- \\
 & WKND & {\cellcolor[HTML]{84A7FC}} \color[HTML]{F1F1F1} \color{black} -0.92 & {\cellcolor[HTML]{E8D6CC}} \color[HTML]{000000} \color{black} 0.09 & {\cellcolor[HTML]{F7AC8E}} \color[HTML]{000000} \color{black} 0.58 & {\cellcolor[HTML]{F1CDBA}} \color[HTML]{000000} \color{black} 0.24 & {\cellcolor[HTML]{000000}} \color[HTML]{F1F1F1} {\cellcolor[HTML]{FFFFFF}} \color{black} --- \\
\cline{1-7}
\multirow[c]{2}{*}{Dijo} & WK & {\cellcolor[HTML]{9ABBFF}} \color[HTML]{000000} \color{black} -0.72 & {\cellcolor[HTML]{C9D7F0}} \color[HTML]{000000} \color{black} -0.28 & {\cellcolor[HTML]{E7745B}} \color[HTML]{F1F1F1} \color{black} 1.01 & {\cellcolor[HTML]{000000}} \color[HTML]{F1F1F1} {\cellcolor[HTML]{FFFFFF}} \color{black} --- & {\cellcolor[HTML]{000000}} \color[HTML]{F1F1F1} {\cellcolor[HTML]{FFFFFF}} \color{black} --- \\
 & WKND & {\cellcolor[HTML]{A2C1FF}} \color[HTML]{000000} \color{black} -0.65 & {\cellcolor[HTML]{D3DBE7}} \color[HTML]{000000} \color{black} -0.17 & {\cellcolor[HTML]{F18F71}} \color[HTML]{F1F1F1} \color{black} 0.82 & {\cellcolor[HTML]{000000}} \color[HTML]{F1F1F1} {\cellcolor[HTML]{FFFFFF}} \color{black} --- & {\cellcolor[HTML]{000000}} \color[HTML]{F1F1F1} {\cellcolor[HTML]{FFFFFF}} \color{black} --- \\
\cline{1-7}
\multirow[c]{2}{*}{Mars} & WK & {\cellcolor[HTML]{96B7FF}} \color[HTML]{000000} \color{black} -0.76 & {\cellcolor[HTML]{C0D4F5}} \color[HTML]{000000} \color{black} -0.36 & {\cellcolor[HTML]{F7A98B}} \color[HTML]{000000} \color{black} 0.61 & {\cellcolor[HTML]{F7B497}} \color[HTML]{000000} \color{black} 0.50 & {\cellcolor[HTML]{000000}} \color[HTML]{F1F1F1} {\cellcolor[HTML]{FFFFFF}} \color{black} --- \\
 & WKND & {\cellcolor[HTML]{A9C6FD}} \color[HTML]{000000} \color{black} -0.59 & {\cellcolor[HTML]{D2DBE8}} \color[HTML]{000000} \color{black} -0.18 & {\cellcolor[HTML]{F39577}} \color[HTML]{000000} \color{black} 0.77 & {\cellcolor[HTML]{000000}} \color[HTML]{F1F1F1} {\cellcolor[HTML]{FFFFFF}} \color{black} --- & {\cellcolor[HTML]{000000}} \color[HTML]{F1F1F1} {\cellcolor[HTML]{FFFFFF}} \color{black} --- \\
\cline{1-7}
\multirow[c]{2}{*}{Nice} & WK & {\cellcolor[HTML]{92B4FE}} \color[HTML]{000000} \color{black} -0.79 & {\cellcolor[HTML]{E2DAD5}} \color[HTML]{000000} \color{black} 0.02 & {\cellcolor[HTML]{F39475}} \color[HTML]{000000} \color{black} 0.78 & {\cellcolor[HTML]{000000}} \color[HTML]{F1F1F1} {\cellcolor[HTML]{FFFFFF}} \color{black} --- & {\cellcolor[HTML]{000000}} \color[HTML]{F1F1F1} {\cellcolor[HTML]{FFFFFF}} \color{black} --- \\
 & WKND & {\cellcolor[HTML]{A5C3FE}} \color[HTML]{000000} \color{black} -0.62 & {\cellcolor[HTML]{F5C0A7}} \color[HTML]{000000} \color{black} 0.38 & {\cellcolor[HTML]{F1CDBA}} \color[HTML]{000000} \color{black} 0.24 & {\cellcolor[HTML]{000000}} \color[HTML]{F1F1F1} {\cellcolor[HTML]{FFFFFF}} \color{black} --- & {\cellcolor[HTML]{000000}} \color[HTML]{F1F1F1} {\cellcolor[HTML]{FFFFFF}} \color{black} --- \\
\cline{1-7}
\multirow[c]{2}{*}{Nanc} & WK & {\cellcolor[HTML]{80A3FA}} \color[HTML]{F1F1F1} \color{black} -0.95 & {\cellcolor[HTML]{D7DCE3}} \color[HTML]{000000} \color{black} -0.11 & {\cellcolor[HTML]{F7BA9F}} \color[HTML]{000000} \color{black} 0.44 & {\cellcolor[HTML]{F7A889}} \color[HTML]{000000} \color{black} 0.62 & {\cellcolor[HTML]{000000}} \color[HTML]{F1F1F1} {\cellcolor[HTML]{FFFFFF}} \color{black} --- \\
 & WKND & {\cellcolor[HTML]{9BBCFF}} \color[HTML]{000000} \color{black} -0.70 & {\cellcolor[HTML]{E7D7CE}} \color[HTML]{000000} \color{black} 0.08 & {\cellcolor[HTML]{F7A889}} \color[HTML]{000000} \color{black} 0.63 & {\cellcolor[HTML]{000000}} \color[HTML]{F1F1F1} {\cellcolor[HTML]{FFFFFF}} \color{black} --- & {\cellcolor[HTML]{000000}} \color[HTML]{F1F1F1} {\cellcolor[HTML]{FFFFFF}} \color{black} --- \\
\cline{1-7}
\bottomrule
\end{tabular}
\end{table}

\begin{table}
\centering
\caption[Commercial service diversity coefficient estimates for local-level multinomial logistic regression predictive model]{Commercial service diversity coefficient estimates for local-level multinomial logistic regression predictive model for the week and weekend and by cluster and by city (Paris--Nancy shown in size order). The figure contains the strength of the association between each of the covariates. Coefficients are coloured according to the strength of the relationship. Data preprocessing was applied to the \texttt{NetMob23} dataset~\citep{martinez-duriveNetMob23DatasetHighresolution2023}. City names and Week/Weekend labels are abbreviated for brevity.}
\label{tab:commercial-service-diversity-coefficients}
\begin{tabular}{llrrrrr}
\toprule
 & Class & 1 & 2 & 3 & 4 & 5 \\
City & WK/WKND &  &  &  &  &  \\
\midrule
\multirow[c]{2}{*}{Pari} & WK & {\cellcolor[HTML]{C7D7F0}} \color[HTML]{000000} \color{black} -0.29 & {\cellcolor[HTML]{D8DCE2}} \color[HTML]{000000} \color{black} -0.11 & {\cellcolor[HTML]{F6BFA6}} \color[HTML]{000000} \color{black} 0.40 & {\cellcolor[HTML]{000000}} \color[HTML]{F1F1F1} {\cellcolor[HTML]{FFFFFF}} \color{black} --- & {\cellcolor[HTML]{000000}} \color[HTML]{F1F1F1} {\cellcolor[HTML]{FFFFFF}} \color{black} --- \\
 & WKND & {\cellcolor[HTML]{D3DBE7}} \color[HTML]{000000} \color{black} -0.16 & {\cellcolor[HTML]{E5D8D1}} \color[HTML]{000000} \color{black} 0.06 & {\cellcolor[HTML]{E9D5CB}} \color[HTML]{000000} \color{black} 0.10 & {\cellcolor[HTML]{000000}} \color[HTML]{F1F1F1} {\cellcolor[HTML]{FFFFFF}} \color{black} --- & {\cellcolor[HTML]{000000}} \color[HTML]{F1F1F1} {\cellcolor[HTML]{FFFFFF}} \color{black} --- \\
\cline{1-7}
\multirow[c]{2}{*}{Renn} & WK & {\cellcolor[HTML]{F39778}} \color[HTML]{000000} \color{black} 0.75 & {\cellcolor[HTML]{C4D5F3}} \color[HTML]{000000} \color{black} -0.32 & {\cellcolor[HTML]{BAD0F8}} \color[HTML]{000000} \color{black} -0.43 & {\cellcolor[HTML]{000000}} \color[HTML]{F1F1F1} {\cellcolor[HTML]{FFFFFF}} \color{black} --- & {\cellcolor[HTML]{000000}} \color[HTML]{F1F1F1} {\cellcolor[HTML]{FFFFFF}} \color{black} --- \\
 & WKND & {\cellcolor[HTML]{F4987A}} \color[HTML]{000000} \color{black} 0.74 & {\cellcolor[HTML]{C3D5F4}} \color[HTML]{000000} \color{black} -0.34 & {\cellcolor[HTML]{BCD2F7}} \color[HTML]{000000} \color{black} -0.40 & {\cellcolor[HTML]{000000}} \color[HTML]{F1F1F1} {\cellcolor[HTML]{FFFFFF}} \color{black} --- & {\cellcolor[HTML]{000000}} \color[HTML]{F1F1F1} {\cellcolor[HTML]{FFFFFF}} \color{black} --- \\
\cline{1-7}
\multirow[c]{2}{*}{Lill} & WK & {\cellcolor[HTML]{ECD3C5}} \color[HTML]{000000} \color{black} 0.15 & {\cellcolor[HTML]{E8D6CC}} \color[HTML]{000000} \color{black} 0.09 & {\cellcolor[HTML]{CCD9ED}} \color[HTML]{000000} \color{black} -0.24 & {\cellcolor[HTML]{000000}} \color[HTML]{F1F1F1} {\cellcolor[HTML]{FFFFFF}} \color{black} --- & {\cellcolor[HTML]{000000}} \color[HTML]{F1F1F1} {\cellcolor[HTML]{FFFFFF}} \color{black} --- \\
 & WKND & {\cellcolor[HTML]{EDD2C3}} \color[HTML]{000000} \color{black} 0.17 & {\cellcolor[HTML]{E6D7CF}} \color[HTML]{000000} \color{black} 0.06 & {\cellcolor[HTML]{CDD9EC}} \color[HTML]{000000} \color{black} -0.23 & {\cellcolor[HTML]{000000}} \color[HTML]{F1F1F1} {\cellcolor[HTML]{FFFFFF}} \color{black} --- & {\cellcolor[HTML]{000000}} \color[HTML]{F1F1F1} {\cellcolor[HTML]{FFFFFF}} \color{black} --- \\
\cline{1-7}
\multirow[c]{2}{*}{Bord} & WK & {\cellcolor[HTML]{F7BCA1}} \color[HTML]{000000} \color{black} 0.44 & {\cellcolor[HTML]{D9DCE1}} \color[HTML]{000000} \color{black} -0.10 & {\cellcolor[HTML]{C3D5F4}} \color[HTML]{000000} \color{black} -0.34 & {\cellcolor[HTML]{000000}} \color[HTML]{F1F1F1} {\cellcolor[HTML]{FFFFFF}} \color{black} --- & {\cellcolor[HTML]{000000}} \color[HTML]{F1F1F1} {\cellcolor[HTML]{FFFFFF}} \color{black} --- \\
 & WKND & {\cellcolor[HTML]{F2CAB5}} \color[HTML]{000000} \color{black} 0.27 & {\cellcolor[HTML]{C7D7F0}} \color[HTML]{000000} \color{black} -0.29 & {\cellcolor[HTML]{E2DAD5}} \color[HTML]{000000} \color{black} 0.02 & {\cellcolor[HTML]{000000}} \color[HTML]{F1F1F1} {\cellcolor[HTML]{FFFFFF}} \color{black} --- & {\cellcolor[HTML]{000000}} \color[HTML]{F1F1F1} {\cellcolor[HTML]{FFFFFF}} \color{black} --- \\
\cline{1-7}
\multirow[c]{2}{*}{Gren} & WK & {\cellcolor[HTML]{F7B599}} \color[HTML]{000000} \color{black} 0.50 & {\cellcolor[HTML]{E1DAD6}} \color[HTML]{000000} \color{black} 0.00 & {\cellcolor[HTML]{B2CCFB}} \color[HTML]{000000} \color{black} -0.50 & {\cellcolor[HTML]{000000}} \color[HTML]{F1F1F1} {\cellcolor[HTML]{FFFFFF}} \color{black} --- & {\cellcolor[HTML]{000000}} \color[HTML]{F1F1F1} {\cellcolor[HTML]{FFFFFF}} \color{black} --- \\
 & WKND & {\cellcolor[HTML]{F6BDA2}} \color[HTML]{000000} \color{black} 0.43 & {\cellcolor[HTML]{D7DCE3}} \color[HTML]{000000} \color{black} -0.12 & {\cellcolor[HTML]{C6D6F1}} \color[HTML]{000000} \color{black} -0.31 & {\cellcolor[HTML]{000000}} \color[HTML]{F1F1F1} {\cellcolor[HTML]{FFFFFF}} \color{black} --- & {\cellcolor[HTML]{000000}} \color[HTML]{F1F1F1} {\cellcolor[HTML]{FFFFFF}} \color{black} --- \\
\cline{1-7}
\multirow[c]{2}{*}{Lyon} & WK & {\cellcolor[HTML]{F4C5AD}} \color[HTML]{000000} \color{black} 0.33 & {\cellcolor[HTML]{DEDCDB}} \color[HTML]{000000} \color{black} -0.03 & {\cellcolor[HTML]{C6D6F1}} \color[HTML]{000000} \color{black} -0.30 & {\cellcolor[HTML]{000000}} \color[HTML]{F1F1F1} {\cellcolor[HTML]{FFFFFF}} \color{black} --- & {\cellcolor[HTML]{000000}} \color[HTML]{F1F1F1} {\cellcolor[HTML]{FFFFFF}} \color{black} --- \\
 & WKND & {\cellcolor[HTML]{F4C6AF}} \color[HTML]{000000} \color{black} 0.33 & {\cellcolor[HTML]{DFDBD9}} \color[HTML]{000000} \color{black} -0.02 & {\cellcolor[HTML]{C6D6F1}} \color[HTML]{000000} \color{black} -0.30 & {\cellcolor[HTML]{000000}} \color[HTML]{F1F1F1} {\cellcolor[HTML]{FFFFFF}} \color{black} --- & {\cellcolor[HTML]{000000}} \color[HTML]{F1F1F1} {\cellcolor[HTML]{FFFFFF}} \color{black} --- \\
\cline{1-7}
\multirow[c]{2}{*}{Nant} & WK & {\cellcolor[HTML]{F7B194}} \color[HTML]{000000} \color{black} 0.53 & {\cellcolor[HTML]{D5DBE5}} \color[HTML]{000000} \color{black} -0.14 & {\cellcolor[HTML]{BED2F6}} \color[HTML]{000000} \color{black} -0.38 & {\cellcolor[HTML]{000000}} \color[HTML]{F1F1F1} {\cellcolor[HTML]{FFFFFF}} \color{black} --- & {\cellcolor[HTML]{000000}} \color[HTML]{F1F1F1} {\cellcolor[HTML]{FFFFFF}} \color{black} --- \\
 & WKND & {\cellcolor[HTML]{F7BCA1}} \color[HTML]{000000} \color{black} 0.44 & {\cellcolor[HTML]{CCD9ED}} \color[HTML]{000000} \color{black} -0.24 & {\cellcolor[HTML]{D1DAE9}} \color[HTML]{000000} \color{black} -0.19 & {\cellcolor[HTML]{000000}} \color[HTML]{F1F1F1} {\cellcolor[HTML]{FFFFFF}} \color{black} --- & {\cellcolor[HTML]{000000}} \color[HTML]{F1F1F1} {\cellcolor[HTML]{FFFFFF}} \color{black} --- \\
\cline{1-7}
\multirow[c]{2}{*}{Toul} & WK & {\cellcolor[HTML]{F5C0A7}} \color[HTML]{000000} \color{black} 0.38 & {\cellcolor[HTML]{D7DCE3}} \color[HTML]{000000} \color{black} -0.11 & {\cellcolor[HTML]{C9D7F0}} \color[HTML]{000000} \color{black} -0.27 & {\cellcolor[HTML]{000000}} \color[HTML]{F1F1F1} {\cellcolor[HTML]{FFFFFF}} \color{black} --- & {\cellcolor[HTML]{000000}} \color[HTML]{F1F1F1} {\cellcolor[HTML]{FFFFFF}} \color{black} --- \\
 & WKND & {\cellcolor[HTML]{F5C0A7}} \color[HTML]{000000} \color{black} 0.39 & {\cellcolor[HTML]{D4DBE6}} \color[HTML]{000000} \color{black} -0.15 & {\cellcolor[HTML]{CCD9ED}} \color[HTML]{000000} \color{black} -0.24 & {\cellcolor[HTML]{000000}} \color[HTML]{F1F1F1} {\cellcolor[HTML]{FFFFFF}} \color{black} --- & {\cellcolor[HTML]{000000}} \color[HTML]{F1F1F1} {\cellcolor[HTML]{FFFFFF}} \color{black} --- \\
\cline{1-7}
\multirow[c]{2}{*}{Mont} & WK & {\cellcolor[HTML]{EDD1C2}} \color[HTML]{000000} \color{black} 0.17 & {\cellcolor[HTML]{C6D6F1}} \color[HTML]{000000} \color{black} -0.30 & {\cellcolor[HTML]{EAD4C8}} \color[HTML]{000000} \color{black} 0.13 & {\cellcolor[HTML]{000000}} \color[HTML]{F1F1F1} {\cellcolor[HTML]{FFFFFF}} \color{black} --- & {\cellcolor[HTML]{000000}} \color[HTML]{F1F1F1} {\cellcolor[HTML]{FFFFFF}} \color{black} --- \\
 & WKND & {\cellcolor[HTML]{ECD3C5}} \color[HTML]{000000} \color{black} 0.15 & {\cellcolor[HTML]{C6D6F1}} \color[HTML]{000000} \color{black} -0.30 & {\cellcolor[HTML]{ECD3C5}} \color[HTML]{000000} \color{black} 0.15 & {\cellcolor[HTML]{000000}} \color[HTML]{F1F1F1} {\cellcolor[HTML]{FFFFFF}} \color{black} --- & {\cellcolor[HTML]{000000}} \color[HTML]{F1F1F1} {\cellcolor[HTML]{FFFFFF}} \color{black} --- \\
\cline{1-7}
\multirow[c]{2}{*}{Tour} & WK & {\cellcolor[HTML]{F7AF91}} \color[HTML]{000000} \color{black} 0.56 & {\cellcolor[HTML]{F5C0A7}} \color[HTML]{000000} \color{black} 0.39 & {\cellcolor[HTML]{7396F5}} \color[HTML]{F1F1F1} \color{black} -1.06 & {\cellcolor[HTML]{F29072}} \color[HTML]{F1F1F1} \color{black} 0.80 & {\cellcolor[HTML]{9EBEFF}} \color[HTML]{000000} \color{black} -0.69 \\
 & WKND & {\cellcolor[HTML]{F7B599}} \color[HTML]{000000} \color{black} 0.49 & {\cellcolor[HTML]{E1DAD6}} \color[HTML]{000000} \color{black} 0.01 & {\cellcolor[HTML]{B2CCFB}} \color[HTML]{000000} \color{black} -0.50 & {\cellcolor[HTML]{000000}} \color[HTML]{F1F1F1} {\cellcolor[HTML]{FFFFFF}} \color{black} --- & {\cellcolor[HTML]{000000}} \color[HTML]{F1F1F1} {\cellcolor[HTML]{FFFFFF}} \color{black} --- \\
\cline{1-7}
\multirow[c]{2}{*}{Stra} & WK & {\cellcolor[HTML]{F0CDBB}} \color[HTML]{000000} \color{black} 0.23 & {\cellcolor[HTML]{C0D4F5}} \color[HTML]{000000} \color{black} -0.37 & {\cellcolor[HTML]{AEC9FC}} \color[HTML]{000000} \color{black} -0.53 & {\cellcolor[HTML]{F6A283}} \color[HTML]{000000} \color{black} 0.67 & {\cellcolor[HTML]{000000}} \color[HTML]{F1F1F1} {\cellcolor[HTML]{FFFFFF}} \color{black} --- \\
 & WKND & {\cellcolor[HTML]{F7B599}} \color[HTML]{000000} \color{black} 0.50 & {\cellcolor[HTML]{CCD9ED}} \color[HTML]{000000} \color{black} -0.23 & {\cellcolor[HTML]{CAD8EF}} \color[HTML]{000000} \color{black} -0.26 & {\cellcolor[HTML]{000000}} \color[HTML]{F1F1F1} {\cellcolor[HTML]{FFFFFF}} \color{black} --- & {\cellcolor[HTML]{000000}} \color[HTML]{F1F1F1} {\cellcolor[HTML]{FFFFFF}} \color{black} --- \\
\cline{1-7}
\multirow[c]{2}{*}{Orle} & WK & {\cellcolor[HTML]{F5C2AA}} \color[HTML]{000000} \color{black} 0.36 & {\cellcolor[HTML]{DFDBD9}} \color[HTML]{000000} \color{black} -0.02 & {\cellcolor[HTML]{C3D5F4}} \color[HTML]{000000} \color{black} -0.34 & {\cellcolor[HTML]{000000}} \color[HTML]{F1F1F1} {\cellcolor[HTML]{FFFFFF}} \color{black} --- & {\cellcolor[HTML]{000000}} \color[HTML]{F1F1F1} {\cellcolor[HTML]{FFFFFF}} \color{black} --- \\
 & WKND & {\cellcolor[HTML]{F6BFA6}} \color[HTML]{000000} \color{black} 0.40 & {\cellcolor[HTML]{DDDCDC}} \color[HTML]{000000} \color{black} -0.05 & {\cellcolor[HTML]{C0D4F5}} \color[HTML]{000000} \color{black} -0.36 & {\cellcolor[HTML]{000000}} \color[HTML]{F1F1F1} {\cellcolor[HTML]{FFFFFF}} \color{black} --- & {\cellcolor[HTML]{000000}} \color[HTML]{F1F1F1} {\cellcolor[HTML]{FFFFFF}} \color{black} --- \\
\cline{1-7}
\multirow[c]{2}{*}{Metz} & WK & {\cellcolor[HTML]{EFCEBD}} \color[HTML]{000000} \color{black} 0.21 & {\cellcolor[HTML]{CBD8EE}} \color[HTML]{000000} \color{black} -0.26 & {\cellcolor[HTML]{E4D9D2}} \color[HTML]{000000} \color{black} 0.04 & {\cellcolor[HTML]{000000}} \color[HTML]{F1F1F1} {\cellcolor[HTML]{FFFFFF}} \color{black} --- & {\cellcolor[HTML]{000000}} \color[HTML]{F1F1F1} {\cellcolor[HTML]{FFFFFF}} \color{black} --- \\
 & WKND & {\cellcolor[HTML]{EAD4C8}} \color[HTML]{000000} \color{black} 0.13 & {\cellcolor[HTML]{CCD9ED}} \color[HTML]{000000} \color{black} -0.23 & {\cellcolor[HTML]{E9D5CB}} \color[HTML]{000000} \color{black} 0.11 & {\cellcolor[HTML]{000000}} \color[HTML]{F1F1F1} {\cellcolor[HTML]{FFFFFF}} \color{black} --- & {\cellcolor[HTML]{000000}} \color[HTML]{F1F1F1} {\cellcolor[HTML]{FFFFFF}} \color{black} --- \\
\cline{1-7}
\multirow[c]{2}{*}{Mans} & WK & {\cellcolor[HTML]{EF886B}} \color[HTML]{F1F1F1} \color{black} 0.86 & {\cellcolor[HTML]{A9C6FD}} \color[HTML]{000000} \color{black} -0.59 & {\cellcolor[HTML]{CCD9ED}} \color[HTML]{000000} \color{black} -0.24 & {\cellcolor[HTML]{DEDCDB}} \color[HTML]{000000} \color{black} -0.03 & {\cellcolor[HTML]{000000}} \color[HTML]{F1F1F1} {\cellcolor[HTML]{FFFFFF}} \color{black} --- \\
 & WKND & {\cellcolor[HTML]{F29274}} \color[HTML]{F1F1F1} \color{black} 0.79 & {\cellcolor[HTML]{98B9FF}} \color[HTML]{000000} \color{black} -0.74 & {\cellcolor[HTML]{C5D6F2}} \color[HTML]{000000} \color{black} -0.32 & {\cellcolor[HTML]{F2CBB7}} \color[HTML]{000000} \color{black} 0.26 & {\cellcolor[HTML]{000000}} \color[HTML]{F1F1F1} {\cellcolor[HTML]{FFFFFF}} \color{black} --- \\
\cline{1-7}
\multirow[c]{2}{*}{Dijo} & WK & {\cellcolor[HTML]{F5A081}} \color[HTML]{000000} \color{black} 0.68 & {\cellcolor[HTML]{B1CBFC}} \color[HTML]{000000} \color{black} -0.51 & {\cellcolor[HTML]{D3DBE7}} \color[HTML]{000000} \color{black} -0.17 & {\cellcolor[HTML]{000000}} \color[HTML]{F1F1F1} {\cellcolor[HTML]{FFFFFF}} \color{black} --- & {\cellcolor[HTML]{000000}} \color[HTML]{F1F1F1} {\cellcolor[HTML]{FFFFFF}} \color{black} --- \\
 & WKND & {\cellcolor[HTML]{F7AF91}} \color[HTML]{000000} \color{black} 0.56 & {\cellcolor[HTML]{B6CEFA}} \color[HTML]{000000} \color{black} -0.46 & {\cellcolor[HTML]{D8DCE2}} \color[HTML]{000000} \color{black} -0.10 & {\cellcolor[HTML]{000000}} \color[HTML]{F1F1F1} {\cellcolor[HTML]{FFFFFF}} \color{black} --- & {\cellcolor[HTML]{000000}} \color[HTML]{F1F1F1} {\cellcolor[HTML]{FFFFFF}} \color{black} --- \\
\cline{1-7}
\multirow[c]{2}{*}{Mars} & WK & {\cellcolor[HTML]{E8D6CC}} \color[HTML]{000000} \color{black} 0.10 & {\cellcolor[HTML]{D5DBE5}} \color[HTML]{000000} \color{black} -0.14 & {\cellcolor[HTML]{EAD4C8}} \color[HTML]{000000} \color{black} 0.13 & {\cellcolor[HTML]{DADCE0}} \color[HTML]{000000} \color{black} -0.08 & {\cellcolor[HTML]{000000}} \color[HTML]{F1F1F1} {\cellcolor[HTML]{FFFFFF}} \color{black} --- \\
 & WKND & {\cellcolor[HTML]{E7D7CE}} \color[HTML]{000000} \color{black} 0.07 & {\cellcolor[HTML]{D2DBE8}} \color[HTML]{000000} \color{black} -0.17 & {\cellcolor[HTML]{E9D5CB}} \color[HTML]{000000} \color{black} 0.10 & {\cellcolor[HTML]{000000}} \color[HTML]{F1F1F1} {\cellcolor[HTML]{FFFFFF}} \color{black} --- & {\cellcolor[HTML]{000000}} \color[HTML]{F1F1F1} {\cellcolor[HTML]{FFFFFF}} \color{black} --- \\
\cline{1-7}
\multirow[c]{2}{*}{Nice} & WK & {\cellcolor[HTML]{F7B194}} \color[HTML]{000000} \color{black} 0.54 & {\cellcolor[HTML]{CAD8EF}} \color[HTML]{000000} \color{black} -0.26 & {\cellcolor[HTML]{C9D7F0}} \color[HTML]{000000} \color{black} -0.27 & {\cellcolor[HTML]{000000}} \color[HTML]{F1F1F1} {\cellcolor[HTML]{FFFFFF}} \color{black} --- & {\cellcolor[HTML]{000000}} \color[HTML]{F1F1F1} {\cellcolor[HTML]{FFFFFF}} \color{black} --- \\
 & WKND & {\cellcolor[HTML]{F7B99E}} \color[HTML]{000000} \color{black} 0.46 & {\cellcolor[HTML]{E0DBD8}} \color[HTML]{000000} \color{black} -0.01 & {\cellcolor[HTML]{B7CFF9}} \color[HTML]{000000} \color{black} -0.45 & {\cellcolor[HTML]{000000}} \color[HTML]{F1F1F1} {\cellcolor[HTML]{FFFFFF}} \color{black} --- & {\cellcolor[HTML]{000000}} \color[HTML]{F1F1F1} {\cellcolor[HTML]{FFFFFF}} \color{black} --- \\
\cline{1-7}
\multirow[c]{2}{*}{Nanc} & WK & {\cellcolor[HTML]{F4987A}} \color[HTML]{000000} \color{black} 0.74 & {\cellcolor[HTML]{E5D8D1}} \color[HTML]{000000} \color{black} 0.05 & {\cellcolor[HTML]{EDD1C2}} \color[HTML]{000000} \color{black} 0.18 & {\cellcolor[HTML]{7DA0F9}} \color[HTML]{F1F1F1} \color{black} -0.97 & {\cellcolor[HTML]{000000}} \color[HTML]{F1F1F1} {\cellcolor[HTML]{FFFFFF}} \color{black} --- \\
 & WKND & {\cellcolor[HTML]{F6BDA2}} \color[HTML]{000000} \color{black} 0.43 & {\cellcolor[HTML]{C9D7F0}} \color[HTML]{000000} \color{black} -0.28 & {\cellcolor[HTML]{D5DBE5}} \color[HTML]{000000} \color{black} -0.15 & {\cellcolor[HTML]{000000}} \color[HTML]{F1F1F1} {\cellcolor[HTML]{FFFFFF}} \color{black} --- & {\cellcolor[HTML]{000000}} \color[HTML]{F1F1F1} {\cellcolor[HTML]{FFFFFF}} \color{black} --- \\
\cline{1-7}
\bottomrule
\end{tabular}
\end{table}

\begin{table}
\centering
\caption[Commercial venue diversity coefficient estimates for local-level multinomial logistic regression predictive model]{Commercial venue diversity coefficient estimates for local-level multinomial logistic regression predictive model for the week and weekend and by cluster and by city (Paris--Nancy shown in size order). The figure contains the strength of the association between each of the covariates. Coefficients are coloured according to the strength of the relationship. Data preprocessing was applied to the \texttt{NetMob23} dataset~\citep{martinez-duriveNetMob23DatasetHighresolution2023}. City names and Week/Weekend labels are abbreviated for brevity.}
\label{tab:commercial-venue-diversity-coefficients}
\begin{tabular}{llrrrrr}
\toprule
 & Class & 1 & 2 & 3 & 4 & 5 \\
City & WK/WKND &  &  &  &  &  \\
\midrule
\multirow[c]{2}{*}{Pari} & WK & {\cellcolor[HTML]{BCD2F7}} \color[HTML]{000000} \color{black} -0.41 & {\cellcolor[HTML]{DEDCDB}} \color[HTML]{000000} \color{black} -0.03 & {\cellcolor[HTML]{F7BCA1}} \color[HTML]{000000} \color{black} 0.43 & {\cellcolor[HTML]{000000}} \color[HTML]{F1F1F1} {\cellcolor[HTML]{FFFFFF}} \color{black} --- & {\cellcolor[HTML]{000000}} \color[HTML]{F1F1F1} {\cellcolor[HTML]{FFFFFF}} \color{black} --- \\
 & WKND & {\cellcolor[HTML]{B2CCFB}} \color[HTML]{000000} \color{black} -0.50 & {\cellcolor[HTML]{DBDCDE}} \color[HTML]{000000} \color{black} -0.06 & {\cellcolor[HTML]{F7AF91}} \color[HTML]{000000} \color{black} 0.56 & {\cellcolor[HTML]{000000}} \color[HTML]{F1F1F1} {\cellcolor[HTML]{FFFFFF}} \color{black} --- & {\cellcolor[HTML]{000000}} \color[HTML]{F1F1F1} {\cellcolor[HTML]{FFFFFF}} \color{black} --- \\
\cline{1-7}
\multirow[c]{2}{*}{Renn} & WK & {\cellcolor[HTML]{CAD8EF}} \color[HTML]{000000} \color{black} -0.27 & {\cellcolor[HTML]{CDD9EC}} \color[HTML]{000000} \color{black} -0.23 & {\cellcolor[HTML]{F7B599}} \color[HTML]{000000} \color{black} 0.50 & {\cellcolor[HTML]{000000}} \color[HTML]{F1F1F1} {\cellcolor[HTML]{FFFFFF}} \color{black} --- & {\cellcolor[HTML]{000000}} \color[HTML]{F1F1F1} {\cellcolor[HTML]{FFFFFF}} \color{black} --- \\
 & WKND & {\cellcolor[HTML]{CAD8EF}} \color[HTML]{000000} \color{black} -0.26 & {\cellcolor[HTML]{CEDAEB}} \color[HTML]{000000} \color{black} -0.22 & {\cellcolor[HTML]{F7B79B}} \color[HTML]{000000} \color{black} 0.48 & {\cellcolor[HTML]{000000}} \color[HTML]{F1F1F1} {\cellcolor[HTML]{FFFFFF}} \color{black} --- & {\cellcolor[HTML]{000000}} \color[HTML]{F1F1F1} {\cellcolor[HTML]{FFFFFF}} \color{black} --- \\
\cline{1-7}
\multirow[c]{2}{*}{Lill} & WK & {\cellcolor[HTML]{D5DBE5}} \color[HTML]{000000} \color{black} -0.14 & {\cellcolor[HTML]{EFCFBF}} \color[HTML]{000000} \color{black} 0.20 & {\cellcolor[HTML]{DCDDDD}} \color[HTML]{000000} \color{black} -0.05 & {\cellcolor[HTML]{000000}} \color[HTML]{F1F1F1} {\cellcolor[HTML]{FFFFFF}} \color{black} --- & {\cellcolor[HTML]{000000}} \color[HTML]{F1F1F1} {\cellcolor[HTML]{FFFFFF}} \color{black} --- \\
 & WKND & {\cellcolor[HTML]{D3DBE7}} \color[HTML]{000000} \color{black} -0.16 & {\cellcolor[HTML]{EFCEBD}} \color[HTML]{000000} \color{black} 0.21 & {\cellcolor[HTML]{DDDCDC}} \color[HTML]{000000} \color{black} -0.05 & {\cellcolor[HTML]{000000}} \color[HTML]{F1F1F1} {\cellcolor[HTML]{FFFFFF}} \color{black} --- & {\cellcolor[HTML]{000000}} \color[HTML]{F1F1F1} {\cellcolor[HTML]{FFFFFF}} \color{black} --- \\
\cline{1-7}
\multirow[c]{2}{*}{Bord} & WK & {\cellcolor[HTML]{BED2F6}} \color[HTML]{000000} \color{black} -0.38 & {\cellcolor[HTML]{CFDAEA}} \color[HTML]{000000} \color{black} -0.20 & {\cellcolor[HTML]{F7AC8E}} \color[HTML]{000000} \color{black} 0.58 & {\cellcolor[HTML]{000000}} \color[HTML]{F1F1F1} {\cellcolor[HTML]{FFFFFF}} \color{black} --- & {\cellcolor[HTML]{000000}} \color[HTML]{F1F1F1} {\cellcolor[HTML]{FFFFFF}} \color{black} --- \\
 & WKND & {\cellcolor[HTML]{B5CDFA}} \color[HTML]{000000} \color{black} -0.47 & {\cellcolor[HTML]{CAD8EF}} \color[HTML]{000000} \color{black} -0.27 & {\cellcolor[HTML]{F4987A}} \color[HTML]{000000} \color{black} 0.74 & {\cellcolor[HTML]{000000}} \color[HTML]{F1F1F1} {\cellcolor[HTML]{FFFFFF}} \color{black} --- & {\cellcolor[HTML]{000000}} \color[HTML]{F1F1F1} {\cellcolor[HTML]{FFFFFF}} \color{black} --- \\
\cline{1-7}
\multirow[c]{2}{*}{Gren} & WK & {\cellcolor[HTML]{F1CCB8}} \color[HTML]{000000} \color{black} 0.25 & {\cellcolor[HTML]{F0CDBB}} \color[HTML]{000000} \color{black} 0.22 & {\cellcolor[HTML]{B6CEFA}} \color[HTML]{000000} \color{black} -0.47 & {\cellcolor[HTML]{000000}} \color[HTML]{F1F1F1} {\cellcolor[HTML]{FFFFFF}} \color{black} --- & {\cellcolor[HTML]{000000}} \color[HTML]{F1F1F1} {\cellcolor[HTML]{FFFFFF}} \color{black} --- \\
 & WKND & {\cellcolor[HTML]{F5C2AA}} \color[HTML]{000000} \color{black} 0.36 & {\cellcolor[HTML]{E1DAD6}} \color[HTML]{000000} \color{black} 0.01 & {\cellcolor[HTML]{C0D4F5}} \color[HTML]{000000} \color{black} -0.36 & {\cellcolor[HTML]{000000}} \color[HTML]{F1F1F1} {\cellcolor[HTML]{FFFFFF}} \color{black} --- & {\cellcolor[HTML]{000000}} \color[HTML]{F1F1F1} {\cellcolor[HTML]{FFFFFF}} \color{black} --- \\
\cline{1-7}
\multirow[c]{2}{*}{Lyon} & WK & {\cellcolor[HTML]{F0CDBB}} \color[HTML]{000000} \color{black} 0.22 & {\cellcolor[HTML]{EDD2C3}} \color[HTML]{000000} \color{black} 0.16 & {\cellcolor[HTML]{BED2F6}} \color[HTML]{000000} \color{black} -0.39 & {\cellcolor[HTML]{000000}} \color[HTML]{F1F1F1} {\cellcolor[HTML]{FFFFFF}} \color{black} --- & {\cellcolor[HTML]{000000}} \color[HTML]{F1F1F1} {\cellcolor[HTML]{FFFFFF}} \color{black} --- \\
 & WKND & {\cellcolor[HTML]{F1CDBA}} \color[HTML]{000000} \color{black} 0.24 & {\cellcolor[HTML]{EDD1C2}} \color[HTML]{000000} \color{black} 0.17 & {\cellcolor[HTML]{BBD1F8}} \color[HTML]{000000} \color{black} -0.42 & {\cellcolor[HTML]{000000}} \color[HTML]{F1F1F1} {\cellcolor[HTML]{FFFFFF}} \color{black} --- & {\cellcolor[HTML]{000000}} \color[HTML]{F1F1F1} {\cellcolor[HTML]{FFFFFF}} \color{black} --- \\
\cline{1-7}
\multirow[c]{2}{*}{Nant} & WK & {\cellcolor[HTML]{90B2FE}} \color[HTML]{000000} \color{black} -0.81 & {\cellcolor[HTML]{BED2F6}} \color[HTML]{000000} \color{black} -0.38 & {\cellcolor[HTML]{D85646}} \color[HTML]{F1F1F1} \color{black} 1.19 & {\cellcolor[HTML]{000000}} \color[HTML]{F1F1F1} {\cellcolor[HTML]{FFFFFF}} \color{black} --- & {\cellcolor[HTML]{000000}} \color[HTML]{F1F1F1} {\cellcolor[HTML]{FFFFFF}} \color{black} --- \\
 & WKND & {\cellcolor[HTML]{C1D4F4}} \color[HTML]{000000} \color{black} -0.35 & {\cellcolor[HTML]{F7B99E}} \color[HTML]{000000} \color{black} 0.46 & {\cellcolor[HTML]{D8DCE2}} \color[HTML]{000000} \color{black} -0.11 & {\cellcolor[HTML]{000000}} \color[HTML]{F1F1F1} {\cellcolor[HTML]{FFFFFF}} \color{black} --- & {\cellcolor[HTML]{000000}} \color[HTML]{F1F1F1} {\cellcolor[HTML]{FFFFFF}} \color{black} --- \\
\cline{1-7}
\multirow[c]{2}{*}{Toul} & WK & {\cellcolor[HTML]{DADCE0}} \color[HTML]{000000} \color{black} -0.09 & {\cellcolor[HTML]{DFDBD9}} \color[HTML]{000000} \color{black} -0.02 & {\cellcolor[HTML]{E9D5CB}} \color[HTML]{000000} \color{black} 0.10 & {\cellcolor[HTML]{000000}} \color[HTML]{F1F1F1} {\cellcolor[HTML]{FFFFFF}} \color{black} --- & {\cellcolor[HTML]{000000}} \color[HTML]{F1F1F1} {\cellcolor[HTML]{FFFFFF}} \color{black} --- \\
 & WKND & {\cellcolor[HTML]{D8DCE2}} \color[HTML]{000000} \color{black} -0.10 & {\cellcolor[HTML]{E0DBD8}} \color[HTML]{000000} \color{black} -0.01 & {\cellcolor[HTML]{EAD5C9}} \color[HTML]{000000} \color{black} 0.11 & {\cellcolor[HTML]{000000}} \color[HTML]{F1F1F1} {\cellcolor[HTML]{FFFFFF}} \color{black} --- & {\cellcolor[HTML]{000000}} \color[HTML]{F1F1F1} {\cellcolor[HTML]{FFFFFF}} \color{black} --- \\
\cline{1-7}
\multirow[c]{2}{*}{Mont} & WK & {\cellcolor[HTML]{E6D7CF}} \color[HTML]{000000} \color{black} 0.07 & {\cellcolor[HTML]{D3DBE7}} \color[HTML]{000000} \color{black} -0.17 & {\cellcolor[HTML]{E9D5CB}} \color[HTML]{000000} \color{black} 0.10 & {\cellcolor[HTML]{000000}} \color[HTML]{F1F1F1} {\cellcolor[HTML]{FFFFFF}} \color{black} --- & {\cellcolor[HTML]{000000}} \color[HTML]{F1F1F1} {\cellcolor[HTML]{FFFFFF}} \color{black} --- \\
 & WKND & {\cellcolor[HTML]{E2DAD5}} \color[HTML]{000000} \color{black} 0.02 & {\cellcolor[HTML]{CFDAEA}} \color[HTML]{000000} \color{black} -0.20 & {\cellcolor[HTML]{EDD1C2}} \color[HTML]{000000} \color{black} 0.18 & {\cellcolor[HTML]{000000}} \color[HTML]{F1F1F1} {\cellcolor[HTML]{FFFFFF}} \color{black} --- & {\cellcolor[HTML]{000000}} \color[HTML]{F1F1F1} {\cellcolor[HTML]{FFFFFF}} \color{black} --- \\
\cline{1-7}
\multirow[c]{2}{*}{Tour} & WK & {\cellcolor[HTML]{E5D8D1}} \color[HTML]{000000} \color{black} 0.06 & {\cellcolor[HTML]{EBD3C6}} \color[HTML]{000000} \color{black} 0.14 & {\cellcolor[HTML]{C5D6F2}} \color[HTML]{000000} \color{black} -0.31 & {\cellcolor[HTML]{F7BA9F}} \color[HTML]{000000} \color{black} 0.45 & {\cellcolor[HTML]{C3D5F4}} \color[HTML]{000000} \color{black} -0.34 \\
 & WKND & {\cellcolor[HTML]{EFCFBF}} \color[HTML]{000000} \color{black} 0.21 & {\cellcolor[HTML]{F2CBB7}} \color[HTML]{000000} \color{black} 0.27 & {\cellcolor[HTML]{B5CDFA}} \color[HTML]{000000} \color{black} -0.48 & {\cellcolor[HTML]{000000}} \color[HTML]{F1F1F1} {\cellcolor[HTML]{FFFFFF}} \color{black} --- & {\cellcolor[HTML]{000000}} \color[HTML]{F1F1F1} {\cellcolor[HTML]{FFFFFF}} \color{black} --- \\
\cline{1-7}
\multirow[c]{2}{*}{Stra} & WK & {\cellcolor[HTML]{E2DAD5}} \color[HTML]{000000} \color{black} 0.01 & {\cellcolor[HTML]{C1D4F4}} \color[HTML]{000000} \color{black} -0.35 & {\cellcolor[HTML]{EFCFBF}} \color[HTML]{000000} \color{black} 0.20 & {\cellcolor[HTML]{EBD3C6}} \color[HTML]{000000} \color{black} 0.14 & {\cellcolor[HTML]{000000}} \color[HTML]{F1F1F1} {\cellcolor[HTML]{FFFFFF}} \color{black} --- \\
 & WKND & {\cellcolor[HTML]{EAD5C9}} \color[HTML]{000000} \color{black} 0.11 & {\cellcolor[HTML]{C7D7F0}} \color[HTML]{000000} \color{black} -0.29 & {\cellcolor[HTML]{EDD1C2}} \color[HTML]{000000} \color{black} 0.18 & {\cellcolor[HTML]{000000}} \color[HTML]{F1F1F1} {\cellcolor[HTML]{FFFFFF}} \color{black} --- & {\cellcolor[HTML]{000000}} \color[HTML]{F1F1F1} {\cellcolor[HTML]{FFFFFF}} \color{black} --- \\
\cline{1-7}
\multirow[c]{2}{*}{Orle} & WK & {\cellcolor[HTML]{E1DAD6}} \color[HTML]{000000} \color{black} 0.01 & {\cellcolor[HTML]{D7DCE3}} \color[HTML]{000000} \color{black} -0.12 & {\cellcolor[HTML]{EAD5C9}} \color[HTML]{000000} \color{black} 0.11 & {\cellcolor[HTML]{000000}} \color[HTML]{F1F1F1} {\cellcolor[HTML]{FFFFFF}} \color{black} --- & {\cellcolor[HTML]{000000}} \color[HTML]{F1F1F1} {\cellcolor[HTML]{FFFFFF}} \color{black} --- \\
 & WKND & {\cellcolor[HTML]{DEDCDB}} \color[HTML]{000000} \color{black} -0.03 & {\cellcolor[HTML]{D4DBE6}} \color[HTML]{000000} \color{black} -0.16 & {\cellcolor[HTML]{EDD1C2}} \color[HTML]{000000} \color{black} 0.18 & {\cellcolor[HTML]{000000}} \color[HTML]{F1F1F1} {\cellcolor[HTML]{FFFFFF}} \color{black} --- & {\cellcolor[HTML]{000000}} \color[HTML]{F1F1F1} {\cellcolor[HTML]{FFFFFF}} \color{black} --- \\
\cline{1-7}
\multirow[c]{2}{*}{Metz} & WK & {\cellcolor[HTML]{DFDBD9}} \color[HTML]{000000} \color{black} -0.03 & {\cellcolor[HTML]{E1DAD6}} \color[HTML]{000000} \color{black} 0.01 & {\cellcolor[HTML]{E2DAD5}} \color[HTML]{000000} \color{black} 0.02 & {\cellcolor[HTML]{000000}} \color[HTML]{F1F1F1} {\cellcolor[HTML]{FFFFFF}} \color{black} --- & {\cellcolor[HTML]{000000}} \color[HTML]{F1F1F1} {\cellcolor[HTML]{FFFFFF}} \color{black} --- \\
 & WKND & {\cellcolor[HTML]{DCDDDD}} \color[HTML]{000000} \color{black} -0.06 & {\cellcolor[HTML]{E1DAD6}} \color[HTML]{000000} \color{black} 0.00 & {\cellcolor[HTML]{E5D8D1}} \color[HTML]{000000} \color{black} 0.06 & {\cellcolor[HTML]{000000}} \color[HTML]{F1F1F1} {\cellcolor[HTML]{FFFFFF}} \color{black} --- & {\cellcolor[HTML]{000000}} \color[HTML]{F1F1F1} {\cellcolor[HTML]{FFFFFF}} \color{black} --- \\
\cline{1-7}
\multirow[c]{2}{*}{Mans} & WK & {\cellcolor[HTML]{F7B89C}} \color[HTML]{000000} \color{black} 0.47 & {\cellcolor[HTML]{D8DCE2}} \color[HTML]{000000} \color{black} -0.11 & {\cellcolor[HTML]{F5C2AA}} \color[HTML]{000000} \color{black} 0.37 & {\cellcolor[HTML]{98B9FF}} \color[HTML]{000000} \color{black} -0.73 & {\cellcolor[HTML]{000000}} \color[HTML]{F1F1F1} {\cellcolor[HTML]{FFFFFF}} \color{black} --- \\
 & WKND & {\cellcolor[HTML]{E9D5CB}} \color[HTML]{000000} \color{black} 0.11 & {\cellcolor[HTML]{EBD3C6}} \color[HTML]{000000} \color{black} 0.14 & {\cellcolor[HTML]{F7B79B}} \color[HTML]{000000} \color{black} 0.48 & {\cellcolor[HTML]{98B9FF}} \color[HTML]{000000} \color{black} -0.73 & {\cellcolor[HTML]{000000}} \color[HTML]{F1F1F1} {\cellcolor[HTML]{FFFFFF}} \color{black} --- \\
\cline{1-7}
\multirow[c]{2}{*}{Dijo} & WK & {\cellcolor[HTML]{8CAFFE}} \color[HTML]{000000} \color{black} -0.85 & {\cellcolor[HTML]{ECD3C5}} \color[HTML]{000000} \color{black} 0.15 & {\cellcolor[HTML]{F59F80}} \color[HTML]{000000} \color{black} 0.70 & {\cellcolor[HTML]{000000}} \color[HTML]{F1F1F1} {\cellcolor[HTML]{FFFFFF}} \color{black} --- & {\cellcolor[HTML]{000000}} \color[HTML]{F1F1F1} {\cellcolor[HTML]{FFFFFF}} \color{black} --- \\
 & WKND & {\cellcolor[HTML]{ADC9FD}} \color[HTML]{000000} \color{black} -0.54 & {\cellcolor[HTML]{F49A7B}} \color[HTML]{000000} \color{black} 0.73 & {\cellcolor[HTML]{D2DBE8}} \color[HTML]{000000} \color{black} -0.18 & {\cellcolor[HTML]{000000}} \color[HTML]{F1F1F1} {\cellcolor[HTML]{FFFFFF}} \color{black} --- & {\cellcolor[HTML]{000000}} \color[HTML]{F1F1F1} {\cellcolor[HTML]{FFFFFF}} \color{black} --- \\
\cline{1-7}
\multirow[c]{2}{*}{Mars} & WK & {\cellcolor[HTML]{A3C2FE}} \color[HTML]{000000} \color{black} -0.63 & {\cellcolor[HTML]{DADCE0}} \color[HTML]{000000} \color{black} -0.08 & {\cellcolor[HTML]{F6A385}} \color[HTML]{000000} \color{black} 0.66 & {\cellcolor[HTML]{E5D8D1}} \color[HTML]{000000} \color{black} 0.06 & {\cellcolor[HTML]{000000}} \color[HTML]{F1F1F1} {\cellcolor[HTML]{FFFFFF}} \color{black} --- \\
 & WKND & {\cellcolor[HTML]{9BBCFF}} \color[HTML]{000000} \color{black} -0.71 & {\cellcolor[HTML]{E0DBD8}} \color[HTML]{000000} \color{black} -0.01 & {\cellcolor[HTML]{F59C7D}} \color[HTML]{000000} \color{black} 0.72 & {\cellcolor[HTML]{000000}} \color[HTML]{F1F1F1} {\cellcolor[HTML]{FFFFFF}} \color{black} --- & {\cellcolor[HTML]{000000}} \color[HTML]{F1F1F1} {\cellcolor[HTML]{FFFFFF}} \color{black} --- \\
\cline{1-7}
\multirow[c]{2}{*}{Nice} & WK & {\cellcolor[HTML]{D5DBE5}} \color[HTML]{000000} \color{black} -0.14 & {\cellcolor[HTML]{D6DCE4}} \color[HTML]{000000} \color{black} -0.13 & {\cellcolor[HTML]{F2CBB7}} \color[HTML]{000000} \color{black} 0.26 & {\cellcolor[HTML]{000000}} \color[HTML]{F1F1F1} {\cellcolor[HTML]{FFFFFF}} \color{black} --- & {\cellcolor[HTML]{000000}} \color[HTML]{F1F1F1} {\cellcolor[HTML]{FFFFFF}} \color{black} --- \\
 & WKND & {\cellcolor[HTML]{E3D9D3}} \color[HTML]{000000} \color{black} 0.03 & {\cellcolor[HTML]{F6BEA4}} \color[HTML]{000000} \color{black} 0.41 & {\cellcolor[HTML]{B7CFF9}} \color[HTML]{000000} \color{black} -0.44 & {\cellcolor[HTML]{000000}} \color[HTML]{F1F1F1} {\cellcolor[HTML]{FFFFFF}} \color{black} --- & {\cellcolor[HTML]{000000}} \color[HTML]{F1F1F1} {\cellcolor[HTML]{FFFFFF}} \color{black} --- \\
\cline{1-7}
\multirow[c]{2}{*}{Nanc} & WK & {\cellcolor[HTML]{ADC9FD}} \color[HTML]{000000} \color{black} -0.55 & {\cellcolor[HTML]{C7D7F0}} \color[HTML]{000000} \color{black} -0.29 & {\cellcolor[HTML]{EE8468}} \color[HTML]{F1F1F1} \color{black} 0.89 & {\cellcolor[HTML]{DCDDDD}} \color[HTML]{000000} \color{black} -0.05 & {\cellcolor[HTML]{000000}} \color[HTML]{F1F1F1} {\cellcolor[HTML]{FFFFFF}} \color{black} --- \\
 & WKND & {\cellcolor[HTML]{C1D4F4}} \color[HTML]{000000} \color{black} -0.35 & {\cellcolor[HTML]{B6CEFA}} \color[HTML]{000000} \color{black} -0.46 & {\cellcolor[HTML]{F29072}} \color[HTML]{F1F1F1} \color{black} 0.81 & {\cellcolor[HTML]{000000}} \color[HTML]{F1F1F1} {\cellcolor[HTML]{FFFFFF}} \color{black} --- & {\cellcolor[HTML]{000000}} \color[HTML]{F1F1F1} {\cellcolor[HTML]{FFFFFF}} \color{black} --- \\
\cline{1-7}
\bottomrule
\end{tabular}
\end{table}

\begin{table}
\centering
\caption[Eating drinking talking diversity coefficient estimates for local-level multinomial logistic regression predictive model]{Eating drinking talking diversity coefficient estimates for local-level multinomial logistic regression predictive model for the week and weekend and by cluster and by city (Paris--Nancy shown in size order). The figure contains the strength of the association between each of the covariates. Coefficients are coloured according to the strength of the relationship. Data preprocessing was applied to the \texttt{NetMob23} dataset~\citep{martinez-duriveNetMob23DatasetHighresolution2023}. City names and Week/Weekend labels are abbreviated for brevity.}
\label{tab:eating-drinking-talking-diversity-coefficients}
\begin{tabular}{llrrrrr}
\toprule
 & Class & 1 & 2 & 3 & 4 & 5 \\
City & WK/WKND &  &  &  &  &  \\
\midrule
\multirow[c]{2}{*}{Pari} & WK & {\cellcolor[HTML]{F1CDBA}} \color[HTML]{000000} \color{black} 0.24 & {\cellcolor[HTML]{F0CDBB}} \color[HTML]{000000} \color{black} 0.23 & {\cellcolor[HTML]{B5CDFA}} \color[HTML]{000000} \color{black} -0.48 & {\cellcolor[HTML]{000000}} \color[HTML]{F1F1F1} {\cellcolor[HTML]{FFFFFF}} \color{black} --- & {\cellcolor[HTML]{000000}} \color[HTML]{F1F1F1} {\cellcolor[HTML]{FFFFFF}} \color{black} --- \\
 & WKND & {\cellcolor[HTML]{E8D6CC}} \color[HTML]{000000} \color{black} 0.09 & {\cellcolor[HTML]{EBD3C6}} \color[HTML]{000000} \color{black} 0.15 & {\cellcolor[HTML]{CDD9EC}} \color[HTML]{000000} \color{black} -0.23 & {\cellcolor[HTML]{000000}} \color[HTML]{F1F1F1} {\cellcolor[HTML]{FFFFFF}} \color{black} --- & {\cellcolor[HTML]{000000}} \color[HTML]{F1F1F1} {\cellcolor[HTML]{FFFFFF}} \color{black} --- \\
\cline{1-7}
\multirow[c]{2}{*}{Renn} & WK & {\cellcolor[HTML]{D44E41}} \color[HTML]{F1F1F1} \color{black} 1.23 & {\cellcolor[HTML]{96B7FF}} \color[HTML]{000000} \color{black} -0.75 & {\cellcolor[HTML]{B5CDFA}} \color[HTML]{000000} \color{black} -0.48 & {\cellcolor[HTML]{000000}} \color[HTML]{F1F1F1} {\cellcolor[HTML]{FFFFFF}} \color{black} --- & {\cellcolor[HTML]{000000}} \color[HTML]{F1F1F1} {\cellcolor[HTML]{FFFFFF}} \color{black} --- \\
 & WKND & {\cellcolor[HTML]{DC5D4A}} \color[HTML]{F1F1F1} \color{black} 1.14 & {\cellcolor[HTML]{9DBDFF}} \color[HTML]{000000} \color{black} -0.69 & {\cellcolor[HTML]{B7CFF9}} \color[HTML]{000000} \color{black} -0.45 & {\cellcolor[HTML]{000000}} \color[HTML]{F1F1F1} {\cellcolor[HTML]{FFFFFF}} \color{black} --- & {\cellcolor[HTML]{000000}} \color[HTML]{F1F1F1} {\cellcolor[HTML]{FFFFFF}} \color{black} --- \\
\cline{1-7}
\multirow[c]{2}{*}{Lill} & WK & {\cellcolor[HTML]{F6BEA4}} \color[HTML]{000000} \color{black} 0.41 & {\cellcolor[HTML]{D7DCE3}} \color[HTML]{000000} \color{black} -0.12 & {\cellcolor[HTML]{C7D7F0}} \color[HTML]{000000} \color{black} -0.29 & {\cellcolor[HTML]{000000}} \color[HTML]{F1F1F1} {\cellcolor[HTML]{FFFFFF}} \color{black} --- & {\cellcolor[HTML]{000000}} \color[HTML]{F1F1F1} {\cellcolor[HTML]{FFFFFF}} \color{black} --- \\
 & WKND & {\cellcolor[HTML]{F5C0A7}} \color[HTML]{000000} \color{black} 0.38 & {\cellcolor[HTML]{D8DCE2}} \color[HTML]{000000} \color{black} -0.11 & {\cellcolor[HTML]{C9D7F0}} \color[HTML]{000000} \color{black} -0.28 & {\cellcolor[HTML]{000000}} \color[HTML]{F1F1F1} {\cellcolor[HTML]{FFFFFF}} \color{black} --- & {\cellcolor[HTML]{000000}} \color[HTML]{F1F1F1} {\cellcolor[HTML]{FFFFFF}} \color{black} --- \\
\cline{1-7}
\multirow[c]{2}{*}{Bord} & WK & {\cellcolor[HTML]{F6BEA4}} \color[HTML]{000000} \color{black} 0.42 & {\cellcolor[HTML]{D2DBE8}} \color[HTML]{000000} \color{black} -0.17 & {\cellcolor[HTML]{CCD9ED}} \color[HTML]{000000} \color{black} -0.24 & {\cellcolor[HTML]{000000}} \color[HTML]{F1F1F1} {\cellcolor[HTML]{FFFFFF}} \color{black} --- & {\cellcolor[HTML]{000000}} \color[HTML]{F1F1F1} {\cellcolor[HTML]{FFFFFF}} \color{black} --- \\
 & WKND & {\cellcolor[HTML]{F2CAB5}} \color[HTML]{000000} \color{black} 0.28 & {\cellcolor[HTML]{D9DCE1}} \color[HTML]{000000} \color{black} -0.10 & {\cellcolor[HTML]{D2DBE8}} \color[HTML]{000000} \color{black} -0.18 & {\cellcolor[HTML]{000000}} \color[HTML]{F1F1F1} {\cellcolor[HTML]{FFFFFF}} \color{black} --- & {\cellcolor[HTML]{000000}} \color[HTML]{F1F1F1} {\cellcolor[HTML]{FFFFFF}} \color{black} --- \\
\cline{1-7}
\multirow[c]{2}{*}{Gren} & WK & {\cellcolor[HTML]{F7B79B}} \color[HTML]{000000} \color{black} 0.49 & {\cellcolor[HTML]{D2DBE8}} \color[HTML]{000000} \color{black} -0.18 & {\cellcolor[HTML]{C6D6F1}} \color[HTML]{000000} \color{black} -0.31 & {\cellcolor[HTML]{000000}} \color[HTML]{F1F1F1} {\cellcolor[HTML]{FFFFFF}} \color{black} --- & {\cellcolor[HTML]{000000}} \color[HTML]{F1F1F1} {\cellcolor[HTML]{FFFFFF}} \color{black} --- \\
 & WKND & {\cellcolor[HTML]{F7AC8E}} \color[HTML]{000000} \color{black} 0.58 & {\cellcolor[HTML]{D6DCE4}} \color[HTML]{000000} \color{black} -0.13 & {\cellcolor[HTML]{B7CFF9}} \color[HTML]{000000} \color{black} -0.45 & {\cellcolor[HTML]{000000}} \color[HTML]{F1F1F1} {\cellcolor[HTML]{FFFFFF}} \color{black} --- & {\cellcolor[HTML]{000000}} \color[HTML]{F1F1F1} {\cellcolor[HTML]{FFFFFF}} \color{black} --- \\
\cline{1-7}
\multirow[c]{2}{*}{Lyon} & WK & {\cellcolor[HTML]{F3C7B1}} \color[HTML]{000000} \color{black} 0.31 & {\cellcolor[HTML]{D2DBE8}} \color[HTML]{000000} \color{black} -0.18 & {\cellcolor[HTML]{D6DCE4}} \color[HTML]{000000} \color{black} -0.13 & {\cellcolor[HTML]{000000}} \color[HTML]{F1F1F1} {\cellcolor[HTML]{FFFFFF}} \color{black} --- & {\cellcolor[HTML]{000000}} \color[HTML]{F1F1F1} {\cellcolor[HTML]{FFFFFF}} \color{black} --- \\
 & WKND & {\cellcolor[HTML]{F1CCB8}} \color[HTML]{000000} \color{black} 0.25 & {\cellcolor[HTML]{D7DCE3}} \color[HTML]{000000} \color{black} -0.11 & {\cellcolor[HTML]{D5DBE5}} \color[HTML]{000000} \color{black} -0.14 & {\cellcolor[HTML]{000000}} \color[HTML]{F1F1F1} {\cellcolor[HTML]{FFFFFF}} \color{black} --- & {\cellcolor[HTML]{000000}} \color[HTML]{F1F1F1} {\cellcolor[HTML]{FFFFFF}} \color{black} --- \\
\cline{1-7}
\multirow[c]{2}{*}{Nant} & WK & {\cellcolor[HTML]{F1CDBA}} \color[HTML]{000000} \color{black} 0.24 & {\cellcolor[HTML]{B2CCFB}} \color[HTML]{000000} \color{black} -0.50 & {\cellcolor[HTML]{F1CCB8}} \color[HTML]{000000} \color{black} 0.25 & {\cellcolor[HTML]{000000}} \color[HTML]{F1F1F1} {\cellcolor[HTML]{FFFFFF}} \color{black} --- & {\cellcolor[HTML]{000000}} \color[HTML]{F1F1F1} {\cellcolor[HTML]{FFFFFF}} \color{black} --- \\
 & WKND & {\cellcolor[HTML]{F6BEA4}} \color[HTML]{000000} \color{black} 0.41 & {\cellcolor[HTML]{D2DBE8}} \color[HTML]{000000} \color{black} -0.18 & {\cellcolor[HTML]{CDD9EC}} \color[HTML]{000000} \color{black} -0.23 & {\cellcolor[HTML]{000000}} \color[HTML]{F1F1F1} {\cellcolor[HTML]{FFFFFF}} \color{black} --- & {\cellcolor[HTML]{000000}} \color[HTML]{F1F1F1} {\cellcolor[HTML]{FFFFFF}} \color{black} --- \\
\cline{1-7}
\multirow[c]{2}{*}{Toul} & WK & {\cellcolor[HTML]{F2CAB5}} \color[HTML]{000000} \color{black} 0.27 & {\cellcolor[HTML]{BCD2F7}} \color[HTML]{000000} \color{black} -0.41 & {\cellcolor[HTML]{EBD3C6}} \color[HTML]{000000} \color{black} 0.14 & {\cellcolor[HTML]{000000}} \color[HTML]{F1F1F1} {\cellcolor[HTML]{FFFFFF}} \color{black} --- & {\cellcolor[HTML]{000000}} \color[HTML]{F1F1F1} {\cellcolor[HTML]{FFFFFF}} \color{black} --- \\
 & WKND & {\cellcolor[HTML]{EDD2C3}} \color[HTML]{000000} \color{black} 0.17 & {\cellcolor[HTML]{C5D6F2}} \color[HTML]{000000} \color{black} -0.31 & {\cellcolor[HTML]{EBD3C6}} \color[HTML]{000000} \color{black} 0.15 & {\cellcolor[HTML]{000000}} \color[HTML]{F1F1F1} {\cellcolor[HTML]{FFFFFF}} \color{black} --- & {\cellcolor[HTML]{000000}} \color[HTML]{F1F1F1} {\cellcolor[HTML]{FFFFFF}} \color{black} --- \\
\cline{1-7}
\multirow[c]{2}{*}{Mont} & WK & {\cellcolor[HTML]{F1CCB8}} \color[HTML]{000000} \color{black} 0.25 & {\cellcolor[HTML]{BED2F6}} \color[HTML]{000000} \color{black} -0.38 & {\cellcolor[HTML]{EBD3C6}} \color[HTML]{000000} \color{black} 0.14 & {\cellcolor[HTML]{000000}} \color[HTML]{F1F1F1} {\cellcolor[HTML]{FFFFFF}} \color{black} --- & {\cellcolor[HTML]{000000}} \color[HTML]{F1F1F1} {\cellcolor[HTML]{FFFFFF}} \color{black} --- \\
 & WKND & {\cellcolor[HTML]{EDD2C3}} \color[HTML]{000000} \color{black} 0.16 & {\cellcolor[HTML]{C1D4F4}} \color[HTML]{000000} \color{black} -0.35 & {\cellcolor[HTML]{EED0C0}} \color[HTML]{000000} \color{black} 0.19 & {\cellcolor[HTML]{000000}} \color[HTML]{F1F1F1} {\cellcolor[HTML]{FFFFFF}} \color{black} --- & {\cellcolor[HTML]{000000}} \color[HTML]{F1F1F1} {\cellcolor[HTML]{FFFFFF}} \color{black} --- \\
\cline{1-7}
\multirow[c]{2}{*}{Tour} & WK & {\cellcolor[HTML]{F7BCA1}} \color[HTML]{000000} \color{black} 0.44 & {\cellcolor[HTML]{CCD9ED}} \color[HTML]{000000} \color{black} -0.24 & {\cellcolor[HTML]{EDD1C2}} \color[HTML]{000000} \color{black} 0.18 & {\cellcolor[HTML]{BFD3F6}} \color[HTML]{000000} \color{black} -0.38 & {\cellcolor[HTML]{E1DAD6}} \color[HTML]{000000} \color{black} 0.00 \\
 & WKND & {\cellcolor[HTML]{E5D8D1}} \color[HTML]{000000} \color{black} 0.05 & {\cellcolor[HTML]{F2CBB7}} \color[HTML]{000000} \color{black} 0.26 & {\cellcolor[HTML]{C5D6F2}} \color[HTML]{000000} \color{black} -0.32 & {\cellcolor[HTML]{000000}} \color[HTML]{F1F1F1} {\cellcolor[HTML]{FFFFFF}} \color{black} --- & {\cellcolor[HTML]{000000}} \color[HTML]{F1F1F1} {\cellcolor[HTML]{FFFFFF}} \color{black} --- \\
\cline{1-7}
\multirow[c]{2}{*}{Stra} & WK & {\cellcolor[HTML]{A7C5FE}} \color[HTML]{000000} \color{black} -0.60 & {\cellcolor[HTML]{DADCE0}} \color[HTML]{000000} \color{black} -0.08 & {\cellcolor[HTML]{DEDCDB}} \color[HTML]{000000} \color{black} -0.03 & {\cellcolor[HTML]{F59C7D}} \color[HTML]{000000} \color{black} 0.72 & {\cellcolor[HTML]{000000}} \color[HTML]{F1F1F1} {\cellcolor[HTML]{FFFFFF}} \color{black} --- \\
 & WKND & {\cellcolor[HTML]{A7C5FE}} \color[HTML]{000000} \color{black} -0.60 & {\cellcolor[HTML]{EDD1C2}} \color[HTML]{000000} \color{black} 0.18 & {\cellcolor[HTML]{F6BDA2}} \color[HTML]{000000} \color{black} 0.42 & {\cellcolor[HTML]{000000}} \color[HTML]{F1F1F1} {\cellcolor[HTML]{FFFFFF}} \color{black} --- & {\cellcolor[HTML]{000000}} \color[HTML]{F1F1F1} {\cellcolor[HTML]{FFFFFF}} \color{black} --- \\
\cline{1-7}
\multirow[c]{2}{*}{Orle} & WK & {\cellcolor[HTML]{EB7D62}} \color[HTML]{F1F1F1} \color{black} 0.94 & {\cellcolor[HTML]{A9C6FD}} \color[HTML]{000000} \color{black} -0.58 & {\cellcolor[HTML]{C1D4F4}} \color[HTML]{000000} \color{black} -0.36 & {\cellcolor[HTML]{000000}} \color[HTML]{F1F1F1} {\cellcolor[HTML]{FFFFFF}} \color{black} --- & {\cellcolor[HTML]{000000}} \color[HTML]{F1F1F1} {\cellcolor[HTML]{FFFFFF}} \color{black} --- \\
 & WKND & {\cellcolor[HTML]{EE8468}} \color[HTML]{F1F1F1} \color{black} 0.89 & {\cellcolor[HTML]{AAC7FD}} \color[HTML]{000000} \color{black} -0.57 & {\cellcolor[HTML]{C4D5F3}} \color[HTML]{000000} \color{black} -0.32 & {\cellcolor[HTML]{000000}} \color[HTML]{F1F1F1} {\cellcolor[HTML]{FFFFFF}} \color{black} --- & {\cellcolor[HTML]{000000}} \color[HTML]{F1F1F1} {\cellcolor[HTML]{FFFFFF}} \color{black} --- \\
\cline{1-7}
\multirow[c]{2}{*}{Metz} & WK & {\cellcolor[HTML]{F7B093}} \color[HTML]{000000} \color{black} 0.54 & {\cellcolor[HTML]{B5CDFA}} \color[HTML]{000000} \color{black} -0.48 & {\cellcolor[HTML]{DBDCDE}} \color[HTML]{000000} \color{black} -0.07 & {\cellcolor[HTML]{000000}} \color[HTML]{F1F1F1} {\cellcolor[HTML]{FFFFFF}} \color{black} --- & {\cellcolor[HTML]{000000}} \color[HTML]{F1F1F1} {\cellcolor[HTML]{FFFFFF}} \color{black} --- \\
 & WKND & {\cellcolor[HTML]{F6BEA4}} \color[HTML]{000000} \color{black} 0.41 & {\cellcolor[HTML]{BBD1F8}} \color[HTML]{000000} \color{black} -0.41 & {\cellcolor[HTML]{E1DAD6}} \color[HTML]{000000} \color{black} 0.00 & {\cellcolor[HTML]{000000}} \color[HTML]{F1F1F1} {\cellcolor[HTML]{FFFFFF}} \color{black} --- & {\cellcolor[HTML]{000000}} \color[HTML]{F1F1F1} {\cellcolor[HTML]{FFFFFF}} \color{black} --- \\
\cline{1-7}
\multirow[c]{2}{*}{Mans} & WK & {\cellcolor[HTML]{E1DAD6}} \color[HTML]{000000} \color{black} 0.00 & {\cellcolor[HTML]{F2CAB5}} \color[HTML]{000000} \color{black} 0.28 & {\cellcolor[HTML]{EFCFBF}} \color[HTML]{000000} \color{black} 0.20 & {\cellcolor[HTML]{B3CDFB}} \color[HTML]{000000} \color{black} -0.48 & {\cellcolor[HTML]{000000}} \color[HTML]{F1F1F1} {\cellcolor[HTML]{FFFFFF}} \color{black} --- \\
 & WKND & {\cellcolor[HTML]{DADCE0}} \color[HTML]{000000} \color{black} -0.08 & {\cellcolor[HTML]{F4C5AD}} \color[HTML]{000000} \color{black} 0.34 & {\cellcolor[HTML]{F2CBB7}} \color[HTML]{000000} \color{black} 0.26 & {\cellcolor[HTML]{AFCAFC}} \color[HTML]{000000} \color{black} -0.53 & {\cellcolor[HTML]{000000}} \color[HTML]{F1F1F1} {\cellcolor[HTML]{FFFFFF}} \color{black} --- \\
\cline{1-7}
\multirow[c]{2}{*}{Dijo} & WK & {\cellcolor[HTML]{F5A081}} \color[HTML]{000000} \color{black} 0.68 & {\cellcolor[HTML]{D7DCE3}} \color[HTML]{000000} \color{black} -0.12 & {\cellcolor[HTML]{ABC8FD}} \color[HTML]{000000} \color{black} -0.56 & {\cellcolor[HTML]{000000}} \color[HTML]{F1F1F1} {\cellcolor[HTML]{FFFFFF}} \color{black} --- & {\cellcolor[HTML]{000000}} \color[HTML]{F1F1F1} {\cellcolor[HTML]{FFFFFF}} \color{black} --- \\
 & WKND & {\cellcolor[HTML]{F7B497}} \color[HTML]{000000} \color{black} 0.50 & {\cellcolor[HTML]{BAD0F8}} \color[HTML]{000000} \color{black} -0.43 & {\cellcolor[HTML]{DBDCDE}} \color[HTML]{000000} \color{black} -0.07 & {\cellcolor[HTML]{000000}} \color[HTML]{F1F1F1} {\cellcolor[HTML]{FFFFFF}} \color{black} --- & {\cellcolor[HTML]{000000}} \color[HTML]{F1F1F1} {\cellcolor[HTML]{FFFFFF}} \color{black} --- \\
\cline{1-7}
\multirow[c]{2}{*}{Mars} & WK & {\cellcolor[HTML]{F4C5AD}} \color[HTML]{000000} \color{black} 0.34 & {\cellcolor[HTML]{C7D7F0}} \color[HTML]{000000} \color{black} -0.29 & {\cellcolor[HTML]{BCD2F7}} \color[HTML]{000000} \color{black} -0.40 & {\cellcolor[HTML]{F5C4AC}} \color[HTML]{000000} \color{black} 0.35 & {\cellcolor[HTML]{000000}} \color[HTML]{F1F1F1} {\cellcolor[HTML]{FFFFFF}} \color{black} --- \\
 & WKND & {\cellcolor[HTML]{F5C1A9}} \color[HTML]{000000} \color{black} 0.38 & {\cellcolor[HTML]{CEDAEB}} \color[HTML]{000000} \color{black} -0.21 & {\cellcolor[HTML]{D3DBE7}} \color[HTML]{000000} \color{black} -0.17 & {\cellcolor[HTML]{000000}} \color[HTML]{F1F1F1} {\cellcolor[HTML]{FFFFFF}} \color{black} --- & {\cellcolor[HTML]{000000}} \color[HTML]{F1F1F1} {\cellcolor[HTML]{FFFFFF}} \color{black} --- \\
\cline{1-7}
\multirow[c]{2}{*}{Nice} & WK & {\cellcolor[HTML]{C5D6F2}} \color[HTML]{000000} \color{black} -0.32 & {\cellcolor[HTML]{F7AF91}} \color[HTML]{000000} \color{black} 0.56 & {\cellcolor[HTML]{CCD9ED}} \color[HTML]{000000} \color{black} -0.24 & {\cellcolor[HTML]{000000}} \color[HTML]{F1F1F1} {\cellcolor[HTML]{FFFFFF}} \color{black} --- & {\cellcolor[HTML]{000000}} \color[HTML]{F1F1F1} {\cellcolor[HTML]{FFFFFF}} \color{black} --- \\
 & WKND & {\cellcolor[HTML]{D4DBE6}} \color[HTML]{000000} \color{black} -0.16 & {\cellcolor[HTML]{EED0C0}} \color[HTML]{000000} \color{black} 0.19 & {\cellcolor[HTML]{DEDCDB}} \color[HTML]{000000} \color{black} -0.03 & {\cellcolor[HTML]{000000}} \color[HTML]{F1F1F1} {\cellcolor[HTML]{FFFFFF}} \color{black} --- & {\cellcolor[HTML]{000000}} \color[HTML]{F1F1F1} {\cellcolor[HTML]{FFFFFF}} \color{black} --- \\
\cline{1-7}
\multirow[c]{2}{*}{Nanc} & WK & {\cellcolor[HTML]{E8765C}} \color[HTML]{F1F1F1} \color{black} 0.99 & {\cellcolor[HTML]{F7B396}} \color[HTML]{000000} \color{black} 0.53 & {\cellcolor[HTML]{B6CEFA}} \color[HTML]{000000} \color{black} -0.46 & {\cellcolor[HTML]{7396F5}} \color[HTML]{F1F1F1} \color{black} -1.06 & {\cellcolor[HTML]{000000}} \color[HTML]{F1F1F1} {\cellcolor[HTML]{FFFFFF}} \color{black} --- \\
 & WKND & {\cellcolor[HTML]{F7AF91}} \color[HTML]{000000} \color{black} 0.56 & {\cellcolor[HTML]{E5D8D1}} \color[HTML]{000000} \color{black} 0.06 & {\cellcolor[HTML]{A5C3FE}} \color[HTML]{000000} \color{black} -0.62 & {\cellcolor[HTML]{000000}} \color[HTML]{F1F1F1} {\cellcolor[HTML]{FFFFFF}} \color{black} --- & {\cellcolor[HTML]{000000}} \color[HTML]{F1F1F1} {\cellcolor[HTML]{FFFFFF}} \color{black} --- \\
\cline{1-7}
\bottomrule
\end{tabular}
\end{table}

\begin{table}
\centering
\caption[Organised activity diversity coefficient estimates for local-level multinomial logistic regression predictive model]{Organised activity diversity coefficient estimates for local-level multinomial logistic regression predictive model for the week and weekend and by cluster and by city (Paris--Nancy shown in size order). The figure contains the strength of the association between each of the covariates. Coefficients are coloured according to the strength of the relationship. Data preprocessing was applied to the \texttt{NetMob23} dataset~\citep{martinez-duriveNetMob23DatasetHighresolution2023}. City names and Week/Weekend labels are abbreviated for brevity.}
\label{tab:organised-activity-diversity-coefficients}
\begin{tabular}{llrrrrr}
\toprule
 & Class & 1 & 2 & 3 & 4 & 5 \\
City & WK/WKND &  &  &  &  &  \\
\midrule
\multirow[c]{2}{*}{Pari} & WK & {\cellcolor[HTML]{F4C5AD}} \color[HTML]{000000} \color{black} 0.33 & {\cellcolor[HTML]{EDD1C2}} \color[HTML]{000000} \color{black} 0.18 & {\cellcolor[HTML]{B1CBFC}} \color[HTML]{000000} \color{black} -0.52 & {\cellcolor[HTML]{000000}} \color[HTML]{F1F1F1} {\cellcolor[HTML]{FFFFFF}} \color{black} --- & {\cellcolor[HTML]{000000}} \color[HTML]{F1F1F1} {\cellcolor[HTML]{FFFFFF}} \color{black} --- \\
 & WKND & {\cellcolor[HTML]{E9D5CB}} \color[HTML]{000000} \color{black} 0.10 & {\cellcolor[HTML]{DDDCDC}} \color[HTML]{000000} \color{black} -0.04 & {\cellcolor[HTML]{DBDCDE}} \color[HTML]{000000} \color{black} -0.07 & {\cellcolor[HTML]{000000}} \color[HTML]{F1F1F1} {\cellcolor[HTML]{FFFFFF}} \color{black} --- & {\cellcolor[HTML]{000000}} \color[HTML]{F1F1F1} {\cellcolor[HTML]{FFFFFF}} \color{black} --- \\
\cline{1-7}
\multirow[c]{2}{*}{Renn} & WK & {\cellcolor[HTML]{E0DBD8}} \color[HTML]{000000} \color{black} -0.01 & {\cellcolor[HTML]{D3DBE7}} \color[HTML]{000000} \color{black} -0.16 & {\cellcolor[HTML]{EDD2C3}} \color[HTML]{000000} \color{black} 0.17 & {\cellcolor[HTML]{000000}} \color[HTML]{F1F1F1} {\cellcolor[HTML]{FFFFFF}} \color{black} --- & {\cellcolor[HTML]{000000}} \color[HTML]{F1F1F1} {\cellcolor[HTML]{FFFFFF}} \color{black} --- \\
 & WKND & {\cellcolor[HTML]{DADCE0}} \color[HTML]{000000} \color{black} -0.08 & {\cellcolor[HTML]{C9D7F0}} \color[HTML]{000000} \color{black} -0.27 & {\cellcolor[HTML]{F5C2AA}} \color[HTML]{000000} \color{black} 0.36 & {\cellcolor[HTML]{000000}} \color[HTML]{F1F1F1} {\cellcolor[HTML]{FFFFFF}} \color{black} --- & {\cellcolor[HTML]{000000}} \color[HTML]{F1F1F1} {\cellcolor[HTML]{FFFFFF}} \color{black} --- \\
\cline{1-7}
\multirow[c]{2}{*}{Lill} & WK & {\cellcolor[HTML]{F1CDBA}} \color[HTML]{000000} \color{black} 0.24 & {\cellcolor[HTML]{D4DBE6}} \color[HTML]{000000} \color{black} -0.16 & {\cellcolor[HTML]{DADCE0}} \color[HTML]{000000} \color{black} -0.08 & {\cellcolor[HTML]{000000}} \color[HTML]{F1F1F1} {\cellcolor[HTML]{FFFFFF}} \color{black} --- & {\cellcolor[HTML]{000000}} \color[HTML]{F1F1F1} {\cellcolor[HTML]{FFFFFF}} \color{black} --- \\
 & WKND & {\cellcolor[HTML]{F1CDBA}} \color[HTML]{000000} \color{black} 0.24 & {\cellcolor[HTML]{CFDAEA}} \color[HTML]{000000} \color{black} -0.21 & {\cellcolor[HTML]{DEDCDB}} \color[HTML]{000000} \color{black} -0.03 & {\cellcolor[HTML]{000000}} \color[HTML]{F1F1F1} {\cellcolor[HTML]{FFFFFF}} \color{black} --- & {\cellcolor[HTML]{000000}} \color[HTML]{F1F1F1} {\cellcolor[HTML]{FFFFFF}} \color{black} --- \\
\cline{1-7}
\multirow[c]{2}{*}{Bord} & WK & {\cellcolor[HTML]{F2C9B4}} \color[HTML]{000000} \color{black} 0.28 & {\cellcolor[HTML]{DBDCDE}} \color[HTML]{000000} \color{black} -0.07 & {\cellcolor[HTML]{CEDAEB}} \color[HTML]{000000} \color{black} -0.21 & {\cellcolor[HTML]{000000}} \color[HTML]{F1F1F1} {\cellcolor[HTML]{FFFFFF}} \color{black} --- & {\cellcolor[HTML]{000000}} \color[HTML]{F1F1F1} {\cellcolor[HTML]{FFFFFF}} \color{black} --- \\
 & WKND & {\cellcolor[HTML]{F3C8B2}} \color[HTML]{000000} \color{black} 0.30 & {\cellcolor[HTML]{D6DCE4}} \color[HTML]{000000} \color{black} -0.12 & {\cellcolor[HTML]{D2DBE8}} \color[HTML]{000000} \color{black} -0.18 & {\cellcolor[HTML]{000000}} \color[HTML]{F1F1F1} {\cellcolor[HTML]{FFFFFF}} \color{black} --- & {\cellcolor[HTML]{000000}} \color[HTML]{F1F1F1} {\cellcolor[HTML]{FFFFFF}} \color{black} --- \\
\cline{1-7}
\multirow[c]{2}{*}{Gren} & WK & {\cellcolor[HTML]{E8D6CC}} \color[HTML]{000000} \color{black} 0.09 & {\cellcolor[HTML]{D1DAE9}} \color[HTML]{000000} \color{black} -0.19 & {\cellcolor[HTML]{E9D5CB}} \color[HTML]{000000} \color{black} 0.10 & {\cellcolor[HTML]{000000}} \color[HTML]{F1F1F1} {\cellcolor[HTML]{FFFFFF}} \color{black} --- & {\cellcolor[HTML]{000000}} \color[HTML]{F1F1F1} {\cellcolor[HTML]{FFFFFF}} \color{black} --- \\
 & WKND & {\cellcolor[HTML]{D9DCE1}} \color[HTML]{000000} \color{black} -0.09 & {\cellcolor[HTML]{C1D4F4}} \color[HTML]{000000} \color{black} -0.35 & {\cellcolor[HTML]{F7BCA1}} \color[HTML]{000000} \color{black} 0.44 & {\cellcolor[HTML]{000000}} \color[HTML]{F1F1F1} {\cellcolor[HTML]{FFFFFF}} \color{black} --- & {\cellcolor[HTML]{000000}} \color[HTML]{F1F1F1} {\cellcolor[HTML]{FFFFFF}} \color{black} --- \\
\cline{1-7}
\multirow[c]{2}{*}{Lyon} & WK & {\cellcolor[HTML]{B3CDFB}} \color[HTML]{000000} \color{black} -0.49 & {\cellcolor[HTML]{D1DAE9}} \color[HTML]{000000} \color{black} -0.19 & {\cellcolor[HTML]{F5A081}} \color[HTML]{000000} \color{black} 0.68 & {\cellcolor[HTML]{000000}} \color[HTML]{F1F1F1} {\cellcolor[HTML]{FFFFFF}} \color{black} --- & {\cellcolor[HTML]{000000}} \color[HTML]{F1F1F1} {\cellcolor[HTML]{FFFFFF}} \color{black} --- \\
 & WKND & {\cellcolor[HTML]{B5CDFA}} \color[HTML]{000000} \color{black} -0.48 & {\cellcolor[HTML]{CFDAEA}} \color[HTML]{000000} \color{black} -0.21 & {\cellcolor[HTML]{F5A081}} \color[HTML]{000000} \color{black} 0.68 & {\cellcolor[HTML]{000000}} \color[HTML]{F1F1F1} {\cellcolor[HTML]{FFFFFF}} \color{black} --- & {\cellcolor[HTML]{000000}} \color[HTML]{F1F1F1} {\cellcolor[HTML]{FFFFFF}} \color{black} --- \\
\cline{1-7}
\multirow[c]{2}{*}{Nant} & WK & {\cellcolor[HTML]{EAD5C9}} \color[HTML]{000000} \color{black} 0.11 & {\cellcolor[HTML]{DFDBD9}} \color[HTML]{000000} \color{black} -0.02 & {\cellcolor[HTML]{D9DCE1}} \color[HTML]{000000} \color{black} -0.09 & {\cellcolor[HTML]{000000}} \color[HTML]{F1F1F1} {\cellcolor[HTML]{FFFFFF}} \color{black} --- & {\cellcolor[HTML]{000000}} \color[HTML]{F1F1F1} {\cellcolor[HTML]{FFFFFF}} \color{black} --- \\
 & WKND & {\cellcolor[HTML]{F0CDBB}} \color[HTML]{000000} \color{black} 0.23 & {\cellcolor[HTML]{E6D7CF}} \color[HTML]{000000} \color{black} 0.07 & {\cellcolor[HTML]{C6D6F1}} \color[HTML]{000000} \color{black} -0.30 & {\cellcolor[HTML]{000000}} \color[HTML]{F1F1F1} {\cellcolor[HTML]{FFFFFF}} \color{black} --- & {\cellcolor[HTML]{000000}} \color[HTML]{F1F1F1} {\cellcolor[HTML]{FFFFFF}} \color{black} --- \\
\cline{1-7}
\multirow[c]{2}{*}{Toul} & WK & {\cellcolor[HTML]{D9DCE1}} \color[HTML]{000000} \color{black} -0.10 & {\cellcolor[HTML]{DBDCDE}} \color[HTML]{000000} \color{black} -0.06 & {\cellcolor[HTML]{EDD2C3}} \color[HTML]{000000} \color{black} 0.16 & {\cellcolor[HTML]{000000}} \color[HTML]{F1F1F1} {\cellcolor[HTML]{FFFFFF}} \color{black} --- & {\cellcolor[HTML]{000000}} \color[HTML]{F1F1F1} {\cellcolor[HTML]{FFFFFF}} \color{black} --- \\
 & WKND & {\cellcolor[HTML]{DEDCDB}} \color[HTML]{000000} \color{black} -0.03 & {\cellcolor[HTML]{DBDCDE}} \color[HTML]{000000} \color{black} -0.07 & {\cellcolor[HTML]{E8D6CC}} \color[HTML]{000000} \color{black} 0.10 & {\cellcolor[HTML]{000000}} \color[HTML]{F1F1F1} {\cellcolor[HTML]{FFFFFF}} \color{black} --- & {\cellcolor[HTML]{000000}} \color[HTML]{F1F1F1} {\cellcolor[HTML]{FFFFFF}} \color{black} --- \\
\cline{1-7}
\multirow[c]{2}{*}{Mont} & WK & {\cellcolor[HTML]{DDDCDC}} \color[HTML]{000000} \color{black} -0.04 & {\cellcolor[HTML]{C3D5F4}} \color[HTML]{000000} \color{black} -0.34 & {\cellcolor[HTML]{F5C1A9}} \color[HTML]{000000} \color{black} 0.38 & {\cellcolor[HTML]{000000}} \color[HTML]{F1F1F1} {\cellcolor[HTML]{FFFFFF}} \color{black} --- & {\cellcolor[HTML]{000000}} \color[HTML]{F1F1F1} {\cellcolor[HTML]{FFFFFF}} \color{black} --- \\
 & WKND & {\cellcolor[HTML]{D4DBE6}} \color[HTML]{000000} \color{black} -0.15 & {\cellcolor[HTML]{BCD2F7}} \color[HTML]{000000} \color{black} -0.40 & {\cellcolor[HTML]{F7AF91}} \color[HTML]{000000} \color{black} 0.55 & {\cellcolor[HTML]{000000}} \color[HTML]{F1F1F1} {\cellcolor[HTML]{FFFFFF}} \color{black} --- & {\cellcolor[HTML]{000000}} \color[HTML]{F1F1F1} {\cellcolor[HTML]{FFFFFF}} \color{black} --- \\
\cline{1-7}
\multirow[c]{2}{*}{Tour} & WK & {\cellcolor[HTML]{F6A385}} \color[HTML]{000000} \color{black} 0.65 & {\cellcolor[HTML]{ECD3C5}} \color[HTML]{000000} \color{black} 0.15 & {\cellcolor[HTML]{E9D5CB}} \color[HTML]{000000} \color{black} 0.10 & {\cellcolor[HTML]{B3CDFB}} \color[HTML]{000000} \color{black} -0.49 & {\cellcolor[HTML]{BAD0F8}} \color[HTML]{000000} \color{black} -0.42 \\
 & WKND & {\cellcolor[HTML]{CAD8EF}} \color[HTML]{000000} \color{black} -0.26 & {\cellcolor[HTML]{D4DBE6}} \color[HTML]{000000} \color{black} -0.16 & {\cellcolor[HTML]{F6BDA2}} \color[HTML]{000000} \color{black} 0.42 & {\cellcolor[HTML]{000000}} \color[HTML]{F1F1F1} {\cellcolor[HTML]{FFFFFF}} \color{black} --- & {\cellcolor[HTML]{000000}} \color[HTML]{F1F1F1} {\cellcolor[HTML]{FFFFFF}} \color{black} --- \\
\cline{1-7}
\multirow[c]{2}{*}{Stra} & WK & {\cellcolor[HTML]{D6DCE4}} \color[HTML]{000000} \color{black} -0.13 & {\cellcolor[HTML]{ECD3C5}} \color[HTML]{000000} \color{black} 0.15 & {\cellcolor[HTML]{F2C9B4}} \color[HTML]{000000} \color{black} 0.29 & {\cellcolor[HTML]{C5D6F2}} \color[HTML]{000000} \color{black} -0.31 & {\cellcolor[HTML]{000000}} \color[HTML]{F1F1F1} {\cellcolor[HTML]{FFFFFF}} \color{black} --- \\
 & WKND & {\cellcolor[HTML]{D5DBE5}} \color[HTML]{000000} \color{black} -0.14 & {\cellcolor[HTML]{E1DAD6}} \color[HTML]{000000} \color{black} 0.01 & {\cellcolor[HTML]{EAD4C8}} \color[HTML]{000000} \color{black} 0.13 & {\cellcolor[HTML]{000000}} \color[HTML]{F1F1F1} {\cellcolor[HTML]{FFFFFF}} \color{black} --- & {\cellcolor[HTML]{000000}} \color[HTML]{F1F1F1} {\cellcolor[HTML]{FFFFFF}} \color{black} --- \\
\cline{1-7}
\multirow[c]{2}{*}{Orle} & WK & {\cellcolor[HTML]{EDD1C2}} \color[HTML]{000000} \color{black} 0.18 & {\cellcolor[HTML]{B2CCFB}} \color[HTML]{000000} \color{black} -0.50 & {\cellcolor[HTML]{F4C6AF}} \color[HTML]{000000} \color{black} 0.32 & {\cellcolor[HTML]{000000}} \color[HTML]{F1F1F1} {\cellcolor[HTML]{FFFFFF}} \color{black} --- & {\cellcolor[HTML]{000000}} \color[HTML]{F1F1F1} {\cellcolor[HTML]{FFFFFF}} \color{black} --- \\
 & WKND & {\cellcolor[HTML]{EFCEBD}} \color[HTML]{000000} \color{black} 0.22 & {\cellcolor[HTML]{C1D4F4}} \color[HTML]{000000} \color{black} -0.35 & {\cellcolor[HTML]{EBD3C6}} \color[HTML]{000000} \color{black} 0.13 & {\cellcolor[HTML]{000000}} \color[HTML]{F1F1F1} {\cellcolor[HTML]{FFFFFF}} \color{black} --- & {\cellcolor[HTML]{000000}} \color[HTML]{F1F1F1} {\cellcolor[HTML]{FFFFFF}} \color{black} --- \\
\cline{1-7}
\multirow[c]{2}{*}{Metz} & WK & {\cellcolor[HTML]{E2DAD5}} \color[HTML]{000000} \color{black} 0.02 & {\cellcolor[HTML]{BED2F6}} \color[HTML]{000000} \color{black} -0.39 & {\cellcolor[HTML]{F5C1A9}} \color[HTML]{000000} \color{black} 0.37 & {\cellcolor[HTML]{000000}} \color[HTML]{F1F1F1} {\cellcolor[HTML]{FFFFFF}} \color{black} --- & {\cellcolor[HTML]{000000}} \color[HTML]{F1F1F1} {\cellcolor[HTML]{FFFFFF}} \color{black} --- \\
 & WKND & {\cellcolor[HTML]{E0DBD8}} \color[HTML]{000000} \color{black} -0.00 & {\cellcolor[HTML]{BCD2F7}} \color[HTML]{000000} \color{black} -0.40 & {\cellcolor[HTML]{F6BFA6}} \color[HTML]{000000} \color{black} 0.40 & {\cellcolor[HTML]{000000}} \color[HTML]{F1F1F1} {\cellcolor[HTML]{FFFFFF}} \color{black} --- & {\cellcolor[HTML]{000000}} \color[HTML]{F1F1F1} {\cellcolor[HTML]{FFFFFF}} \color{black} --- \\
\cline{1-7}
\multirow[c]{2}{*}{Mans} & WK & {\cellcolor[HTML]{DEDCDB}} \color[HTML]{000000} \color{black} -0.04 & {\cellcolor[HTML]{F2CBB7}} \color[HTML]{000000} \color{black} 0.26 & {\cellcolor[HTML]{ECD3C5}} \color[HTML]{000000} \color{black} 0.16 & {\cellcolor[HTML]{BED2F6}} \color[HTML]{000000} \color{black} -0.38 & {\cellcolor[HTML]{000000}} \color[HTML]{F1F1F1} {\cellcolor[HTML]{FFFFFF}} \color{black} --- \\
 & WKND & {\cellcolor[HTML]{E0DBD8}} \color[HTML]{000000} \color{black} -0.00 & {\cellcolor[HTML]{E5D8D1}} \color[HTML]{000000} \color{black} 0.05 & {\cellcolor[HTML]{EFCFBF}} \color[HTML]{000000} \color{black} 0.20 & {\cellcolor[HTML]{CBD8EE}} \color[HTML]{000000} \color{black} -0.25 & {\cellcolor[HTML]{000000}} \color[HTML]{F1F1F1} {\cellcolor[HTML]{FFFFFF}} \color{black} --- \\
\cline{1-7}
\multirow[c]{2}{*}{Dijo} & WK & {\cellcolor[HTML]{F5C0A7}} \color[HTML]{000000} \color{black} 0.38 & {\cellcolor[HTML]{BAD0F8}} \color[HTML]{000000} \color{black} -0.43 & {\cellcolor[HTML]{E4D9D2}} \color[HTML]{000000} \color{black} 0.05 & {\cellcolor[HTML]{000000}} \color[HTML]{F1F1F1} {\cellcolor[HTML]{FFFFFF}} \color{black} --- & {\cellcolor[HTML]{000000}} \color[HTML]{F1F1F1} {\cellcolor[HTML]{FFFFFF}} \color{black} --- \\
 & WKND & {\cellcolor[HTML]{F2C9B4}} \color[HTML]{000000} \color{black} 0.29 & {\cellcolor[HTML]{A7C5FE}} \color[HTML]{000000} \color{black} -0.60 & {\cellcolor[HTML]{F3C8B2}} \color[HTML]{000000} \color{black} 0.31 & {\cellcolor[HTML]{000000}} \color[HTML]{F1F1F1} {\cellcolor[HTML]{FFFFFF}} \color{black} --- & {\cellcolor[HTML]{000000}} \color[HTML]{F1F1F1} {\cellcolor[HTML]{FFFFFF}} \color{black} --- \\
\cline{1-7}
\multirow[c]{2}{*}{Mars} & WK & {\cellcolor[HTML]{F7B79B}} \color[HTML]{000000} \color{black} 0.48 & {\cellcolor[HTML]{F7B99E}} \color[HTML]{000000} \color{black} 0.46 & {\cellcolor[HTML]{EFCFBF}} \color[HTML]{000000} \color{black} 0.20 & {\cellcolor[HTML]{6B8DF0}} \color[HTML]{F1F1F1} \color{black} -1.14 & {\cellcolor[HTML]{000000}} \color[HTML]{F1F1F1} {\cellcolor[HTML]{FFFFFF}} \color{black} --- \\
 & WKND & {\cellcolor[HTML]{EED0C0}} \color[HTML]{000000} \color{black} 0.18 & {\cellcolor[HTML]{E5D8D1}} \color[HTML]{000000} \color{black} 0.05 & {\cellcolor[HTML]{CCD9ED}} \color[HTML]{000000} \color{black} -0.24 & {\cellcolor[HTML]{000000}} \color[HTML]{F1F1F1} {\cellcolor[HTML]{FFFFFF}} \color{black} --- & {\cellcolor[HTML]{000000}} \color[HTML]{F1F1F1} {\cellcolor[HTML]{FFFFFF}} \color{black} --- \\
\cline{1-7}
\multirow[c]{2}{*}{Nice} & WK & {\cellcolor[HTML]{F1CDBA}} \color[HTML]{000000} \color{black} 0.24 & {\cellcolor[HTML]{D2DBE8}} \color[HTML]{000000} \color{black} -0.17 & {\cellcolor[HTML]{DBDCDE}} \color[HTML]{000000} \color{black} -0.07 & {\cellcolor[HTML]{000000}} \color[HTML]{F1F1F1} {\cellcolor[HTML]{FFFFFF}} \color{black} --- & {\cellcolor[HTML]{000000}} \color[HTML]{F1F1F1} {\cellcolor[HTML]{FFFFFF}} \color{black} --- \\
 & WKND & {\cellcolor[HTML]{F6BEA4}} \color[HTML]{000000} \color{black} 0.41 & {\cellcolor[HTML]{DEDCDB}} \color[HTML]{000000} \color{black} -0.03 & {\cellcolor[HTML]{BFD3F6}} \color[HTML]{000000} \color{black} -0.38 & {\cellcolor[HTML]{000000}} \color[HTML]{F1F1F1} {\cellcolor[HTML]{FFFFFF}} \color{black} --- & {\cellcolor[HTML]{000000}} \color[HTML]{F1F1F1} {\cellcolor[HTML]{FFFFFF}} \color{black} --- \\
\cline{1-7}
\multirow[c]{2}{*}{Nanc} & WK & {\cellcolor[HTML]{EED0C0}} \color[HTML]{000000} \color{black} 0.19 & {\cellcolor[HTML]{B6CEFA}} \color[HTML]{000000} \color{black} -0.46 & {\cellcolor[HTML]{DEDCDB}} \color[HTML]{000000} \color{black} -0.03 & {\cellcolor[HTML]{F3C8B2}} \color[HTML]{000000} \color{black} 0.30 & {\cellcolor[HTML]{000000}} \color[HTML]{F1F1F1} {\cellcolor[HTML]{FFFFFF}} \color{black} --- \\
 & WKND & {\cellcolor[HTML]{EFCFBF}} \color[HTML]{000000} \color{black} 0.20 & {\cellcolor[HTML]{C0D4F5}} \color[HTML]{000000} \color{black} -0.36 & {\cellcolor[HTML]{ECD3C5}} \color[HTML]{000000} \color{black} 0.15 & {\cellcolor[HTML]{000000}} \color[HTML]{F1F1F1} {\cellcolor[HTML]{FFFFFF}} \color{black} --- & {\cellcolor[HTML]{000000}} \color[HTML]{F1F1F1} {\cellcolor[HTML]{FFFFFF}} \color{black} --- \\
\cline{1-7}
\bottomrule
\end{tabular}
\end{table}

\begin{table}
\centering
\caption[Outdoor diversity coefficient estimates for local-level multinomial logistic regression predictive model]{Outdoor diversity coefficient estimates for local-level multinomial logistic regression predictive model for the week and weekend and by cluster and by city (Paris--Nancy shown in size order). The figure contains the strength of the association between each of the covariates. Coefficients are coloured according to the strength of the relationship. Data preprocessing was applied to the \texttt{NetMob23} dataset~\citep{martinez-duriveNetMob23DatasetHighresolution2023}. City names and Week/Weekend labels are abbreviated for brevity.}
\label{tab:outdoor-diversity-coefficients}
\begin{tabular}{llrrrrr}
\toprule
 & Class & 1 & 2 & 3 & 4 & 5 \\
City & WK/WKND &  &  &  &  &  \\
\midrule
\multirow[c]{2}{*}{Pari} & WK & {\cellcolor[HTML]{E8D6CC}} \color[HTML]{000000} \color{black} 0.09 & {\cellcolor[HTML]{DBDCDE}} \color[HTML]{000000} \color{black} -0.07 & {\cellcolor[HTML]{DFDBD9}} \color[HTML]{000000} \color{black} -0.02 & {\cellcolor[HTML]{000000}} \color[HTML]{F1F1F1} {\cellcolor[HTML]{FFFFFF}} \color{black} --- & {\cellcolor[HTML]{000000}} \color[HTML]{F1F1F1} {\cellcolor[HTML]{FFFFFF}} \color{black} --- \\
 & WKND & {\cellcolor[HTML]{E0DBD8}} \color[HTML]{000000} \color{black} -0.01 & {\cellcolor[HTML]{CFDAEA}} \color[HTML]{000000} \color{black} -0.21 & {\cellcolor[HTML]{EFCEBD}} \color[HTML]{000000} \color{black} 0.22 & {\cellcolor[HTML]{000000}} \color[HTML]{F1F1F1} {\cellcolor[HTML]{FFFFFF}} \color{black} --- & {\cellcolor[HTML]{000000}} \color[HTML]{F1F1F1} {\cellcolor[HTML]{FFFFFF}} \color{black} --- \\
\cline{1-7}
\multirow[c]{2}{*}{Renn} & WK & {\cellcolor[HTML]{DBDCDE}} \color[HTML]{000000} \color{black} -0.07 & {\cellcolor[HTML]{E8D6CC}} \color[HTML]{000000} \color{black} 0.09 & {\cellcolor[HTML]{DFDBD9}} \color[HTML]{000000} \color{black} -0.02 & {\cellcolor[HTML]{000000}} \color[HTML]{F1F1F1} {\cellcolor[HTML]{FFFFFF}} \color{black} --- & {\cellcolor[HTML]{000000}} \color[HTML]{F1F1F1} {\cellcolor[HTML]{FFFFFF}} \color{black} --- \\
 & WKND & {\cellcolor[HTML]{DADCE0}} \color[HTML]{000000} \color{black} -0.08 & {\cellcolor[HTML]{E3D9D3}} \color[HTML]{000000} \color{black} 0.03 & {\cellcolor[HTML]{E5D8D1}} \color[HTML]{000000} \color{black} 0.05 & {\cellcolor[HTML]{000000}} \color[HTML]{F1F1F1} {\cellcolor[HTML]{FFFFFF}} \color{black} --- & {\cellcolor[HTML]{000000}} \color[HTML]{F1F1F1} {\cellcolor[HTML]{FFFFFF}} \color{black} --- \\
\cline{1-7}
\multirow[c]{2}{*}{Lill} & WK & {\cellcolor[HTML]{DBDCDE}} \color[HTML]{000000} \color{black} -0.07 & {\cellcolor[HTML]{F2CAB5}} \color[HTML]{000000} \color{black} 0.28 & {\cellcolor[HTML]{CEDAEB}} \color[HTML]{000000} \color{black} -0.21 & {\cellcolor[HTML]{000000}} \color[HTML]{F1F1F1} {\cellcolor[HTML]{FFFFFF}} \color{black} --- & {\cellcolor[HTML]{000000}} \color[HTML]{F1F1F1} {\cellcolor[HTML]{FFFFFF}} \color{black} --- \\
 & WKND & {\cellcolor[HTML]{DADCE0}} \color[HTML]{000000} \color{black} -0.08 & {\cellcolor[HTML]{F3C7B1}} \color[HTML]{000000} \color{black} 0.32 & {\cellcolor[HTML]{CDD9EC}} \color[HTML]{000000} \color{black} -0.23 & {\cellcolor[HTML]{000000}} \color[HTML]{F1F1F1} {\cellcolor[HTML]{FFFFFF}} \color{black} --- & {\cellcolor[HTML]{000000}} \color[HTML]{F1F1F1} {\cellcolor[HTML]{FFFFFF}} \color{black} --- \\
\cline{1-7}
\multirow[c]{2}{*}{Bord} & WK & {\cellcolor[HTML]{DFDBD9}} \color[HTML]{000000} \color{black} -0.02 & {\cellcolor[HTML]{D2DBE8}} \color[HTML]{000000} \color{black} -0.18 & {\cellcolor[HTML]{EFCFBF}} \color[HTML]{000000} \color{black} 0.20 & {\cellcolor[HTML]{000000}} \color[HTML]{F1F1F1} {\cellcolor[HTML]{FFFFFF}} \color{black} --- & {\cellcolor[HTML]{000000}} \color[HTML]{F1F1F1} {\cellcolor[HTML]{FFFFFF}} \color{black} --- \\
 & WKND & {\cellcolor[HTML]{D5DBE5}} \color[HTML]{000000} \color{black} -0.15 & {\cellcolor[HTML]{E4D9D2}} \color[HTML]{000000} \color{black} 0.04 & {\cellcolor[HTML]{E9D5CB}} \color[HTML]{000000} \color{black} 0.11 & {\cellcolor[HTML]{000000}} \color[HTML]{F1F1F1} {\cellcolor[HTML]{FFFFFF}} \color{black} --- & {\cellcolor[HTML]{000000}} \color[HTML]{F1F1F1} {\cellcolor[HTML]{FFFFFF}} \color{black} --- \\
\cline{1-7}
\multirow[c]{2}{*}{Gren} & WK & {\cellcolor[HTML]{C0D4F5}} \color[HTML]{000000} \color{black} -0.36 & {\cellcolor[HTML]{E8D6CC}} \color[HTML]{000000} \color{black} 0.10 & {\cellcolor[HTML]{F2CBB7}} \color[HTML]{000000} \color{black} 0.26 & {\cellcolor[HTML]{000000}} \color[HTML]{F1F1F1} {\cellcolor[HTML]{FFFFFF}} \color{black} --- & {\cellcolor[HTML]{000000}} \color[HTML]{F1F1F1} {\cellcolor[HTML]{FFFFFF}} \color{black} --- \\
 & WKND & {\cellcolor[HTML]{C0D4F5}} \color[HTML]{000000} \color{black} -0.36 & {\cellcolor[HTML]{E8D6CC}} \color[HTML]{000000} \color{black} 0.09 & {\cellcolor[HTML]{F2CBB7}} \color[HTML]{000000} \color{black} 0.27 & {\cellcolor[HTML]{000000}} \color[HTML]{F1F1F1} {\cellcolor[HTML]{FFFFFF}} \color{black} --- & {\cellcolor[HTML]{000000}} \color[HTML]{F1F1F1} {\cellcolor[HTML]{FFFFFF}} \color{black} --- \\
\cline{1-7}
\multirow[c]{2}{*}{Lyon} & WK & {\cellcolor[HTML]{E2DAD5}} \color[HTML]{000000} \color{black} 0.02 & {\cellcolor[HTML]{F2C9B4}} \color[HTML]{000000} \color{black} 0.29 & {\cellcolor[HTML]{C6D6F1}} \color[HTML]{000000} \color{black} -0.31 & {\cellcolor[HTML]{000000}} \color[HTML]{F1F1F1} {\cellcolor[HTML]{FFFFFF}} \color{black} --- & {\cellcolor[HTML]{000000}} \color[HTML]{F1F1F1} {\cellcolor[HTML]{FFFFFF}} \color{black} --- \\
 & WKND & {\cellcolor[HTML]{E2DAD5}} \color[HTML]{000000} \color{black} 0.01 & {\cellcolor[HTML]{F2C9B4}} \color[HTML]{000000} \color{black} 0.29 & {\cellcolor[HTML]{C6D6F1}} \color[HTML]{000000} \color{black} -0.30 & {\cellcolor[HTML]{000000}} \color[HTML]{F1F1F1} {\cellcolor[HTML]{FFFFFF}} \color{black} --- & {\cellcolor[HTML]{000000}} \color[HTML]{F1F1F1} {\cellcolor[HTML]{FFFFFF}} \color{black} --- \\
\cline{1-7}
\multirow[c]{2}{*}{Nant} & WK & {\cellcolor[HTML]{DDDCDC}} \color[HTML]{000000} \color{black} -0.05 & {\cellcolor[HTML]{DADCE0}} \color[HTML]{000000} \color{black} -0.08 & {\cellcolor[HTML]{EAD4C8}} \color[HTML]{000000} \color{black} 0.12 & {\cellcolor[HTML]{000000}} \color[HTML]{F1F1F1} {\cellcolor[HTML]{FFFFFF}} \color{black} --- & {\cellcolor[HTML]{000000}} \color[HTML]{F1F1F1} {\cellcolor[HTML]{FFFFFF}} \color{black} --- \\
 & WKND & {\cellcolor[HTML]{DEDCDB}} \color[HTML]{000000} \color{black} -0.04 & {\cellcolor[HTML]{D6DCE4}} \color[HTML]{000000} \color{black} -0.13 & {\cellcolor[HTML]{EDD2C3}} \color[HTML]{000000} \color{black} 0.17 & {\cellcolor[HTML]{000000}} \color[HTML]{F1F1F1} {\cellcolor[HTML]{FFFFFF}} \color{black} --- & {\cellcolor[HTML]{000000}} \color[HTML]{F1F1F1} {\cellcolor[HTML]{FFFFFF}} \color{black} --- \\
\cline{1-7}
\multirow[c]{2}{*}{Toul} & WK & {\cellcolor[HTML]{DDDCDC}} \color[HTML]{000000} \color{black} -0.04 & {\cellcolor[HTML]{E3D9D3}} \color[HTML]{000000} \color{black} 0.03 & {\cellcolor[HTML]{E2DAD5}} \color[HTML]{000000} \color{black} 0.02 & {\cellcolor[HTML]{000000}} \color[HTML]{F1F1F1} {\cellcolor[HTML]{FFFFFF}} \color{black} --- & {\cellcolor[HTML]{000000}} \color[HTML]{F1F1F1} {\cellcolor[HTML]{FFFFFF}} \color{black} --- \\
 & WKND & {\cellcolor[HTML]{E2DAD5}} \color[HTML]{000000} \color{black} 0.02 & {\cellcolor[HTML]{EAD4C8}} \color[HTML]{000000} \color{black} 0.12 & {\cellcolor[HTML]{D5DBE5}} \color[HTML]{000000} \color{black} -0.14 & {\cellcolor[HTML]{000000}} \color[HTML]{F1F1F1} {\cellcolor[HTML]{FFFFFF}} \color{black} --- & {\cellcolor[HTML]{000000}} \color[HTML]{F1F1F1} {\cellcolor[HTML]{FFFFFF}} \color{black} --- \\
\cline{1-7}
\multirow[c]{2}{*}{Mont} & WK & {\cellcolor[HTML]{E6D7CF}} \color[HTML]{000000} \color{black} 0.07 & {\cellcolor[HTML]{E9D5CB}} \color[HTML]{000000} \color{black} 0.10 & {\cellcolor[HTML]{D3DBE7}} \color[HTML]{000000} \color{black} -0.17 & {\cellcolor[HTML]{000000}} \color[HTML]{F1F1F1} {\cellcolor[HTML]{FFFFFF}} \color{black} --- & {\cellcolor[HTML]{000000}} \color[HTML]{F1F1F1} {\cellcolor[HTML]{FFFFFF}} \color{black} --- \\
 & WKND & {\cellcolor[HTML]{E4D9D2}} \color[HTML]{000000} \color{black} 0.05 & {\cellcolor[HTML]{EFCFBF}} \color[HTML]{000000} \color{black} 0.20 & {\cellcolor[HTML]{CBD8EE}} \color[HTML]{000000} \color{black} -0.25 & {\cellcolor[HTML]{000000}} \color[HTML]{F1F1F1} {\cellcolor[HTML]{FFFFFF}} \color{black} --- & {\cellcolor[HTML]{000000}} \color[HTML]{F1F1F1} {\cellcolor[HTML]{FFFFFF}} \color{black} --- \\
\cline{1-7}
\multirow[c]{2}{*}{Tour} & WK & {\cellcolor[HTML]{ADC9FD}} \color[HTML]{000000} \color{black} -0.55 & {\cellcolor[HTML]{E3D9D3}} \color[HTML]{000000} \color{black} 0.03 & {\cellcolor[HTML]{F2CBB7}} \color[HTML]{000000} \color{black} 0.27 & {\cellcolor[HTML]{CBD8EE}} \color[HTML]{000000} \color{black} -0.25 & {\cellcolor[HTML]{F7B599}} \color[HTML]{000000} \color{black} 0.50 \\
 & WKND & {\cellcolor[HTML]{DEDCDB}} \color[HTML]{000000} \color{black} -0.03 & {\cellcolor[HTML]{EAD4C8}} \color[HTML]{000000} \color{black} 0.13 & {\cellcolor[HTML]{D8DCE2}} \color[HTML]{000000} \color{black} -0.10 & {\cellcolor[HTML]{000000}} \color[HTML]{F1F1F1} {\cellcolor[HTML]{FFFFFF}} \color{black} --- & {\cellcolor[HTML]{000000}} \color[HTML]{F1F1F1} {\cellcolor[HTML]{FFFFFF}} \color{black} --- \\
\cline{1-7}
\multirow[c]{2}{*}{Stra} & WK & {\cellcolor[HTML]{E6D7CF}} \color[HTML]{000000} \color{black} 0.07 & {\cellcolor[HTML]{EAD4C8}} \color[HTML]{000000} \color{black} 0.13 & {\cellcolor[HTML]{F2CAB5}} \color[HTML]{000000} \color{black} 0.27 & {\cellcolor[HTML]{B5CDFA}} \color[HTML]{000000} \color{black} -0.47 & {\cellcolor[HTML]{000000}} \color[HTML]{F1F1F1} {\cellcolor[HTML]{FFFFFF}} \color{black} --- \\
 & WKND & {\cellcolor[HTML]{DEDCDB}} \color[HTML]{000000} \color{black} -0.03 & {\cellcolor[HTML]{DADCE0}} \color[HTML]{000000} \color{black} -0.08 & {\cellcolor[HTML]{E9D5CB}} \color[HTML]{000000} \color{black} 0.11 & {\cellcolor[HTML]{000000}} \color[HTML]{F1F1F1} {\cellcolor[HTML]{FFFFFF}} \color{black} --- & {\cellcolor[HTML]{000000}} \color[HTML]{F1F1F1} {\cellcolor[HTML]{FFFFFF}} \color{black} --- \\
\cline{1-7}
\multirow[c]{2}{*}{Orle} & WK & {\cellcolor[HTML]{D9DCE1}} \color[HTML]{000000} \color{black} -0.09 & {\cellcolor[HTML]{D6DCE4}} \color[HTML]{000000} \color{black} -0.13 & {\cellcolor[HTML]{EFCEBD}} \color[HTML]{000000} \color{black} 0.22 & {\cellcolor[HTML]{000000}} \color[HTML]{F1F1F1} {\cellcolor[HTML]{FFFFFF}} \color{black} --- & {\cellcolor[HTML]{000000}} \color[HTML]{F1F1F1} {\cellcolor[HTML]{FFFFFF}} \color{black} --- \\
 & WKND & {\cellcolor[HTML]{D2DBE8}} \color[HTML]{000000} \color{black} -0.18 & {\cellcolor[HTML]{CFDAEA}} \color[HTML]{000000} \color{black} -0.20 & {\cellcolor[HTML]{F5C0A7}} \color[HTML]{000000} \color{black} 0.38 & {\cellcolor[HTML]{000000}} \color[HTML]{F1F1F1} {\cellcolor[HTML]{FFFFFF}} \color{black} --- & {\cellcolor[HTML]{000000}} \color[HTML]{F1F1F1} {\cellcolor[HTML]{FFFFFF}} \color{black} --- \\
\cline{1-7}
\multirow[c]{2}{*}{Metz} & WK & {\cellcolor[HTML]{E0DBD8}} \color[HTML]{000000} \color{black} -0.01 & {\cellcolor[HTML]{D5DBE5}} \color[HTML]{000000} \color{black} -0.14 & {\cellcolor[HTML]{ECD3C5}} \color[HTML]{000000} \color{black} 0.15 & {\cellcolor[HTML]{000000}} \color[HTML]{F1F1F1} {\cellcolor[HTML]{FFFFFF}} \color{black} --- & {\cellcolor[HTML]{000000}} \color[HTML]{F1F1F1} {\cellcolor[HTML]{FFFFFF}} \color{black} --- \\
 & WKND & {\cellcolor[HTML]{E2DAD5}} \color[HTML]{000000} \color{black} 0.02 & {\cellcolor[HTML]{D5DBE5}} \color[HTML]{000000} \color{black} -0.14 & {\cellcolor[HTML]{EAD5C9}} \color[HTML]{000000} \color{black} 0.12 & {\cellcolor[HTML]{000000}} \color[HTML]{F1F1F1} {\cellcolor[HTML]{FFFFFF}} \color{black} --- & {\cellcolor[HTML]{000000}} \color[HTML]{F1F1F1} {\cellcolor[HTML]{FFFFFF}} \color{black} --- \\
\cline{1-7}
\multirow[c]{2}{*}{Mans} & WK & {\cellcolor[HTML]{EAD5C9}} \color[HTML]{000000} \color{black} 0.12 & {\cellcolor[HTML]{EBD3C6}} \color[HTML]{000000} \color{black} 0.14 & {\cellcolor[HTML]{E6D7CF}} \color[HTML]{000000} \color{black} 0.06 & {\cellcolor[HTML]{C4D5F3}} \color[HTML]{000000} \color{black} -0.33 & {\cellcolor[HTML]{000000}} \color[HTML]{F1F1F1} {\cellcolor[HTML]{FFFFFF}} \color{black} --- \\
 & WKND & {\cellcolor[HTML]{E0DBD8}} \color[HTML]{000000} \color{black} -0.00 & {\cellcolor[HTML]{F2CAB5}} \color[HTML]{000000} \color{black} 0.27 & {\cellcolor[HTML]{EED0C0}} \color[HTML]{000000} \color{black} 0.20 & {\cellcolor[HTML]{B6CEFA}} \color[HTML]{000000} \color{black} -0.46 & {\cellcolor[HTML]{000000}} \color[HTML]{F1F1F1} {\cellcolor[HTML]{FFFFFF}} \color{black} --- \\
\cline{1-7}
\multirow[c]{2}{*}{Dijo} & WK & {\cellcolor[HTML]{F2C9B4}} \color[HTML]{000000} \color{black} 0.29 & {\cellcolor[HTML]{D3DBE7}} \color[HTML]{000000} \color{black} -0.16 & {\cellcolor[HTML]{D7DCE3}} \color[HTML]{000000} \color{black} -0.12 & {\cellcolor[HTML]{000000}} \color[HTML]{F1F1F1} {\cellcolor[HTML]{FFFFFF}} \color{black} --- & {\cellcolor[HTML]{000000}} \color[HTML]{F1F1F1} {\cellcolor[HTML]{FFFFFF}} \color{black} --- \\
 & WKND & {\cellcolor[HTML]{E8D6CC}} \color[HTML]{000000} \color{black} 0.09 & {\cellcolor[HTML]{CFDAEA}} \color[HTML]{000000} \color{black} -0.21 & {\cellcolor[HTML]{EAD5C9}} \color[HTML]{000000} \color{black} 0.12 & {\cellcolor[HTML]{000000}} \color[HTML]{F1F1F1} {\cellcolor[HTML]{FFFFFF}} \color{black} --- & {\cellcolor[HTML]{000000}} \color[HTML]{F1F1F1} {\cellcolor[HTML]{FFFFFF}} \color{black} --- \\
\cline{1-7}
\multirow[c]{2}{*}{Mars} & WK & {\cellcolor[HTML]{ADC9FD}} \color[HTML]{000000} \color{black} -0.55 & {\cellcolor[HTML]{F1CCB8}} \color[HTML]{000000} \color{black} 0.25 & {\cellcolor[HTML]{D9DCE1}} \color[HTML]{000000} \color{black} -0.09 & {\cellcolor[HTML]{F6BFA6}} \color[HTML]{000000} \color{black} 0.40 & {\cellcolor[HTML]{000000}} \color[HTML]{F1F1F1} {\cellcolor[HTML]{FFFFFF}} \color{black} --- \\
 & WKND & {\cellcolor[HTML]{B6CEFA}} \color[HTML]{000000} \color{black} -0.46 & {\cellcolor[HTML]{F2CBB7}} \color[HTML]{000000} \color{black} 0.27 & {\cellcolor[HTML]{EED0C0}} \color[HTML]{000000} \color{black} 0.20 & {\cellcolor[HTML]{000000}} \color[HTML]{F1F1F1} {\cellcolor[HTML]{FFFFFF}} \color{black} --- & {\cellcolor[HTML]{000000}} \color[HTML]{F1F1F1} {\cellcolor[HTML]{FFFFFF}} \color{black} --- \\
\cline{1-7}
\multirow[c]{2}{*}{Nice} & WK & {\cellcolor[HTML]{D6DCE4}} \color[HTML]{000000} \color{black} -0.12 & {\cellcolor[HTML]{CBD8EE}} \color[HTML]{000000} \color{black} -0.25 & {\cellcolor[HTML]{F5C1A9}} \color[HTML]{000000} \color{black} 0.37 & {\cellcolor[HTML]{000000}} \color[HTML]{F1F1F1} {\cellcolor[HTML]{FFFFFF}} \color{black} --- & {\cellcolor[HTML]{000000}} \color[HTML]{F1F1F1} {\cellcolor[HTML]{FFFFFF}} \color{black} --- \\
 & WKND & {\cellcolor[HTML]{E2DAD5}} \color[HTML]{000000} \color{black} 0.02 & {\cellcolor[HTML]{EDD1C2}} \color[HTML]{000000} \color{black} 0.17 & {\cellcolor[HTML]{D1DAE9}} \color[HTML]{000000} \color{black} -0.20 & {\cellcolor[HTML]{000000}} \color[HTML]{F1F1F1} {\cellcolor[HTML]{FFFFFF}} \color{black} --- & {\cellcolor[HTML]{000000}} \color[HTML]{F1F1F1} {\cellcolor[HTML]{FFFFFF}} \color{black} --- \\
\cline{1-7}
\multirow[c]{2}{*}{Nanc} & WK & {\cellcolor[HTML]{F7B89C}} \color[HTML]{000000} \color{black} 0.47 & {\cellcolor[HTML]{E0DBD8}} \color[HTML]{000000} \color{black} -0.01 & {\cellcolor[HTML]{CEDAEB}} \color[HTML]{000000} \color{black} -0.22 & {\cellcolor[HTML]{CCD9ED}} \color[HTML]{000000} \color{black} -0.25 & {\cellcolor[HTML]{000000}} \color[HTML]{F1F1F1} {\cellcolor[HTML]{FFFFFF}} \color{black} --- \\
 & WKND & {\cellcolor[HTML]{F3C8B2}} \color[HTML]{000000} \color{black} 0.30 & {\cellcolor[HTML]{D5DBE5}} \color[HTML]{000000} \color{black} -0.15 & {\cellcolor[HTML]{D4DBE6}} \color[HTML]{000000} \color{black} -0.15 & {\cellcolor[HTML]{000000}} \color[HTML]{F1F1F1} {\cellcolor[HTML]{FFFFFF}} \color{black} --- & {\cellcolor[HTML]{000000}} \color[HTML]{F1F1F1} {\cellcolor[HTML]{FFFFFF}} \color{black} --- \\
\cline{1-7}
\bottomrule
\end{tabular}
\end{table}

\begin{table}
\centering
\caption[Total count coefficient estimates for local-level multinomial logistic regression predictive model]{Total count coefficient estimates for local-level multinomial logistic regression predictive model for the week and weekend and by cluster and by city (Paris--Nancy shown in size order). The figure contains the strength of the association between each of the covariates. Coefficients are coloured according to the strength of the relationship. Data preprocessing was applied to the \texttt{NetMob23} dataset~\citep{martinez-duriveNetMob23DatasetHighresolution2023}. City names and Week/Weekend labels are abbreviated for brevity.}
\label{tab:total-count-coefficients}
\begin{tabular}{llrrrrr}
\toprule
 & Class & 1 & 2 & 3 & 4 & 5 \\
City & WK/WKND &  &  &  &  &  \\
\midrule
\multirow[c]{2}{*}{Pari} & WK & {\cellcolor[HTML]{E3D9D3}} \color[HTML]{000000} \color{black} 0.00 & {\cellcolor[HTML]{EAD5C9}} \color[HTML]{000000} \color{black} 0.04 & {\cellcolor[HTML]{DCDDDD}} \color[HTML]{000000} \color{black} -0.04 & {\cellcolor[HTML]{000000}} \color[HTML]{F1F1F1} {\cellcolor[HTML]{FFFFFF}} \color{black} --- & {\cellcolor[HTML]{000000}} \color[HTML]{F1F1F1} {\cellcolor[HTML]{FFFFFF}} \color{black} --- \\
 & WKND & {\cellcolor[HTML]{E2DAD5}} \color[HTML]{000000} \color{black} -0.01 & {\cellcolor[HTML]{E7D7CE}} \color[HTML]{000000} \color{black} 0.02 & {\cellcolor[HTML]{E0DBD8}} \color[HTML]{000000} \color{black} -0.02 & {\cellcolor[HTML]{000000}} \color[HTML]{F1F1F1} {\cellcolor[HTML]{FFFFFF}} \color{black} --- & {\cellcolor[HTML]{000000}} \color[HTML]{F1F1F1} {\cellcolor[HTML]{FFFFFF}} \color{black} --- \\
\cline{1-7}
\multirow[c]{2}{*}{Renn} & WK & {\cellcolor[HTML]{C9D7F0}} \color[HTML]{000000} \color{black} -0.16 & {\cellcolor[HTML]{F0CDBB}} \color[HTML]{000000} \color{black} 0.10 & {\cellcolor[HTML]{ECD3C5}} \color[HTML]{000000} \color{black} 0.06 & {\cellcolor[HTML]{000000}} \color[HTML]{F1F1F1} {\cellcolor[HTML]{FFFFFF}} \color{black} --- & {\cellcolor[HTML]{000000}} \color[HTML]{F1F1F1} {\cellcolor[HTML]{FFFFFF}} \color{black} --- \\
 & WKND & {\cellcolor[HTML]{CBD8EE}} \color[HTML]{000000} \color{black} -0.14 & {\cellcolor[HTML]{F0CDBB}} \color[HTML]{000000} \color{black} 0.10 & {\cellcolor[HTML]{EAD5C9}} \color[HTML]{000000} \color{black} 0.05 & {\cellcolor[HTML]{000000}} \color[HTML]{F1F1F1} {\cellcolor[HTML]{FFFFFF}} \color{black} --- & {\cellcolor[HTML]{000000}} \color[HTML]{F1F1F1} {\cellcolor[HTML]{FFFFFF}} \color{black} --- \\
\cline{1-7}
\multirow[c]{2}{*}{Lill} & WK & {\cellcolor[HTML]{EDD2C3}} \color[HTML]{000000} \color{black} 0.07 & {\cellcolor[HTML]{F6BFA6}} \color[HTML]{000000} \color{black} 0.19 & {\cellcolor[HTML]{B5CDFA}} \color[HTML]{000000} \color{black} -0.25 & {\cellcolor[HTML]{000000}} \color[HTML]{F1F1F1} {\cellcolor[HTML]{FFFFFF}} \color{black} --- & {\cellcolor[HTML]{000000}} \color[HTML]{F1F1F1} {\cellcolor[HTML]{FFFFFF}} \color{black} --- \\
 & WKND & {\cellcolor[HTML]{F1CDBA}} \color[HTML]{000000} \color{black} 0.10 & {\cellcolor[HTML]{F7B599}} \color[HTML]{000000} \color{black} 0.24 & {\cellcolor[HTML]{A3C2FE}} \color[HTML]{000000} \color{black} -0.34 & {\cellcolor[HTML]{000000}} \color[HTML]{F1F1F1} {\cellcolor[HTML]{FFFFFF}} \color{black} --- & {\cellcolor[HTML]{000000}} \color[HTML]{F1F1F1} {\cellcolor[HTML]{FFFFFF}} \color{black} --- \\
\cline{1-7}
\multirow[c]{2}{*}{Bord} & WK & {\cellcolor[HTML]{C5D6F2}} \color[HTML]{000000} \color{black} -0.17 & {\cellcolor[HTML]{EAD4C8}} \color[HTML]{000000} \color{black} 0.05 & {\cellcolor[HTML]{F2CAB5}} \color[HTML]{000000} \color{black} 0.12 & {\cellcolor[HTML]{000000}} \color[HTML]{F1F1F1} {\cellcolor[HTML]{FFFFFF}} \color{black} --- & {\cellcolor[HTML]{000000}} \color[HTML]{F1F1F1} {\cellcolor[HTML]{FFFFFF}} \color{black} --- \\
 & WKND & {\cellcolor[HTML]{CEDAEB}} \color[HTML]{000000} \color{black} -0.12 & {\cellcolor[HTML]{EBD3C6}} \color[HTML]{000000} \color{black} 0.06 & {\cellcolor[HTML]{EDD2C3}} \color[HTML]{000000} \color{black} 0.07 & {\cellcolor[HTML]{000000}} \color[HTML]{F1F1F1} {\cellcolor[HTML]{FFFFFF}} \color{black} --- & {\cellcolor[HTML]{000000}} \color[HTML]{F1F1F1} {\cellcolor[HTML]{FFFFFF}} \color{black} --- \\
\cline{1-7}
\multirow[c]{2}{*}{Gren} & WK & {\cellcolor[HTML]{D3DBE7}} \color[HTML]{000000} \color{black} -0.10 & {\cellcolor[HTML]{E5D8D1}} \color[HTML]{000000} \color{black} 0.01 & {\cellcolor[HTML]{EFCEBD}} \color[HTML]{000000} \color{black} 0.09 & {\cellcolor[HTML]{000000}} \color[HTML]{F1F1F1} {\cellcolor[HTML]{FFFFFF}} \color{black} --- & {\cellcolor[HTML]{000000}} \color[HTML]{F1F1F1} {\cellcolor[HTML]{FFFFFF}} \color{black} --- \\
 & WKND & {\cellcolor[HTML]{D5DBE5}} \color[HTML]{000000} \color{black} -0.09 & {\cellcolor[HTML]{E7D7CE}} \color[HTML]{000000} \color{black} 0.02 & {\cellcolor[HTML]{ECD3C5}} \color[HTML]{000000} \color{black} 0.06 & {\cellcolor[HTML]{000000}} \color[HTML]{F1F1F1} {\cellcolor[HTML]{FFFFFF}} \color{black} --- & {\cellcolor[HTML]{000000}} \color[HTML]{F1F1F1} {\cellcolor[HTML]{FFFFFF}} \color{black} --- \\
\cline{1-7}
\multirow[c]{2}{*}{Lyon} & WK & {\cellcolor[HTML]{E6D7CF}} \color[HTML]{000000} \color{black} 0.02 & {\cellcolor[HTML]{EFCFBF}} \color[HTML]{000000} \color{black} 0.09 & {\cellcolor[HTML]{D2DBE8}} \color[HTML]{000000} \color{black} -0.10 & {\cellcolor[HTML]{000000}} \color[HTML]{F1F1F1} {\cellcolor[HTML]{FFFFFF}} \color{black} --- & {\cellcolor[HTML]{000000}} \color[HTML]{F1F1F1} {\cellcolor[HTML]{FFFFFF}} \color{black} --- \\
 & WKND & {\cellcolor[HTML]{E6D7CF}} \color[HTML]{000000} \color{black} 0.02 & {\cellcolor[HTML]{EED0C0}} \color[HTML]{000000} \color{black} 0.08 & {\cellcolor[HTML]{D3DBE7}} \color[HTML]{000000} \color{black} -0.10 & {\cellcolor[HTML]{000000}} \color[HTML]{F1F1F1} {\cellcolor[HTML]{FFFFFF}} \color{black} --- & {\cellcolor[HTML]{000000}} \color[HTML]{F1F1F1} {\cellcolor[HTML]{FFFFFF}} \color{black} --- \\
\cline{1-7}
\multirow[c]{2}{*}{Nant} & WK & {\cellcolor[HTML]{D8DCE2}} \color[HTML]{000000} \color{black} -0.07 & {\cellcolor[HTML]{F1CDBA}} \color[HTML]{000000} \color{black} 0.11 & {\cellcolor[HTML]{DDDCDC}} \color[HTML]{000000} \color{black} -0.04 & {\cellcolor[HTML]{000000}} \color[HTML]{F1F1F1} {\cellcolor[HTML]{FFFFFF}} \color{black} --- & {\cellcolor[HTML]{000000}} \color[HTML]{F1F1F1} {\cellcolor[HTML]{FFFFFF}} \color{black} --- \\
 & WKND & {\cellcolor[HTML]{CEDAEB}} \color[HTML]{000000} \color{black} -0.12 & {\cellcolor[HTML]{EAD5C9}} \color[HTML]{000000} \color{black} 0.04 & {\cellcolor[HTML]{EED0C0}} \color[HTML]{000000} \color{black} 0.08 & {\cellcolor[HTML]{000000}} \color[HTML]{F1F1F1} {\cellcolor[HTML]{FFFFFF}} \color{black} --- & {\cellcolor[HTML]{000000}} \color[HTML]{F1F1F1} {\cellcolor[HTML]{FFFFFF}} \color{black} --- \\
\cline{1-7}
\multirow[c]{2}{*}{Toul} & WK & {\cellcolor[HTML]{D3DBE7}} \color[HTML]{000000} \color{black} -0.10 & {\cellcolor[HTML]{E7D7CE}} \color[HTML]{000000} \color{black} 0.02 & {\cellcolor[HTML]{EED0C0}} \color[HTML]{000000} \color{black} 0.08 & {\cellcolor[HTML]{000000}} \color[HTML]{F1F1F1} {\cellcolor[HTML]{FFFFFF}} \color{black} --- & {\cellcolor[HTML]{000000}} \color[HTML]{F1F1F1} {\cellcolor[HTML]{FFFFFF}} \color{black} --- \\
 & WKND & {\cellcolor[HTML]{D2DBE8}} \color[HTML]{000000} \color{black} -0.11 & {\cellcolor[HTML]{E7D7CE}} \color[HTML]{000000} \color{black} 0.02 & {\cellcolor[HTML]{EFCFBF}} \color[HTML]{000000} \color{black} 0.08 & {\cellcolor[HTML]{000000}} \color[HTML]{F1F1F1} {\cellcolor[HTML]{FFFFFF}} \color{black} --- & {\cellcolor[HTML]{000000}} \color[HTML]{F1F1F1} {\cellcolor[HTML]{FFFFFF}} \color{black} --- \\
\cline{1-7}
\multirow[c]{2}{*}{Mont} & WK & {\cellcolor[HTML]{DDDCDC}} \color[HTML]{000000} \color{black} -0.04 & {\cellcolor[HTML]{ECD3C5}} \color[HTML]{000000} \color{black} 0.06 & {\cellcolor[HTML]{DFDBD9}} \color[HTML]{000000} \color{black} -0.02 & {\cellcolor[HTML]{000000}} \color[HTML]{F1F1F1} {\cellcolor[HTML]{FFFFFF}} \color{black} --- & {\cellcolor[HTML]{000000}} \color[HTML]{F1F1F1} {\cellcolor[HTML]{FFFFFF}} \color{black} --- \\
 & WKND & {\cellcolor[HTML]{DFDBD9}} \color[HTML]{000000} \color{black} -0.02 & {\cellcolor[HTML]{EDD2C3}} \color[HTML]{000000} \color{black} 0.07 & {\cellcolor[HTML]{DCDDDD}} \color[HTML]{000000} \color{black} -0.04 & {\cellcolor[HTML]{000000}} \color[HTML]{F1F1F1} {\cellcolor[HTML]{FFFFFF}} \color{black} --- & {\cellcolor[HTML]{000000}} \color[HTML]{F1F1F1} {\cellcolor[HTML]{FFFFFF}} \color{black} --- \\
\cline{1-7}
\multirow[c]{2}{*}{Tour} & WK & {\cellcolor[HTML]{D8DCE2}} \color[HTML]{000000} \color{black} -0.07 & {\cellcolor[HTML]{ECD3C5}} \color[HTML]{000000} \color{black} 0.06 & {\cellcolor[HTML]{D3DBE7}} \color[HTML]{000000} \color{black} -0.10 & {\cellcolor[HTML]{EAD5C9}} \color[HTML]{000000} \color{black} 0.05 & {\cellcolor[HTML]{EBD3C6}} \color[HTML]{000000} \color{black} 0.06 \\
 & WKND & {\cellcolor[HTML]{E2DAD5}} \color[HTML]{000000} \color{black} -0.01 & {\cellcolor[HTML]{EAD4C8}} \color[HTML]{000000} \color{black} 0.05 & {\cellcolor[HTML]{DCDDDD}} \color[HTML]{000000} \color{black} -0.04 & {\cellcolor[HTML]{000000}} \color[HTML]{F1F1F1} {\cellcolor[HTML]{FFFFFF}} \color{black} --- & {\cellcolor[HTML]{000000}} \color[HTML]{F1F1F1} {\cellcolor[HTML]{FFFFFF}} \color{black} --- \\
\cline{1-7}
\multirow[c]{2}{*}{Stra} & WK & {\cellcolor[HTML]{E1DAD6}} \color[HTML]{000000} \color{black} -0.01 & {\cellcolor[HTML]{F0CDBB}} \color[HTML]{000000} \color{black} 0.10 & {\cellcolor[HTML]{F0CDBB}} \color[HTML]{000000} \color{black} 0.10 & {\cellcolor[HTML]{C3D5F4}} \color[HTML]{000000} \color{black} -0.19 & {\cellcolor[HTML]{000000}} \color[HTML]{F1F1F1} {\cellcolor[HTML]{FFFFFF}} \color{black} --- \\
 & WKND & {\cellcolor[HTML]{D6DCE4}} \color[HTML]{000000} \color{black} -0.08 & {\cellcolor[HTML]{EAD4C8}} \color[HTML]{000000} \color{black} 0.05 & {\cellcolor[HTML]{E8D6CC}} \color[HTML]{000000} \color{black} 0.03 & {\cellcolor[HTML]{000000}} \color[HTML]{F1F1F1} {\cellcolor[HTML]{FFFFFF}} \color{black} --- & {\cellcolor[HTML]{000000}} \color[HTML]{F1F1F1} {\cellcolor[HTML]{FFFFFF}} \color{black} --- \\
\cline{1-7}
\multirow[c]{2}{*}{Orle} & WK & {\cellcolor[HTML]{C4D5F3}} \color[HTML]{000000} \color{black} -0.18 & {\cellcolor[HTML]{F0CDBB}} \color[HTML]{000000} \color{black} 0.10 & {\cellcolor[HTML]{EFCFBF}} \color[HTML]{000000} \color{black} 0.08 & {\cellcolor[HTML]{000000}} \color[HTML]{F1F1F1} {\cellcolor[HTML]{FFFFFF}} \color{black} --- & {\cellcolor[HTML]{000000}} \color[HTML]{F1F1F1} {\cellcolor[HTML]{FFFFFF}} \color{black} --- \\
 & WKND & {\cellcolor[HTML]{C1D4F4}} \color[HTML]{000000} \color{black} -0.19 & {\cellcolor[HTML]{EFCFBF}} \color[HTML]{000000} \color{black} 0.09 & {\cellcolor[HTML]{F1CDBA}} \color[HTML]{000000} \color{black} 0.11 & {\cellcolor[HTML]{000000}} \color[HTML]{F1F1F1} {\cellcolor[HTML]{FFFFFF}} \color{black} --- & {\cellcolor[HTML]{000000}} \color[HTML]{F1F1F1} {\cellcolor[HTML]{FFFFFF}} \color{black} --- \\
\cline{1-7}
\multirow[c]{2}{*}{Metz} & WK & {\cellcolor[HTML]{D1DAE9}} \color[HTML]{000000} \color{black} -0.11 & {\cellcolor[HTML]{F1CDBA}} \color[HTML]{000000} \color{black} 0.11 & {\cellcolor[HTML]{E4D9D2}} \color[HTML]{000000} \color{black} 0.01 & {\cellcolor[HTML]{000000}} \color[HTML]{F1F1F1} {\cellcolor[HTML]{FFFFFF}} \color{black} --- & {\cellcolor[HTML]{000000}} \color[HTML]{F1F1F1} {\cellcolor[HTML]{FFFFFF}} \color{black} --- \\
 & WKND & {\cellcolor[HTML]{D3DBE7}} \color[HTML]{000000} \color{black} -0.10 & {\cellcolor[HTML]{F0CDBB}} \color[HTML]{000000} \color{black} 0.10 & {\cellcolor[HTML]{E3D9D3}} \color[HTML]{000000} \color{black} -0.00 & {\cellcolor[HTML]{000000}} \color[HTML]{F1F1F1} {\cellcolor[HTML]{FFFFFF}} \color{black} --- & {\cellcolor[HTML]{000000}} \color[HTML]{F1F1F1} {\cellcolor[HTML]{FFFFFF}} \color{black} --- \\
\cline{1-7}
\multirow[c]{2}{*}{Mans} & WK & {\cellcolor[HTML]{C4D5F3}} \color[HTML]{000000} \color{black} -0.18 & {\cellcolor[HTML]{E9D5CB}} \color[HTML]{000000} \color{black} 0.04 & {\cellcolor[HTML]{EDD1C2}} \color[HTML]{000000} \color{black} 0.07 & {\cellcolor[HTML]{EDD2C3}} \color[HTML]{000000} \color{black} 0.07 & {\cellcolor[HTML]{000000}} \color[HTML]{F1F1F1} {\cellcolor[HTML]{FFFFFF}} \color{black} --- \\
 & WKND & {\cellcolor[HTML]{CAD8EF}} \color[HTML]{000000} \color{black} -0.15 & {\cellcolor[HTML]{ECD3C5}} \color[HTML]{000000} \color{black} 0.06 & {\cellcolor[HTML]{EAD4C8}} \color[HTML]{000000} \color{black} 0.05 & {\cellcolor[HTML]{EAD5C9}} \color[HTML]{000000} \color{black} 0.04 & {\cellcolor[HTML]{000000}} \color[HTML]{F1F1F1} {\cellcolor[HTML]{FFFFFF}} \color{black} --- \\
\cline{1-7}
\multirow[c]{2}{*}{Dijo} & WK & {\cellcolor[HTML]{C3D5F4}} \color[HTML]{000000} \color{black} -0.19 & {\cellcolor[HTML]{F6BFA6}} \color[HTML]{000000} \color{black} 0.18 & {\cellcolor[HTML]{E4D9D2}} \color[HTML]{000000} \color{black} 0.00 & {\cellcolor[HTML]{000000}} \color[HTML]{F1F1F1} {\cellcolor[HTML]{FFFFFF}} \color{black} --- & {\cellcolor[HTML]{000000}} \color[HTML]{F1F1F1} {\cellcolor[HTML]{FFFFFF}} \color{black} --- \\
 & WKND & {\cellcolor[HTML]{BED2F6}} \color[HTML]{000000} \color{black} -0.21 & {\cellcolor[HTML]{F3C8B2}} \color[HTML]{000000} \color{black} 0.14 & {\cellcolor[HTML]{EED0C0}} \color[HTML]{000000} \color{black} 0.08 & {\cellcolor[HTML]{000000}} \color[HTML]{F1F1F1} {\cellcolor[HTML]{FFFFFF}} \color{black} --- & {\cellcolor[HTML]{000000}} \color[HTML]{F1F1F1} {\cellcolor[HTML]{FFFFFF}} \color{black} --- \\
\cline{1-7}
\multirow[c]{2}{*}{Mars} & WK & {\cellcolor[HTML]{D9DCE1}} \color[HTML]{000000} \color{black} -0.06 & {\cellcolor[HTML]{EAD4C8}} \color[HTML]{000000} \color{black} 0.05 & {\cellcolor[HTML]{E9D5CB}} \color[HTML]{000000} \color{black} 0.04 & {\cellcolor[HTML]{DFDBD9}} \color[HTML]{000000} \color{black} -0.03 & {\cellcolor[HTML]{000000}} \color[HTML]{F1F1F1} {\cellcolor[HTML]{FFFFFF}} \color{black} --- \\
 & WKND & {\cellcolor[HTML]{D6DCE4}} \color[HTML]{000000} \color{black} -0.08 & {\cellcolor[HTML]{EAD5C9}} \color[HTML]{000000} \color{black} 0.05 & {\cellcolor[HTML]{E9D5CB}} \color[HTML]{000000} \color{black} 0.03 & {\cellcolor[HTML]{000000}} \color[HTML]{F1F1F1} {\cellcolor[HTML]{FFFFFF}} \color{black} --- & {\cellcolor[HTML]{000000}} \color[HTML]{F1F1F1} {\cellcolor[HTML]{FFFFFF}} \color{black} --- \\
\cline{1-7}
\multirow[c]{2}{*}{Nice} & WK & {\cellcolor[HTML]{D6DCE4}} \color[HTML]{000000} \color{black} -0.08 & {\cellcolor[HTML]{E5D8D1}} \color[HTML]{000000} \color{black} 0.01 & {\cellcolor[HTML]{EDD2C3}} \color[HTML]{000000} \color{black} 0.07 & {\cellcolor[HTML]{000000}} \color[HTML]{F1F1F1} {\cellcolor[HTML]{FFFFFF}} \color{black} --- & {\cellcolor[HTML]{000000}} \color[HTML]{F1F1F1} {\cellcolor[HTML]{FFFFFF}} \color{black} --- \\
 & WKND & {\cellcolor[HTML]{ECD3C5}} \color[HTML]{000000} \color{black} 0.06 & {\cellcolor[HTML]{F5C0A7}} \color[HTML]{000000} \color{black} 0.18 & {\cellcolor[HTML]{B7CFF9}} \color[HTML]{000000} \color{black} -0.24 & {\cellcolor[HTML]{000000}} \color[HTML]{F1F1F1} {\cellcolor[HTML]{FFFFFF}} \color{black} --- & {\cellcolor[HTML]{000000}} \color[HTML]{F1F1F1} {\cellcolor[HTML]{FFFFFF}} \color{black} --- \\
\cline{1-7}
\multirow[c]{2}{*}{Nanc} & WK & {\cellcolor[HTML]{CEDAEB}} \color[HTML]{000000} \color{black} -0.12 & {\cellcolor[HTML]{EDD1C2}} \color[HTML]{000000} \color{black} 0.07 & {\cellcolor[HTML]{F5C2AA}} \color[HTML]{000000} \color{black} 0.17 & {\cellcolor[HTML]{CFDAEA}} \color[HTML]{000000} \color{black} -0.12 & {\cellcolor[HTML]{000000}} \color[HTML]{F1F1F1} {\cellcolor[HTML]{FFFFFF}} \color{black} --- \\
 & WKND & {\cellcolor[HTML]{C7D7F0}} \color[HTML]{000000} \color{black} -0.16 & {\cellcolor[HTML]{EAD4C8}} \color[HTML]{000000} \color{black} 0.05 & {\cellcolor[HTML]{F1CCB8}} \color[HTML]{000000} \color{black} 0.11 & {\cellcolor[HTML]{000000}} \color[HTML]{F1F1F1} {\cellcolor[HTML]{FFFFFF}} \color{black} --- & {\cellcolor[HTML]{000000}} \color[HTML]{F1F1F1} {\cellcolor[HTML]{FFFFFF}} \color{black} --- \\
\cline{1-7}
\bottomrule
\end{tabular}
\end{table}

\begin{table}
\centering
\caption[Commercial service count coefficient estimates for local-level multinomial logistic regression predictive model]{Commercial service count coefficient estimates for local-level multinomial logistic regression predictive model for the week and weekend and by cluster and by city (Paris--Nancy shown in size order). The figure contains the strength of the association between each of the covariates. Coefficients are coloured according to the strength of the relationship. Data preprocessing was applied to the \texttt{NetMob23} dataset~\citep{martinez-duriveNetMob23DatasetHighresolution2023}. City names and Week/Weekend labels are abbreviated for brevity.}
\label{tab:commercial-service-count-coefficients}
\begin{tabular}{llrrrrr}
\toprule
 & Class & 1 & 2 & 3 & 4 & 5 \\
City & WK/WKND &  &  &  &  &  \\
\midrule
\multirow[c]{2}{*}{Pari} & WK & {\cellcolor[HTML]{F5C2AA}} \color[HTML]{000000} \color{black} 0.17 & {\cellcolor[HTML]{F1CDBA}} \color[HTML]{000000} \color{black} 0.10 & {\cellcolor[HTML]{B2CCFB}} \color[HTML]{000000} \color{black} -0.27 & {\cellcolor[HTML]{000000}} \color[HTML]{F1F1F1} {\cellcolor[HTML]{FFFFFF}} \color{black} --- & {\cellcolor[HTML]{000000}} \color[HTML]{F1F1F1} {\cellcolor[HTML]{FFFFFF}} \color{black} --- \\
 & WKND & {\cellcolor[HTML]{E7D7CE}} \color[HTML]{000000} \color{black} 0.02 & {\cellcolor[HTML]{DEDCDB}} \color[HTML]{000000} \color{black} -0.03 & {\cellcolor[HTML]{E4D9D2}} \color[HTML]{000000} \color{black} 0.01 & {\cellcolor[HTML]{000000}} \color[HTML]{F1F1F1} {\cellcolor[HTML]{FFFFFF}} \color{black} --- & {\cellcolor[HTML]{000000}} \color[HTML]{F1F1F1} {\cellcolor[HTML]{FFFFFF}} \color{black} --- \\
\cline{1-7}
\multirow[c]{2}{*}{Renn} & WK & {\cellcolor[HTML]{D4DBE6}} \color[HTML]{000000} \color{black} -0.10 & {\cellcolor[HTML]{E4D9D2}} \color[HTML]{000000} \color{black} 0.01 & {\cellcolor[HTML]{EFCEBD}} \color[HTML]{000000} \color{black} 0.09 & {\cellcolor[HTML]{000000}} \color[HTML]{F1F1F1} {\cellcolor[HTML]{FFFFFF}} \color{black} --- & {\cellcolor[HTML]{000000}} \color[HTML]{F1F1F1} {\cellcolor[HTML]{FFFFFF}} \color{black} --- \\
 & WKND & {\cellcolor[HTML]{CDD9EC}} \color[HTML]{000000} \color{black} -0.13 & {\cellcolor[HTML]{E5D8D1}} \color[HTML]{000000} \color{black} 0.01 & {\cellcolor[HTML]{F2CBB7}} \color[HTML]{000000} \color{black} 0.12 & {\cellcolor[HTML]{000000}} \color[HTML]{F1F1F1} {\cellcolor[HTML]{FFFFFF}} \color{black} --- & {\cellcolor[HTML]{000000}} \color[HTML]{F1F1F1} {\cellcolor[HTML]{FFFFFF}} \color{black} --- \\
\cline{1-7}
\multirow[c]{2}{*}{Lill} & WK & {\cellcolor[HTML]{BED2F6}} \color[HTML]{000000} \color{black} -0.21 & {\cellcolor[HTML]{86A9FC}} \color[HTML]{F1F1F1} \color{black} -0.47 & {\cellcolor[HTML]{C43032}} \color[HTML]{F1F1F1} \color{black} 0.68 & {\cellcolor[HTML]{000000}} \color[HTML]{F1F1F1} {\cellcolor[HTML]{FFFFFF}} \color{black} --- & {\cellcolor[HTML]{000000}} \color[HTML]{F1F1F1} {\cellcolor[HTML]{FFFFFF}} \color{black} --- \\
 & WKND & {\cellcolor[HTML]{B3CDFB}} \color[HTML]{000000} \color{black} -0.26 & {\cellcolor[HTML]{80A3FA}} \color[HTML]{F1F1F1} \color{black} -0.50 & {\cellcolor[HTML]{B40426}} \color[HTML]{F1F1F1} \color{black} 0.76 & {\cellcolor[HTML]{000000}} \color[HTML]{F1F1F1} {\cellcolor[HTML]{FFFFFF}} \color{black} --- & {\cellcolor[HTML]{000000}} \color[HTML]{F1F1F1} {\cellcolor[HTML]{FFFFFF}} \color{black} --- \\
\cline{1-7}
\multirow[c]{2}{*}{Bord} & WK & {\cellcolor[HTML]{DADCE0}} \color[HTML]{000000} \color{black} -0.06 & {\cellcolor[HTML]{E6D7CF}} \color[HTML]{000000} \color{black} 0.02 & {\cellcolor[HTML]{E9D5CB}} \color[HTML]{000000} \color{black} 0.04 & {\cellcolor[HTML]{000000}} \color[HTML]{F1F1F1} {\cellcolor[HTML]{FFFFFF}} \color{black} --- & {\cellcolor[HTML]{000000}} \color[HTML]{F1F1F1} {\cellcolor[HTML]{FFFFFF}} \color{black} --- \\
 & WKND & {\cellcolor[HTML]{DEDCDB}} \color[HTML]{000000} \color{black} -0.03 & {\cellcolor[HTML]{F1CDBA}} \color[HTML]{000000} \color{black} 0.11 & {\cellcolor[HTML]{D7DCE3}} \color[HTML]{000000} \color{black} -0.07 & {\cellcolor[HTML]{000000}} \color[HTML]{F1F1F1} {\cellcolor[HTML]{FFFFFF}} \color{black} --- & {\cellcolor[HTML]{000000}} \color[HTML]{F1F1F1} {\cellcolor[HTML]{FFFFFF}} \color{black} --- \\
\cline{1-7}
\multirow[c]{2}{*}{Gren} & WK & {\cellcolor[HTML]{CBD8EE}} \color[HTML]{000000} \color{black} -0.14 & {\cellcolor[HTML]{EAD4C8}} \color[HTML]{000000} \color{black} 0.05 & {\cellcolor[HTML]{EFCFBF}} \color[HTML]{000000} \color{black} 0.09 & {\cellcolor[HTML]{000000}} \color[HTML]{F1F1F1} {\cellcolor[HTML]{FFFFFF}} \color{black} --- & {\cellcolor[HTML]{000000}} \color[HTML]{F1F1F1} {\cellcolor[HTML]{FFFFFF}} \color{black} --- \\
 & WKND & {\cellcolor[HTML]{D1DAE9}} \color[HTML]{000000} \color{black} -0.11 & {\cellcolor[HTML]{EDD1C2}} \color[HTML]{000000} \color{black} 0.07 & {\cellcolor[HTML]{E9D5CB}} \color[HTML]{000000} \color{black} 0.04 & {\cellcolor[HTML]{000000}} \color[HTML]{F1F1F1} {\cellcolor[HTML]{FFFFFF}} \color{black} --- & {\cellcolor[HTML]{000000}} \color[HTML]{F1F1F1} {\cellcolor[HTML]{FFFFFF}} \color{black} --- \\
\cline{1-7}
\multirow[c]{2}{*}{Lyon} & WK & {\cellcolor[HTML]{DADCE0}} \color[HTML]{000000} \color{black} -0.05 & {\cellcolor[HTML]{E6D7CF}} \color[HTML]{000000} \color{black} 0.02 & {\cellcolor[HTML]{E8D6CC}} \color[HTML]{000000} \color{black} 0.03 & {\cellcolor[HTML]{000000}} \color[HTML]{F1F1F1} {\cellcolor[HTML]{FFFFFF}} \color{black} --- & {\cellcolor[HTML]{000000}} \color[HTML]{F1F1F1} {\cellcolor[HTML]{FFFFFF}} \color{black} --- \\
 & WKND & {\cellcolor[HTML]{DADCE0}} \color[HTML]{000000} \color{black} -0.06 & {\cellcolor[HTML]{E8D6CC}} \color[HTML]{000000} \color{black} 0.03 & {\cellcolor[HTML]{E8D6CC}} \color[HTML]{000000} \color{black} 0.03 & {\cellcolor[HTML]{000000}} \color[HTML]{F1F1F1} {\cellcolor[HTML]{FFFFFF}} \color{black} --- & {\cellcolor[HTML]{000000}} \color[HTML]{F1F1F1} {\cellcolor[HTML]{FFFFFF}} \color{black} --- \\
\cline{1-7}
\multirow[c]{2}{*}{Nant} & WK & {\cellcolor[HTML]{D9DCE1}} \color[HTML]{000000} \color{black} -0.06 & {\cellcolor[HTML]{E1DAD6}} \color[HTML]{000000} \color{black} -0.02 & {\cellcolor[HTML]{EED0C0}} \color[HTML]{000000} \color{black} 0.08 & {\cellcolor[HTML]{000000}} \color[HTML]{F1F1F1} {\cellcolor[HTML]{FFFFFF}} \color{black} --- & {\cellcolor[HTML]{000000}} \color[HTML]{F1F1F1} {\cellcolor[HTML]{FFFFFF}} \color{black} --- \\
 & WKND & {\cellcolor[HTML]{DFDBD9}} \color[HTML]{000000} \color{black} -0.03 & {\cellcolor[HTML]{F4C6AF}} \color[HTML]{000000} \color{black} 0.15 & {\cellcolor[HTML]{CEDAEB}} \color[HTML]{000000} \color{black} -0.12 & {\cellcolor[HTML]{000000}} \color[HTML]{F1F1F1} {\cellcolor[HTML]{FFFFFF}} \color{black} --- & {\cellcolor[HTML]{000000}} \color[HTML]{F1F1F1} {\cellcolor[HTML]{FFFFFF}} \color{black} --- \\
\cline{1-7}
\multirow[c]{2}{*}{Toul} & WK & {\cellcolor[HTML]{C7D7F0}} \color[HTML]{000000} \color{black} -0.16 & {\cellcolor[HTML]{F0CDBB}} \color[HTML]{000000} \color{black} 0.10 & {\cellcolor[HTML]{ECD3C5}} \color[HTML]{000000} \color{black} 0.06 & {\cellcolor[HTML]{000000}} \color[HTML]{F1F1F1} {\cellcolor[HTML]{FFFFFF}} \color{black} --- & {\cellcolor[HTML]{000000}} \color[HTML]{F1F1F1} {\cellcolor[HTML]{FFFFFF}} \color{black} --- \\
 & WKND & {\cellcolor[HTML]{C4D5F3}} \color[HTML]{000000} \color{black} -0.18 & {\cellcolor[HTML]{F1CDBA}} \color[HTML]{000000} \color{black} 0.11 & {\cellcolor[HTML]{EDD1C2}} \color[HTML]{000000} \color{black} 0.07 & {\cellcolor[HTML]{000000}} \color[HTML]{F1F1F1} {\cellcolor[HTML]{FFFFFF}} \color{black} --- & {\cellcolor[HTML]{000000}} \color[HTML]{F1F1F1} {\cellcolor[HTML]{FFFFFF}} \color{black} --- \\
\cline{1-7}
\multirow[c]{2}{*}{Mont} & WK & {\cellcolor[HTML]{D7DCE3}} \color[HTML]{000000} \color{black} -0.07 & {\cellcolor[HTML]{EDD2C3}} \color[HTML]{000000} \color{black} 0.07 & {\cellcolor[HTML]{E5D8D1}} \color[HTML]{000000} \color{black} 0.01 & {\cellcolor[HTML]{000000}} \color[HTML]{F1F1F1} {\cellcolor[HTML]{FFFFFF}} \color{black} --- & {\cellcolor[HTML]{000000}} \color[HTML]{F1F1F1} {\cellcolor[HTML]{FFFFFF}} \color{black} --- \\
 & WKND & {\cellcolor[HTML]{D5DBE5}} \color[HTML]{000000} \color{black} -0.09 & {\cellcolor[HTML]{EDD2C3}} \color[HTML]{000000} \color{black} 0.07 & {\cellcolor[HTML]{E6D7CF}} \color[HTML]{000000} \color{black} 0.02 & {\cellcolor[HTML]{000000}} \color[HTML]{F1F1F1} {\cellcolor[HTML]{FFFFFF}} \color{black} --- & {\cellcolor[HTML]{000000}} \color[HTML]{F1F1F1} {\cellcolor[HTML]{FFFFFF}} \color{black} --- \\
\cline{1-7}
\multirow[c]{2}{*}{Tour} & WK & {\cellcolor[HTML]{B2CCFB}} \color[HTML]{000000} \color{black} -0.27 & {\cellcolor[HTML]{B6CEFA}} \color[HTML]{000000} \color{black} -0.25 & {\cellcolor[HTML]{F59C7D}} \color[HTML]{000000} \color{black} 0.35 & {\cellcolor[HTML]{DDDCDC}} \color[HTML]{000000} \color{black} -0.04 & {\cellcolor[HTML]{F7BCA1}} \color[HTML]{000000} \color{black} 0.21 \\
 & WKND & {\cellcolor[HTML]{A9C6FD}} \color[HTML]{000000} \color{black} -0.31 & {\cellcolor[HTML]{D2DBE8}} \color[HTML]{000000} \color{black} -0.11 & {\cellcolor[HTML]{EF886B}} \color[HTML]{F1F1F1} \color{black} 0.42 & {\cellcolor[HTML]{000000}} \color[HTML]{F1F1F1} {\cellcolor[HTML]{FFFFFF}} \color{black} --- & {\cellcolor[HTML]{000000}} \color[HTML]{F1F1F1} {\cellcolor[HTML]{FFFFFF}} \color{black} --- \\
\cline{1-7}
\multirow[c]{2}{*}{Stra} & WK & {\cellcolor[HTML]{E3D9D3}} \color[HTML]{000000} \color{black} -0.00 & {\cellcolor[HTML]{F2CAB5}} \color[HTML]{000000} \color{black} 0.12 & {\cellcolor[HTML]{EBD3C6}} \color[HTML]{000000} \color{black} 0.05 & {\cellcolor[HTML]{C5D6F2}} \color[HTML]{000000} \color{black} -0.18 & {\cellcolor[HTML]{000000}} \color[HTML]{F1F1F1} {\cellcolor[HTML]{FFFFFF}} \color{black} --- \\
 & WKND & {\cellcolor[HTML]{D5DBE5}} \color[HTML]{000000} \color{black} -0.08 & {\cellcolor[HTML]{EFCEBD}} \color[HTML]{000000} \color{black} 0.09 & {\cellcolor[HTML]{E2DAD5}} \color[HTML]{000000} \color{black} -0.01 & {\cellcolor[HTML]{000000}} \color[HTML]{F1F1F1} {\cellcolor[HTML]{FFFFFF}} \color{black} --- & {\cellcolor[HTML]{000000}} \color[HTML]{F1F1F1} {\cellcolor[HTML]{FFFFFF}} \color{black} --- \\
\cline{1-7}
\multirow[c]{2}{*}{Orle} & WK & {\cellcolor[HTML]{D1DAE9}} \color[HTML]{000000} \color{black} -0.12 & {\cellcolor[HTML]{E3D9D3}} \color[HTML]{000000} \color{black} 0.00 & {\cellcolor[HTML]{F1CCB8}} \color[HTML]{000000} \color{black} 0.11 & {\cellcolor[HTML]{000000}} \color[HTML]{F1F1F1} {\cellcolor[HTML]{FFFFFF}} \color{black} --- & {\cellcolor[HTML]{000000}} \color[HTML]{F1F1F1} {\cellcolor[HTML]{FFFFFF}} \color{black} --- \\
 & WKND & {\cellcolor[HTML]{D2DBE8}} \color[HTML]{000000} \color{black} -0.11 & {\cellcolor[HTML]{E6D7CF}} \color[HTML]{000000} \color{black} 0.02 & {\cellcolor[HTML]{EFCFBF}} \color[HTML]{000000} \color{black} 0.09 & {\cellcolor[HTML]{000000}} \color[HTML]{F1F1F1} {\cellcolor[HTML]{FFFFFF}} \color{black} --- & {\cellcolor[HTML]{000000}} \color[HTML]{F1F1F1} {\cellcolor[HTML]{FFFFFF}} \color{black} --- \\
\cline{1-7}
\multirow[c]{2}{*}{Metz} & WK & {\cellcolor[HTML]{CAD8EF}} \color[HTML]{000000} \color{black} -0.15 & {\cellcolor[HTML]{EAD5C9}} \color[HTML]{000000} \color{black} 0.04 & {\cellcolor[HTML]{F1CDBA}} \color[HTML]{000000} \color{black} 0.11 & {\cellcolor[HTML]{000000}} \color[HTML]{F1F1F1} {\cellcolor[HTML]{FFFFFF}} \color{black} --- & {\cellcolor[HTML]{000000}} \color[HTML]{F1F1F1} {\cellcolor[HTML]{FFFFFF}} \color{black} --- \\
 & WKND & {\cellcolor[HTML]{D5DBE5}} \color[HTML]{000000} \color{black} -0.09 & {\cellcolor[HTML]{E6D7CF}} \color[HTML]{000000} \color{black} 0.02 & {\cellcolor[HTML]{EDD1C2}} \color[HTML]{000000} \color{black} 0.07 & {\cellcolor[HTML]{000000}} \color[HTML]{F1F1F1} {\cellcolor[HTML]{FFFFFF}} \color{black} --- & {\cellcolor[HTML]{000000}} \color[HTML]{F1F1F1} {\cellcolor[HTML]{FFFFFF}} \color{black} --- \\
\cline{1-7}
\multirow[c]{2}{*}{Mans} & WK & {\cellcolor[HTML]{A2C1FF}} \color[HTML]{000000} \color{black} -0.34 & {\cellcolor[HTML]{F7AC8E}} \color[HTML]{000000} \color{black} 0.28 & {\cellcolor[HTML]{F5C2AA}} \color[HTML]{000000} \color{black} 0.17 & {\cellcolor[HTML]{D3DBE7}} \color[HTML]{000000} \color{black} -0.10 & {\cellcolor[HTML]{000000}} \color[HTML]{F1F1F1} {\cellcolor[HTML]{FFFFFF}} \color{black} --- \\
 & WKND & {\cellcolor[HTML]{ABC8FD}} \color[HTML]{000000} \color{black} -0.30 & {\cellcolor[HTML]{F7A889}} \color[HTML]{000000} \color{black} 0.30 & {\cellcolor[HTML]{F7B99E}} \color[HTML]{000000} \color{black} 0.22 & {\cellcolor[HTML]{BED2F6}} \color[HTML]{000000} \color{black} -0.21 & {\cellcolor[HTML]{000000}} \color[HTML]{F1F1F1} {\cellcolor[HTML]{FFFFFF}} \color{black} --- \\
\cline{1-7}
\multirow[c]{2}{*}{Dijo} & WK & {\cellcolor[HTML]{C0D4F5}} \color[HTML]{000000} \color{black} -0.20 & {\cellcolor[HTML]{F6BDA2}} \color[HTML]{000000} \color{black} 0.20 & {\cellcolor[HTML]{E3D9D3}} \color[HTML]{000000} \color{black} 0.00 & {\cellcolor[HTML]{000000}} \color[HTML]{F1F1F1} {\cellcolor[HTML]{FFFFFF}} \color{black} --- & {\cellcolor[HTML]{000000}} \color[HTML]{F1F1F1} {\cellcolor[HTML]{FFFFFF}} \color{black} --- \\
 & WKND & {\cellcolor[HTML]{AFCAFC}} \color[HTML]{000000} \color{black} -0.28 & {\cellcolor[HTML]{EED0C0}} \color[HTML]{000000} \color{black} 0.08 & {\cellcolor[HTML]{F6BDA2}} \color[HTML]{000000} \color{black} 0.20 & {\cellcolor[HTML]{000000}} \color[HTML]{F1F1F1} {\cellcolor[HTML]{FFFFFF}} \color{black} --- & {\cellcolor[HTML]{000000}} \color[HTML]{F1F1F1} {\cellcolor[HTML]{FFFFFF}} \color{black} --- \\
\cline{1-7}
\multirow[c]{2}{*}{Mars} & WK & {\cellcolor[HTML]{BFD3F6}} \color[HTML]{000000} \color{black} -0.21 & {\cellcolor[HTML]{E7D7CE}} \color[HTML]{000000} \color{black} 0.03 & {\cellcolor[HTML]{E2DAD5}} \color[HTML]{000000} \color{black} -0.01 & {\cellcolor[HTML]{F6BEA4}} \color[HTML]{000000} \color{black} 0.19 & {\cellcolor[HTML]{000000}} \color[HTML]{F1F1F1} {\cellcolor[HTML]{FFFFFF}} \color{black} --- \\
 & WKND & {\cellcolor[HTML]{CBD8EE}} \color[HTML]{000000} \color{black} -0.15 & {\cellcolor[HTML]{F1CDBA}} \color[HTML]{000000} \color{black} 0.11 & {\cellcolor[HTML]{EAD5C9}} \color[HTML]{000000} \color{black} 0.04 & {\cellcolor[HTML]{000000}} \color[HTML]{F1F1F1} {\cellcolor[HTML]{FFFFFF}} \color{black} --- & {\cellcolor[HTML]{000000}} \color[HTML]{F1F1F1} {\cellcolor[HTML]{FFFFFF}} \color{black} --- \\
\cline{1-7}
\multirow[c]{2}{*}{Nice} & WK & {\cellcolor[HTML]{ABC8FD}} \color[HTML]{000000} \color{black} -0.30 & {\cellcolor[HTML]{F7BA9F}} \color[HTML]{000000} \color{black} 0.21 & {\cellcolor[HTML]{EFCEBD}} \color[HTML]{000000} \color{black} 0.09 & {\cellcolor[HTML]{000000}} \color[HTML]{F1F1F1} {\cellcolor[HTML]{FFFFFF}} \color{black} --- & {\cellcolor[HTML]{000000}} \color[HTML]{F1F1F1} {\cellcolor[HTML]{FFFFFF}} \color{black} --- \\
 & WKND & {\cellcolor[HTML]{9DBDFF}} \color[HTML]{000000} \color{black} -0.37 & {\cellcolor[HTML]{D1DAE9}} \color[HTML]{000000} \color{black} -0.12 & {\cellcolor[HTML]{E9785D}} \color[HTML]{F1F1F1} \color{black} 0.48 & {\cellcolor[HTML]{000000}} \color[HTML]{F1F1F1} {\cellcolor[HTML]{FFFFFF}} \color{black} --- & {\cellcolor[HTML]{000000}} \color[HTML]{F1F1F1} {\cellcolor[HTML]{FFFFFF}} \color{black} --- \\
\cline{1-7}
\multirow[c]{2}{*}{Nanc} & WK & {\cellcolor[HTML]{C6D6F1}} \color[HTML]{000000} \color{black} -0.17 & {\cellcolor[HTML]{DFDBD9}} \color[HTML]{000000} \color{black} -0.03 & {\cellcolor[HTML]{B7CFF9}} \color[HTML]{000000} \color{black} -0.24 & {\cellcolor[HTML]{EE8468}} \color[HTML]{F1F1F1} \color{black} 0.44 & {\cellcolor[HTML]{000000}} \color[HTML]{F1F1F1} {\cellcolor[HTML]{FFFFFF}} \color{black} --- \\
 & WKND & {\cellcolor[HTML]{DBDCDE}} \color[HTML]{000000} \color{black} -0.05 & {\cellcolor[HTML]{F2CAB5}} \color[HTML]{000000} \color{black} 0.12 & {\cellcolor[HTML]{D7DCE3}} \color[HTML]{000000} \color{black} -0.07 & {\cellcolor[HTML]{000000}} \color[HTML]{F1F1F1} {\cellcolor[HTML]{FFFFFF}} \color{black} --- & {\cellcolor[HTML]{000000}} \color[HTML]{F1F1F1} {\cellcolor[HTML]{FFFFFF}} \color{black} --- \\
\cline{1-7}
\bottomrule
\end{tabular}
\end{table}

\begin{table}
\centering
\caption[Commercial venue count coefficient estimates for local-level multinomial logistic regression predictive model]{Commercial venue count coefficient estimates for local-level multinomial logistic regression predictive model for the week and weekend and by cluster and by city (Paris--Nancy shown in size order). The figure contains the strength of the association between each of the covariates. Coefficients are coloured according to the strength of the relationship. Data preprocessing was applied to the \texttt{NetMob23} dataset~\citep{martinez-duriveNetMob23DatasetHighresolution2023}. City names and Week/Weekend labels are abbreviated for brevity.}
\label{tab:commercial-venue-count-coefficients}
\begin{tabular}{llrrrrr}
\toprule
 & Class & 1 & 2 & 3 & 4 & 5 \\
City & WK/WKND &  &  &  &  &  \\
\midrule
\multirow[c]{2}{*}{Pari} & WK & {\cellcolor[HTML]{F3C7B1}} \color[HTML]{000000} \color{black} 0.14 & {\cellcolor[HTML]{F1CDBA}} \color[HTML]{000000} \color{black} 0.11 & {\cellcolor[HTML]{B6CEFA}} \color[HTML]{000000} \color{black} -0.25 & {\cellcolor[HTML]{000000}} \color[HTML]{F1F1F1} {\cellcolor[HTML]{FFFFFF}} \color{black} --- & {\cellcolor[HTML]{000000}} \color[HTML]{F1F1F1} {\cellcolor[HTML]{FFFFFF}} \color{black} --- \\
 & WKND & {\cellcolor[HTML]{F2C9B4}} \color[HTML]{000000} \color{black} 0.13 & {\cellcolor[HTML]{F0CDBB}} \color[HTML]{000000} \color{black} 0.10 & {\cellcolor[HTML]{BAD0F8}} \color[HTML]{000000} \color{black} -0.23 & {\cellcolor[HTML]{000000}} \color[HTML]{F1F1F1} {\cellcolor[HTML]{FFFFFF}} \color{black} --- & {\cellcolor[HTML]{000000}} \color[HTML]{F1F1F1} {\cellcolor[HTML]{FFFFFF}} \color{black} --- \\
\cline{1-7}
\multirow[c]{2}{*}{Renn} & WK & {\cellcolor[HTML]{F5A081}} \color[HTML]{000000} \color{black} 0.33 & {\cellcolor[HTML]{C9D7F0}} \color[HTML]{000000} \color{black} -0.15 & {\cellcolor[HTML]{C5D6F2}} \color[HTML]{000000} \color{black} -0.18 & {\cellcolor[HTML]{000000}} \color[HTML]{F1F1F1} {\cellcolor[HTML]{FFFFFF}} \color{black} --- & {\cellcolor[HTML]{000000}} \color[HTML]{F1F1F1} {\cellcolor[HTML]{FFFFFF}} \color{black} --- \\
 & WKND & {\cellcolor[HTML]{F7A889}} \color[HTML]{000000} \color{black} 0.30 & {\cellcolor[HTML]{C0D4F5}} \color[HTML]{000000} \color{black} -0.20 & {\cellcolor[HTML]{D4DBE6}} \color[HTML]{000000} \color{black} -0.10 & {\cellcolor[HTML]{000000}} \color[HTML]{F1F1F1} {\cellcolor[HTML]{FFFFFF}} \color{black} --- & {\cellcolor[HTML]{000000}} \color[HTML]{F1F1F1} {\cellcolor[HTML]{FFFFFF}} \color{black} --- \\
\cline{1-7}
\multirow[c]{2}{*}{Lill} & WK & {\cellcolor[HTML]{F08A6C}} \color[HTML]{F1F1F1} \color{black} 0.42 & {\cellcolor[HTML]{F7B497}} \color[HTML]{000000} \color{black} 0.24 & {\cellcolor[HTML]{5D7CE6}} \color[HTML]{F1F1F1} \color{black} -0.66 & {\cellcolor[HTML]{000000}} \color[HTML]{F1F1F1} {\cellcolor[HTML]{FFFFFF}} \color{black} --- & {\cellcolor[HTML]{000000}} \color[HTML]{F1F1F1} {\cellcolor[HTML]{FFFFFF}} \color{black} --- \\
 & WKND & {\cellcolor[HTML]{F29072}} \color[HTML]{F1F1F1} \color{black} 0.39 & {\cellcolor[HTML]{F7BCA1}} \color[HTML]{000000} \color{black} 0.21 & {\cellcolor[HTML]{6A8BEF}} \color[HTML]{F1F1F1} \color{black} -0.60 & {\cellcolor[HTML]{000000}} \color[HTML]{F1F1F1} {\cellcolor[HTML]{FFFFFF}} \color{black} --- & {\cellcolor[HTML]{000000}} \color[HTML]{F1F1F1} {\cellcolor[HTML]{FFFFFF}} \color{black} --- \\
\cline{1-7}
\multirow[c]{2}{*}{Bord} & WK & {\cellcolor[HTML]{F39778}} \color[HTML]{000000} \color{black} 0.37 & {\cellcolor[HTML]{D8DCE2}} \color[HTML]{000000} \color{black} -0.07 & {\cellcolor[HTML]{ADC9FD}} \color[HTML]{000000} \color{black} -0.30 & {\cellcolor[HTML]{000000}} \color[HTML]{F1F1F1} {\cellcolor[HTML]{FFFFFF}} \color{black} --- & {\cellcolor[HTML]{000000}} \color[HTML]{F1F1F1} {\cellcolor[HTML]{FFFFFF}} \color{black} --- \\
 & WKND & {\cellcolor[HTML]{F6A283}} \color[HTML]{000000} \color{black} 0.32 & {\cellcolor[HTML]{D1DAE9}} \color[HTML]{000000} \color{black} -0.11 & {\cellcolor[HTML]{BFD3F6}} \color[HTML]{000000} \color{black} -0.21 & {\cellcolor[HTML]{000000}} \color[HTML]{F1F1F1} {\cellcolor[HTML]{FFFFFF}} \color{black} --- & {\cellcolor[HTML]{000000}} \color[HTML]{F1F1F1} {\cellcolor[HTML]{FFFFFF}} \color{black} --- \\
\cline{1-7}
\multirow[c]{2}{*}{Gren} & WK & {\cellcolor[HTML]{F7BCA1}} \color[HTML]{000000} \color{black} 0.21 & {\cellcolor[HTML]{DBDCDE}} \color[HTML]{000000} \color{black} -0.05 & {\cellcolor[HTML]{CAD8EF}} \color[HTML]{000000} \color{black} -0.15 & {\cellcolor[HTML]{000000}} \color[HTML]{F1F1F1} {\cellcolor[HTML]{FFFFFF}} \color{black} --- & {\cellcolor[HTML]{000000}} \color[HTML]{F1F1F1} {\cellcolor[HTML]{FFFFFF}} \color{black} --- \\
 & WKND & {\cellcolor[HTML]{F5C4AC}} \color[HTML]{000000} \color{black} 0.16 & {\cellcolor[HTML]{DEDCDB}} \color[HTML]{000000} \color{black} -0.03 & {\cellcolor[HTML]{CEDAEB}} \color[HTML]{000000} \color{black} -0.12 & {\cellcolor[HTML]{000000}} \color[HTML]{F1F1F1} {\cellcolor[HTML]{FFFFFF}} \color{black} --- & {\cellcolor[HTML]{000000}} \color[HTML]{F1F1F1} {\cellcolor[HTML]{FFFFFF}} \color{black} --- \\
\cline{1-7}
\multirow[c]{2}{*}{Lyon} & WK & {\cellcolor[HTML]{DDDCDC}} \color[HTML]{000000} \color{black} -0.04 & {\cellcolor[HTML]{D9DCE1}} \color[HTML]{000000} \color{black} -0.07 & {\cellcolor[HTML]{F0CDBB}} \color[HTML]{000000} \color{black} 0.10 & {\cellcolor[HTML]{000000}} \color[HTML]{F1F1F1} {\cellcolor[HTML]{FFFFFF}} \color{black} --- & {\cellcolor[HTML]{000000}} \color[HTML]{F1F1F1} {\cellcolor[HTML]{FFFFFF}} \color{black} --- \\
 & WKND & {\cellcolor[HTML]{DDDCDC}} \color[HTML]{000000} \color{black} -0.04 & {\cellcolor[HTML]{D8DCE2}} \color[HTML]{000000} \color{black} -0.07 & {\cellcolor[HTML]{F1CDBA}} \color[HTML]{000000} \color{black} 0.11 & {\cellcolor[HTML]{000000}} \color[HTML]{F1F1F1} {\cellcolor[HTML]{FFFFFF}} \color{black} --- & {\cellcolor[HTML]{000000}} \color[HTML]{F1F1F1} {\cellcolor[HTML]{FFFFFF}} \color{black} --- \\
\cline{1-7}
\multirow[c]{2}{*}{Nant} & WK & {\cellcolor[HTML]{EC8165}} \color[HTML]{F1F1F1} \color{black} 0.45 & {\cellcolor[HTML]{F5C4AC}} \color[HTML]{000000} \color{black} 0.16 & {\cellcolor[HTML]{6788EE}} \color[HTML]{F1F1F1} \color{black} -0.61 & {\cellcolor[HTML]{000000}} \color[HTML]{F1F1F1} {\cellcolor[HTML]{FFFFFF}} \color{black} --- & {\cellcolor[HTML]{000000}} \color[HTML]{F1F1F1} {\cellcolor[HTML]{FFFFFF}} \color{black} --- \\
 & WKND & {\cellcolor[HTML]{F7B99E}} \color[HTML]{000000} \color{black} 0.22 & {\cellcolor[HTML]{BAD0F8}} \color[HTML]{000000} \color{black} -0.23 & {\cellcolor[HTML]{E5D8D1}} \color[HTML]{000000} \color{black} 0.01 & {\cellcolor[HTML]{000000}} \color[HTML]{F1F1F1} {\cellcolor[HTML]{FFFFFF}} \color{black} --- & {\cellcolor[HTML]{000000}} \color[HTML]{F1F1F1} {\cellcolor[HTML]{FFFFFF}} \color{black} --- \\
\cline{1-7}
\multirow[c]{2}{*}{Toul} & WK & {\cellcolor[HTML]{F5C1A9}} \color[HTML]{000000} \color{black} 0.18 & {\cellcolor[HTML]{D3DBE7}} \color[HTML]{000000} \color{black} -0.10 & {\cellcolor[HTML]{D7DCE3}} \color[HTML]{000000} \color{black} -0.08 & {\cellcolor[HTML]{000000}} \color[HTML]{F1F1F1} {\cellcolor[HTML]{FFFFFF}} \color{black} --- & {\cellcolor[HTML]{000000}} \color[HTML]{F1F1F1} {\cellcolor[HTML]{FFFFFF}} \color{black} --- \\
 & WKND & {\cellcolor[HTML]{F6BEA4}} \color[HTML]{000000} \color{black} 0.19 & {\cellcolor[HTML]{CEDAEB}} \color[HTML]{000000} \color{black} -0.12 & {\cellcolor[HTML]{D8DCE2}} \color[HTML]{000000} \color{black} -0.07 & {\cellcolor[HTML]{000000}} \color[HTML]{F1F1F1} {\cellcolor[HTML]{FFFFFF}} \color{black} --- & {\cellcolor[HTML]{000000}} \color[HTML]{F1F1F1} {\cellcolor[HTML]{FFFFFF}} \color{black} --- \\
\cline{1-7}
\multirow[c]{2}{*}{Mont} & WK & {\cellcolor[HTML]{E7D7CE}} \color[HTML]{000000} \color{black} 0.03 & {\cellcolor[HTML]{C4D5F3}} \color[HTML]{000000} \color{black} -0.18 & {\cellcolor[HTML]{F4C5AD}} \color[HTML]{000000} \color{black} 0.15 & {\cellcolor[HTML]{000000}} \color[HTML]{F1F1F1} {\cellcolor[HTML]{FFFFFF}} \color{black} --- & {\cellcolor[HTML]{000000}} \color[HTML]{F1F1F1} {\cellcolor[HTML]{FFFFFF}} \color{black} --- \\
 & WKND & {\cellcolor[HTML]{EAD5C9}} \color[HTML]{000000} \color{black} 0.04 & {\cellcolor[HTML]{CBD8EE}} \color[HTML]{000000} \color{black} -0.14 & {\cellcolor[HTML]{F0CDBB}} \color[HTML]{000000} \color{black} 0.10 & {\cellcolor[HTML]{000000}} \color[HTML]{F1F1F1} {\cellcolor[HTML]{FFFFFF}} \color{black} --- & {\cellcolor[HTML]{000000}} \color[HTML]{F1F1F1} {\cellcolor[HTML]{FFFFFF}} \color{black} --- \\
\cline{1-7}
\multirow[c]{2}{*}{Tour} & WK & {\cellcolor[HTML]{F7A688}} \color[HTML]{000000} \color{black} 0.30 & {\cellcolor[HTML]{F3C7B1}} \color[HTML]{000000} \color{black} 0.14 & {\cellcolor[HTML]{BBD1F8}} \color[HTML]{000000} \color{black} -0.23 & {\cellcolor[HTML]{C1D4F4}} \color[HTML]{000000} \color{black} -0.19 & {\cellcolor[HTML]{DFDBD9}} \color[HTML]{000000} \color{black} -0.03 \\
 & WKND & {\cellcolor[HTML]{F4C5AD}} \color[HTML]{000000} \color{black} 0.16 & {\cellcolor[HTML]{EFCFBF}} \color[HTML]{000000} \color{black} 0.09 & {\cellcolor[HTML]{B7CFF9}} \color[HTML]{000000} \color{black} -0.24 & {\cellcolor[HTML]{000000}} \color[HTML]{F1F1F1} {\cellcolor[HTML]{FFFFFF}} \color{black} --- & {\cellcolor[HTML]{000000}} \color[HTML]{F1F1F1} {\cellcolor[HTML]{FFFFFF}} \color{black} --- \\
\cline{1-7}
\multirow[c]{2}{*}{Stra} & WK & {\cellcolor[HTML]{F5C0A7}} \color[HTML]{000000} \color{black} 0.18 & {\cellcolor[HTML]{E9D5CB}} \color[HTML]{000000} \color{black} 0.04 & {\cellcolor[HTML]{E2DAD5}} \color[HTML]{000000} \color{black} -0.01 & {\cellcolor[HTML]{BED2F6}} \color[HTML]{000000} \color{black} -0.21 & {\cellcolor[HTML]{000000}} \color[HTML]{F1F1F1} {\cellcolor[HTML]{FFFFFF}} \color{black} --- \\
 & WKND & {\cellcolor[HTML]{EED0C0}} \color[HTML]{000000} \color{black} 0.08 & {\cellcolor[HTML]{E0DBD8}} \color[HTML]{000000} \color{black} -0.02 & {\cellcolor[HTML]{DADCE0}} \color[HTML]{000000} \color{black} -0.06 & {\cellcolor[HTML]{000000}} \color[HTML]{F1F1F1} {\cellcolor[HTML]{FFFFFF}} \color{black} --- & {\cellcolor[HTML]{000000}} \color[HTML]{F1F1F1} {\cellcolor[HTML]{FFFFFF}} \color{black} --- \\
\cline{1-7}
\multirow[c]{2}{*}{Orle} & WK & {\cellcolor[HTML]{F6BFA6}} \color[HTML]{000000} \color{black} 0.19 & {\cellcolor[HTML]{C5D6F2}} \color[HTML]{000000} \color{black} -0.18 & {\cellcolor[HTML]{E1DAD6}} \color[HTML]{000000} \color{black} -0.01 & {\cellcolor[HTML]{000000}} \color[HTML]{F1F1F1} {\cellcolor[HTML]{FFFFFF}} \color{black} --- & {\cellcolor[HTML]{000000}} \color[HTML]{F1F1F1} {\cellcolor[HTML]{FFFFFF}} \color{black} --- \\
 & WKND & {\cellcolor[HTML]{F6BDA2}} \color[HTML]{000000} \color{black} 0.20 & {\cellcolor[HTML]{C5D6F2}} \color[HTML]{000000} \color{black} -0.17 & {\cellcolor[HTML]{DFDBD9}} \color[HTML]{000000} \color{black} -0.03 & {\cellcolor[HTML]{000000}} \color[HTML]{F1F1F1} {\cellcolor[HTML]{FFFFFF}} \color{black} --- & {\cellcolor[HTML]{000000}} \color[HTML]{F1F1F1} {\cellcolor[HTML]{FFFFFF}} \color{black} --- \\
\cline{1-7}
\multirow[c]{2}{*}{Metz} & WK & {\cellcolor[HTML]{F6A385}} \color[HTML]{000000} \color{black} 0.32 & {\cellcolor[HTML]{DEDCDB}} \color[HTML]{000000} \color{black} -0.03 & {\cellcolor[HTML]{AEC9FC}} \color[HTML]{000000} \color{black} -0.28 & {\cellcolor[HTML]{000000}} \color[HTML]{F1F1F1} {\cellcolor[HTML]{FFFFFF}} \color{black} --- & {\cellcolor[HTML]{000000}} \color[HTML]{F1F1F1} {\cellcolor[HTML]{FFFFFF}} \color{black} --- \\
 & WKND & {\cellcolor[HTML]{F6A385}} \color[HTML]{000000} \color{black} 0.32 & {\cellcolor[HTML]{DEDCDB}} \color[HTML]{000000} \color{black} -0.03 & {\cellcolor[HTML]{AEC9FC}} \color[HTML]{000000} \color{black} -0.29 & {\cellcolor[HTML]{000000}} \color[HTML]{F1F1F1} {\cellcolor[HTML]{FFFFFF}} \color{black} --- & {\cellcolor[HTML]{000000}} \color[HTML]{F1F1F1} {\cellcolor[HTML]{FFFFFF}} \color{black} --- \\
\cline{1-7}
\multirow[c]{2}{*}{Mans} & WK & {\cellcolor[HTML]{EBD3C6}} \color[HTML]{000000} \color{black} 0.05 & {\cellcolor[HTML]{E4D9D2}} \color[HTML]{000000} \color{black} 0.01 & {\cellcolor[HTML]{D2DBE8}} \color[HTML]{000000} \color{black} -0.11 & {\cellcolor[HTML]{EAD4C8}} \color[HTML]{000000} \color{black} 0.05 & {\cellcolor[HTML]{000000}} \color[HTML]{F1F1F1} {\cellcolor[HTML]{FFFFFF}} \color{black} --- \\
 & WKND & {\cellcolor[HTML]{F6BDA2}} \color[HTML]{000000} \color{black} 0.20 & {\cellcolor[HTML]{CDD9EC}} \color[HTML]{000000} \color{black} -0.13 & {\cellcolor[HTML]{C9D7F0}} \color[HTML]{000000} \color{black} -0.16 & {\cellcolor[HTML]{EFCFBF}} \color[HTML]{000000} \color{black} 0.09 & {\cellcolor[HTML]{000000}} \color[HTML]{F1F1F1} {\cellcolor[HTML]{FFFFFF}} \color{black} --- \\
\cline{1-7}
\multirow[c]{2}{*}{Dijo} & WK & {\cellcolor[HTML]{E46E56}} \color[HTML]{F1F1F1} \color{black} 0.51 & {\cellcolor[HTML]{B3CDFB}} \color[HTML]{000000} \color{black} -0.26 & {\cellcolor[HTML]{B5CDFA}} \color[HTML]{000000} \color{black} -0.25 & {\cellcolor[HTML]{000000}} \color[HTML]{F1F1F1} {\cellcolor[HTML]{FFFFFF}} \color{black} --- & {\cellcolor[HTML]{000000}} \color[HTML]{F1F1F1} {\cellcolor[HTML]{FFFFFF}} \color{black} --- \\
 & WKND & {\cellcolor[HTML]{E46E56}} \color[HTML]{F1F1F1} \color{black} 0.51 & {\cellcolor[HTML]{B2CCFB}} \color[HTML]{000000} \color{black} -0.27 & {\cellcolor[HTML]{B9D0F9}} \color[HTML]{000000} \color{black} -0.24 & {\cellcolor[HTML]{000000}} \color[HTML]{F1F1F1} {\cellcolor[HTML]{FFFFFF}} \color{black} --- & {\cellcolor[HTML]{000000}} \color[HTML]{F1F1F1} {\cellcolor[HTML]{FFFFFF}} \color{black} --- \\
\cline{1-7}
\multirow[c]{2}{*}{Mars} & WK & {\cellcolor[HTML]{F7AC8E}} \color[HTML]{000000} \color{black} 0.28 & {\cellcolor[HTML]{E0DBD8}} \color[HTML]{000000} \color{black} -0.02 & {\cellcolor[HTML]{CEDAEB}} \color[HTML]{000000} \color{black} -0.12 & {\cellcolor[HTML]{CCD9ED}} \color[HTML]{000000} \color{black} -0.14 & {\cellcolor[HTML]{000000}} \color[HTML]{F1F1F1} {\cellcolor[HTML]{FFFFFF}} \color{black} --- \\
 & WKND & {\cellcolor[HTML]{F7B194}} \color[HTML]{000000} \color{black} 0.26 & {\cellcolor[HTML]{D6DCE4}} \color[HTML]{000000} \color{black} -0.08 & {\cellcolor[HTML]{C5D6F2}} \color[HTML]{000000} \color{black} -0.18 & {\cellcolor[HTML]{000000}} \color[HTML]{F1F1F1} {\cellcolor[HTML]{FFFFFF}} \color{black} --- & {\cellcolor[HTML]{000000}} \color[HTML]{F1F1F1} {\cellcolor[HTML]{FFFFFF}} \color{black} --- \\
\cline{1-7}
\multirow[c]{2}{*}{Nice} & WK & {\cellcolor[HTML]{F7B89C}} \color[HTML]{000000} \color{black} 0.23 & {\cellcolor[HTML]{CCD9ED}} \color[HTML]{000000} \color{black} -0.14 & {\cellcolor[HTML]{D5DBE5}} \color[HTML]{000000} \color{black} -0.09 & {\cellcolor[HTML]{000000}} \color[HTML]{F1F1F1} {\cellcolor[HTML]{FFFFFF}} \color{black} --- & {\cellcolor[HTML]{000000}} \color[HTML]{F1F1F1} {\cellcolor[HTML]{FFFFFF}} \color{black} --- \\
 & WKND & {\cellcolor[HTML]{EE8669}} \color[HTML]{F1F1F1} \color{black} 0.43 & {\cellcolor[HTML]{F5C2AA}} \color[HTML]{000000} \color{black} 0.17 & {\cellcolor[HTML]{6B8DF0}} \color[HTML]{F1F1F1} \color{black} -0.59 & {\cellcolor[HTML]{000000}} \color[HTML]{F1F1F1} {\cellcolor[HTML]{FFFFFF}} \color{black} --- & {\cellcolor[HTML]{000000}} \color[HTML]{F1F1F1} {\cellcolor[HTML]{FFFFFF}} \color{black} --- \\
\cline{1-7}
\multirow[c]{2}{*}{Nanc} & WK & {\cellcolor[HTML]{F7B396}} \color[HTML]{000000} \color{black} 0.25 & {\cellcolor[HTML]{C1D4F4}} \color[HTML]{000000} \color{black} -0.19 & {\cellcolor[HTML]{98B9FF}} \color[HTML]{000000} \color{black} -0.39 & {\cellcolor[HTML]{F5A081}} \color[HTML]{000000} \color{black} 0.33 & {\cellcolor[HTML]{000000}} \color[HTML]{F1F1F1} {\cellcolor[HTML]{FFFFFF}} \color{black} --- \\
 & WKND & {\cellcolor[HTML]{F59D7E}} \color[HTML]{000000} \color{black} 0.34 & {\cellcolor[HTML]{D7DCE3}} \color[HTML]{000000} \color{black} -0.08 & {\cellcolor[HTML]{B2CCFB}} \color[HTML]{000000} \color{black} -0.27 & {\cellcolor[HTML]{000000}} \color[HTML]{F1F1F1} {\cellcolor[HTML]{FFFFFF}} \color{black} --- & {\cellcolor[HTML]{000000}} \color[HTML]{F1F1F1} {\cellcolor[HTML]{FFFFFF}} \color{black} --- \\
\cline{1-7}
\bottomrule
\end{tabular}
\end{table}

\begin{table}
\centering
\caption[Eating drinking talking count coefficient estimates for local-level multinomial logistic regression predictive model]{Eating drinking talking count coefficient estimates for local-level multinomial logistic regression predictive model for the week and weekend and by cluster and by city (Paris--Nancy shown in size order). The figure contains the strength of the association between each of the covariates. Coefficients are coloured according to the strength of the relationship. Data preprocessing was applied to the \texttt{NetMob23} dataset~\citep{martinez-duriveNetMob23DatasetHighresolution2023}. City names and Week/Weekend labels are abbreviated for brevity.}
\label{tab:eating-drinking-talking-count-coefficients}
\begin{tabular}{llrrrrr}
\toprule
 & Class & 1 & 2 & 3 & 4 & 5 \\
City & WK/WKND &  &  &  &  &  \\
\midrule
\multirow[c]{2}{*}{Pari} & WK & {\cellcolor[HTML]{C4D5F3}} \color[HTML]{000000} \color{black} -0.18 & {\cellcolor[HTML]{E6D7CF}} \color[HTML]{000000} \color{black} 0.02 & {\cellcolor[HTML]{F5C4AC}} \color[HTML]{000000} \color{black} 0.16 & {\cellcolor[HTML]{000000}} \color[HTML]{F1F1F1} {\cellcolor[HTML]{FFFFFF}} \color{black} --- & {\cellcolor[HTML]{000000}} \color[HTML]{F1F1F1} {\cellcolor[HTML]{FFFFFF}} \color{black} --- \\
 & WKND & {\cellcolor[HTML]{D1DAE9}} \color[HTML]{000000} \color{black} -0.11 & {\cellcolor[HTML]{EAD4C8}} \color[HTML]{000000} \color{black} 0.05 & {\cellcolor[HTML]{ECD3C5}} \color[HTML]{000000} \color{black} 0.06 & {\cellcolor[HTML]{000000}} \color[HTML]{F1F1F1} {\cellcolor[HTML]{FFFFFF}} \color{black} --- & {\cellcolor[HTML]{000000}} \color[HTML]{F1F1F1} {\cellcolor[HTML]{FFFFFF}} \color{black} --- \\
\cline{1-7}
\multirow[c]{2}{*}{Renn} & WK & {\cellcolor[HTML]{6687ED}} \color[HTML]{F1F1F1} \color{black} -0.62 & {\cellcolor[HTML]{F6BFA6}} \color[HTML]{000000} \color{black} 0.19 & {\cellcolor[HTML]{EE8669}} \color[HTML]{F1F1F1} \color{black} 0.43 & {\cellcolor[HTML]{000000}} \color[HTML]{F1F1F1} {\cellcolor[HTML]{FFFFFF}} \color{black} --- & {\cellcolor[HTML]{000000}} \color[HTML]{F1F1F1} {\cellcolor[HTML]{FFFFFF}} \color{black} --- \\
 & WKND & {\cellcolor[HTML]{7597F6}} \color[HTML]{F1F1F1} \color{black} -0.55 & {\cellcolor[HTML]{F4C6AF}} \color[HTML]{000000} \color{black} 0.15 & {\cellcolor[HTML]{F18F71}} \color[HTML]{F1F1F1} \color{black} 0.40 & {\cellcolor[HTML]{000000}} \color[HTML]{F1F1F1} {\cellcolor[HTML]{FFFFFF}} \color{black} --- & {\cellcolor[HTML]{000000}} \color[HTML]{F1F1F1} {\cellcolor[HTML]{FFFFFF}} \color{black} --- \\
\cline{1-7}
\multirow[c]{2}{*}{Lill} & WK & {\cellcolor[HTML]{B1CBFC}} \color[HTML]{000000} \color{black} -0.28 & {\cellcolor[HTML]{E67259}} \color[HTML]{F1F1F1} \color{black} 0.50 & {\cellcolor[HTML]{BCD2F7}} \color[HTML]{000000} \color{black} -0.22 & {\cellcolor[HTML]{000000}} \color[HTML]{F1F1F1} {\cellcolor[HTML]{FFFFFF}} \color{black} --- & {\cellcolor[HTML]{000000}} \color[HTML]{F1F1F1} {\cellcolor[HTML]{FFFFFF}} \color{black} --- \\
 & WKND & {\cellcolor[HTML]{ADC9FD}} \color[HTML]{000000} \color{black} -0.29 & {\cellcolor[HTML]{EA7B60}} \color[HTML]{F1F1F1} \color{black} 0.46 & {\cellcolor[HTML]{C5D6F2}} \color[HTML]{000000} \color{black} -0.17 & {\cellcolor[HTML]{000000}} \color[HTML]{F1F1F1} {\cellcolor[HTML]{FFFFFF}} \color{black} --- & {\cellcolor[HTML]{000000}} \color[HTML]{F1F1F1} {\cellcolor[HTML]{FFFFFF}} \color{black} --- \\
\cline{1-7}
\multirow[c]{2}{*}{Bord} & WK & {\cellcolor[HTML]{8CAFFE}} \color[HTML]{000000} \color{black} -0.44 & {\cellcolor[HTML]{E5D8D1}} \color[HTML]{000000} \color{black} 0.01 & {\cellcolor[HTML]{EE8669}} \color[HTML]{F1F1F1} \color{black} 0.43 & {\cellcolor[HTML]{000000}} \color[HTML]{F1F1F1} {\cellcolor[HTML]{FFFFFF}} \color{black} --- & {\cellcolor[HTML]{000000}} \color[HTML]{F1F1F1} {\cellcolor[HTML]{FFFFFF}} \color{black} --- \\
 & WKND & {\cellcolor[HTML]{96B7FF}} \color[HTML]{000000} \color{black} -0.40 & {\cellcolor[HTML]{E3D9D3}} \color[HTML]{000000} \color{black} 0.00 & {\cellcolor[HTML]{F29072}} \color[HTML]{F1F1F1} \color{black} 0.39 & {\cellcolor[HTML]{000000}} \color[HTML]{F1F1F1} {\cellcolor[HTML]{FFFFFF}} \color{black} --- & {\cellcolor[HTML]{000000}} \color[HTML]{F1F1F1} {\cellcolor[HTML]{FFFFFF}} \color{black} --- \\
\cline{1-7}
\multirow[c]{2}{*}{Gren} & WK & {\cellcolor[HTML]{B7CFF9}} \color[HTML]{000000} \color{black} -0.25 & {\cellcolor[HTML]{C9D7F0}} \color[HTML]{000000} \color{black} -0.16 & {\cellcolor[HTML]{F18D6F}} \color[HTML]{F1F1F1} \color{black} 0.41 & {\cellcolor[HTML]{000000}} \color[HTML]{F1F1F1} {\cellcolor[HTML]{FFFFFF}} \color{black} --- & {\cellcolor[HTML]{000000}} \color[HTML]{F1F1F1} {\cellcolor[HTML]{FFFFFF}} \color{black} --- \\
 & WKND & {\cellcolor[HTML]{ABC8FD}} \color[HTML]{000000} \color{black} -0.30 & {\cellcolor[HTML]{BFD3F6}} \color[HTML]{000000} \color{black} -0.21 & {\cellcolor[HTML]{E57058}} \color[HTML]{F1F1F1} \color{black} 0.51 & {\cellcolor[HTML]{000000}} \color[HTML]{F1F1F1} {\cellcolor[HTML]{FFFFFF}} \color{black} --- & {\cellcolor[HTML]{000000}} \color[HTML]{F1F1F1} {\cellcolor[HTML]{FFFFFF}} \color{black} --- \\
\cline{1-7}
\multirow[c]{2}{*}{Lyon} & WK & {\cellcolor[HTML]{BCD2F7}} \color[HTML]{000000} \color{black} -0.22 & {\cellcolor[HTML]{F5C4AC}} \color[HTML]{000000} \color{black} 0.16 & {\cellcolor[HTML]{EBD3C6}} \color[HTML]{000000} \color{black} 0.06 & {\cellcolor[HTML]{000000}} \color[HTML]{F1F1F1} {\cellcolor[HTML]{FFFFFF}} \color{black} --- & {\cellcolor[HTML]{000000}} \color[HTML]{F1F1F1} {\cellcolor[HTML]{FFFFFF}} \color{black} --- \\
 & WKND & {\cellcolor[HTML]{BED2F6}} \color[HTML]{000000} \color{black} -0.21 & {\cellcolor[HTML]{F4C6AF}} \color[HTML]{000000} \color{black} 0.15 & {\cellcolor[HTML]{ECD3C5}} \color[HTML]{000000} \color{black} 0.06 & {\cellcolor[HTML]{000000}} \color[HTML]{F1F1F1} {\cellcolor[HTML]{FFFFFF}} \color{black} --- & {\cellcolor[HTML]{000000}} \color[HTML]{F1F1F1} {\cellcolor[HTML]{FFFFFF}} \color{black} --- \\
\cline{1-7}
\multirow[c]{2}{*}{Nant} & WK & {\cellcolor[HTML]{A9C6FD}} \color[HTML]{000000} \color{black} -0.31 & {\cellcolor[HTML]{F6BFA6}} \color[HTML]{000000} \color{black} 0.19 & {\cellcolor[HTML]{F2CAB5}} \color[HTML]{000000} \color{black} 0.12 & {\cellcolor[HTML]{000000}} \color[HTML]{F1F1F1} {\cellcolor[HTML]{FFFFFF}} \color{black} --- & {\cellcolor[HTML]{000000}} \color[HTML]{F1F1F1} {\cellcolor[HTML]{FFFFFF}} \color{black} --- \\
 & WKND & {\cellcolor[HTML]{ADC9FD}} \color[HTML]{000000} \color{black} -0.30 & {\cellcolor[HTML]{E1DAD6}} \color[HTML]{000000} \color{black} -0.01 & {\cellcolor[HTML]{F7A688}} \color[HTML]{000000} \color{black} 0.31 & {\cellcolor[HTML]{000000}} \color[HTML]{F1F1F1} {\cellcolor[HTML]{FFFFFF}} \color{black} --- & {\cellcolor[HTML]{000000}} \color[HTML]{F1F1F1} {\cellcolor[HTML]{FFFFFF}} \color{black} --- \\
\cline{1-7}
\multirow[c]{2}{*}{Toul} & WK & {\cellcolor[HTML]{B7CFF9}} \color[HTML]{000000} \color{black} -0.24 & {\cellcolor[HTML]{E4D9D2}} \color[HTML]{000000} \color{black} 0.01 & {\cellcolor[HTML]{F7B599}} \color[HTML]{000000} \color{black} 0.23 & {\cellcolor[HTML]{000000}} \color[HTML]{F1F1F1} {\cellcolor[HTML]{FFFFFF}} \color{black} --- & {\cellcolor[HTML]{000000}} \color[HTML]{F1F1F1} {\cellcolor[HTML]{FFFFFF}} \color{black} --- \\
 & WKND & {\cellcolor[HTML]{BFD3F6}} \color[HTML]{000000} \color{black} -0.21 & {\cellcolor[HTML]{E6D7CF}} \color[HTML]{000000} \color{black} 0.02 & {\cellcolor[HTML]{F6BFA6}} \color[HTML]{000000} \color{black} 0.19 & {\cellcolor[HTML]{000000}} \color[HTML]{F1F1F1} {\cellcolor[HTML]{FFFFFF}} \color{black} --- & {\cellcolor[HTML]{000000}} \color[HTML]{F1F1F1} {\cellcolor[HTML]{FFFFFF}} \color{black} --- \\
\cline{1-7}
\multirow[c]{2}{*}{Mont} & WK & {\cellcolor[HTML]{C7D7F0}} \color[HTML]{000000} \color{black} -0.16 & {\cellcolor[HTML]{E6D7CF}} \color[HTML]{000000} \color{black} 0.02 & {\cellcolor[HTML]{F3C7B1}} \color[HTML]{000000} \color{black} 0.14 & {\cellcolor[HTML]{000000}} \color[HTML]{F1F1F1} {\cellcolor[HTML]{FFFFFF}} \color{black} --- & {\cellcolor[HTML]{000000}} \color[HTML]{F1F1F1} {\cellcolor[HTML]{FFFFFF}} \color{black} --- \\
 & WKND & {\cellcolor[HTML]{CAD8EF}} \color[HTML]{000000} \color{black} -0.15 & {\cellcolor[HTML]{E0DBD8}} \color[HTML]{000000} \color{black} -0.02 & {\cellcolor[HTML]{F5C2AA}} \color[HTML]{000000} \color{black} 0.17 & {\cellcolor[HTML]{000000}} \color[HTML]{F1F1F1} {\cellcolor[HTML]{FFFFFF}} \color{black} --- & {\cellcolor[HTML]{000000}} \color[HTML]{F1F1F1} {\cellcolor[HTML]{FFFFFF}} \color{black} --- \\
\cline{1-7}
\multirow[c]{2}{*}{Tour} & WK & {\cellcolor[HTML]{CEDAEB}} \color[HTML]{000000} \color{black} -0.13 & {\cellcolor[HTML]{EAD5C9}} \color[HTML]{000000} \color{black} 0.04 & {\cellcolor[HTML]{E5D8D1}} \color[HTML]{000000} \color{black} 0.01 & {\cellcolor[HTML]{EED0C0}} \color[HTML]{000000} \color{black} 0.08 & {\cellcolor[HTML]{E2DAD5}} \color[HTML]{000000} \color{black} -0.01 \\
 & WKND & {\cellcolor[HTML]{DFDBD9}} \color[HTML]{000000} \color{black} -0.03 & {\cellcolor[HTML]{EAD4C8}} \color[HTML]{000000} \color{black} 0.05 & {\cellcolor[HTML]{E0DBD8}} \color[HTML]{000000} \color{black} -0.02 & {\cellcolor[HTML]{000000}} \color[HTML]{F1F1F1} {\cellcolor[HTML]{FFFFFF}} \color{black} --- & {\cellcolor[HTML]{000000}} \color[HTML]{F1F1F1} {\cellcolor[HTML]{FFFFFF}} \color{black} --- \\
\cline{1-7}
\multirow[c]{2}{*}{Stra} & WK & {\cellcolor[HTML]{C0D4F5}} \color[HTML]{000000} \color{black} -0.20 & {\cellcolor[HTML]{D5DBE5}} \color[HTML]{000000} \color{black} -0.08 & {\cellcolor[HTML]{F2C9B4}} \color[HTML]{000000} \color{black} 0.13 & {\cellcolor[HTML]{F4C6AF}} \color[HTML]{000000} \color{black} 0.15 & {\cellcolor[HTML]{000000}} \color[HTML]{F1F1F1} {\cellcolor[HTML]{FFFFFF}} \color{black} --- \\
 & WKND & {\cellcolor[HTML]{D3DBE7}} \color[HTML]{000000} \color{black} -0.10 & {\cellcolor[HTML]{DADCE0}} \color[HTML]{000000} \color{black} -0.05 & {\cellcolor[HTML]{F4C6AF}} \color[HTML]{000000} \color{black} 0.15 & {\cellcolor[HTML]{000000}} \color[HTML]{F1F1F1} {\cellcolor[HTML]{FFFFFF}} \color{black} --- & {\cellcolor[HTML]{000000}} \color[HTML]{F1F1F1} {\cellcolor[HTML]{FFFFFF}} \color{black} --- \\
\cline{1-7}
\multirow[c]{2}{*}{Orle} & WK & {\cellcolor[HTML]{90B2FE}} \color[HTML]{000000} \color{black} -0.42 & {\cellcolor[HTML]{F3C8B2}} \color[HTML]{000000} \color{black} 0.14 & {\cellcolor[HTML]{F7AA8C}} \color[HTML]{000000} \color{black} 0.29 & {\cellcolor[HTML]{000000}} \color[HTML]{F1F1F1} {\cellcolor[HTML]{FFFFFF}} \color{black} --- & {\cellcolor[HTML]{000000}} \color[HTML]{F1F1F1} {\cellcolor[HTML]{FFFFFF}} \color{black} --- \\
 & WKND & {\cellcolor[HTML]{92B4FE}} \color[HTML]{000000} \color{black} -0.42 & {\cellcolor[HTML]{F1CDBA}} \color[HTML]{000000} \color{black} 0.10 & {\cellcolor[HTML]{F6A385}} \color[HTML]{000000} \color{black} 0.32 & {\cellcolor[HTML]{000000}} \color[HTML]{F1F1F1} {\cellcolor[HTML]{FFFFFF}} \color{black} --- & {\cellcolor[HTML]{000000}} \color[HTML]{F1F1F1} {\cellcolor[HTML]{FFFFFF}} \color{black} --- \\
\cline{1-7}
\multirow[c]{2}{*}{Metz} & WK & {\cellcolor[HTML]{B6CEFA}} \color[HTML]{000000} \color{black} -0.25 & {\cellcolor[HTML]{E1DAD6}} \color[HTML]{000000} \color{black} -0.01 & {\cellcolor[HTML]{F7B093}} \color[HTML]{000000} \color{black} 0.26 & {\cellcolor[HTML]{000000}} \color[HTML]{F1F1F1} {\cellcolor[HTML]{FFFFFF}} \color{black} --- & {\cellcolor[HTML]{000000}} \color[HTML]{F1F1F1} {\cellcolor[HTML]{FFFFFF}} \color{black} --- \\
 & WKND & {\cellcolor[HTML]{AFCAFC}} \color[HTML]{000000} \color{black} -0.28 & {\cellcolor[HTML]{E5D8D1}} \color[HTML]{000000} \color{black} 0.01 & {\cellcolor[HTML]{F7AD90}} \color[HTML]{000000} \color{black} 0.27 & {\cellcolor[HTML]{000000}} \color[HTML]{F1F1F1} {\cellcolor[HTML]{FFFFFF}} \color{black} --- & {\cellcolor[HTML]{000000}} \color[HTML]{F1F1F1} {\cellcolor[HTML]{FFFFFF}} \color{black} --- \\
\cline{1-7}
\multirow[c]{2}{*}{Mans} & WK & {\cellcolor[HTML]{F6BEA4}} \color[HTML]{000000} \color{black} 0.19 & {\cellcolor[HTML]{C1D4F4}} \color[HTML]{000000} \color{black} -0.19 & {\cellcolor[HTML]{F1CCB8}} \color[HTML]{000000} \color{black} 0.11 & {\cellcolor[HTML]{D1DAE9}} \color[HTML]{000000} \color{black} -0.11 & {\cellcolor[HTML]{000000}} \color[HTML]{F1F1F1} {\cellcolor[HTML]{FFFFFF}} \color{black} --- \\
 & WKND & {\cellcolor[HTML]{F1CDBA}} \color[HTML]{000000} \color{black} 0.11 & {\cellcolor[HTML]{C5D6F2}} \color[HTML]{000000} \color{black} -0.18 & {\cellcolor[HTML]{F1CDBA}} \color[HTML]{000000} \color{black} 0.11 & {\cellcolor[HTML]{DDDCDC}} \color[HTML]{000000} \color{black} -0.03 & {\cellcolor[HTML]{000000}} \color[HTML]{F1F1F1} {\cellcolor[HTML]{FFFFFF}} \color{black} --- \\
\cline{1-7}
\multirow[c]{2}{*}{Dijo} & WK & {\cellcolor[HTML]{94B6FF}} \color[HTML]{000000} \color{black} -0.41 & {\cellcolor[HTML]{DFDBD9}} \color[HTML]{000000} \color{black} -0.03 & {\cellcolor[HTML]{EE8468}} \color[HTML]{F1F1F1} \color{black} 0.44 & {\cellcolor[HTML]{000000}} \color[HTML]{F1F1F1} {\cellcolor[HTML]{FFFFFF}} \color{black} --- & {\cellcolor[HTML]{000000}} \color[HTML]{F1F1F1} {\cellcolor[HTML]{FFFFFF}} \color{black} --- \\
 & WKND & {\cellcolor[HTML]{98B9FF}} \color[HTML]{000000} \color{black} -0.38 & {\cellcolor[HTML]{EAD4C8}} \color[HTML]{000000} \color{black} 0.05 & {\cellcolor[HTML]{F59F80}} \color[HTML]{000000} \color{black} 0.33 & {\cellcolor[HTML]{000000}} \color[HTML]{F1F1F1} {\cellcolor[HTML]{FFFFFF}} \color{black} --- & {\cellcolor[HTML]{000000}} \color[HTML]{F1F1F1} {\cellcolor[HTML]{FFFFFF}} \color{black} --- \\
\cline{1-7}
\multirow[c]{2}{*}{Mars} & WK & {\cellcolor[HTML]{BBD1F8}} \color[HTML]{000000} \color{black} -0.23 & {\cellcolor[HTML]{EED0C0}} \color[HTML]{000000} \color{black} 0.08 & {\cellcolor[HTML]{F7AC8E}} \color[HTML]{000000} \color{black} 0.28 & {\cellcolor[HTML]{CCD9ED}} \color[HTML]{000000} \color{black} -0.14 & {\cellcolor[HTML]{000000}} \color[HTML]{F1F1F1} {\cellcolor[HTML]{FFFFFF}} \color{black} --- \\
 & WKND & {\cellcolor[HTML]{B6CEFA}} \color[HTML]{000000} \color{black} -0.25 & {\cellcolor[HTML]{E9D5CB}} \color[HTML]{000000} \color{black} 0.04 & {\cellcolor[HTML]{F7B99E}} \color[HTML]{000000} \color{black} 0.22 & {\cellcolor[HTML]{000000}} \color[HTML]{F1F1F1} {\cellcolor[HTML]{FFFFFF}} \color{black} --- & {\cellcolor[HTML]{000000}} \color[HTML]{F1F1F1} {\cellcolor[HTML]{FFFFFF}} \color{black} --- \\
\cline{1-7}
\multirow[c]{2}{*}{Nice} & WK & {\cellcolor[HTML]{ECD3C5}} \color[HTML]{000000} \color{black} 0.06 & {\cellcolor[HTML]{A9C6FD}} \color[HTML]{000000} \color{black} -0.31 & {\cellcolor[HTML]{F7B396}} \color[HTML]{000000} \color{black} 0.25 & {\cellcolor[HTML]{000000}} \color[HTML]{F1F1F1} {\cellcolor[HTML]{FFFFFF}} \color{black} --- & {\cellcolor[HTML]{000000}} \color[HTML]{F1F1F1} {\cellcolor[HTML]{FFFFFF}} \color{black} --- \\
 & WKND & {\cellcolor[HTML]{F18D6F}} \color[HTML]{F1F1F1} \color{black} 0.40 & {\cellcolor[HTML]{EE8468}} \color[HTML]{F1F1F1} \color{black} 0.44 & {\cellcolor[HTML]{3B4CC0}} \color[HTML]{F1F1F1} \color{black} -0.84 & {\cellcolor[HTML]{000000}} \color[HTML]{F1F1F1} {\cellcolor[HTML]{FFFFFF}} \color{black} --- & {\cellcolor[HTML]{000000}} \color[HTML]{F1F1F1} {\cellcolor[HTML]{FFFFFF}} \color{black} --- \\
\cline{1-7}
\multirow[c]{2}{*}{Nanc} & WK & {\cellcolor[HTML]{799CF8}} \color[HTML]{F1F1F1} \color{black} -0.53 & {\cellcolor[HTML]{CFDAEA}} \color[HTML]{000000} \color{black} -0.12 & {\cellcolor[HTML]{C12B30}} \color[HTML]{F1F1F1} \color{black} 0.70 & {\cellcolor[HTML]{DBDCDE}} \color[HTML]{000000} \color{black} -0.05 & {\cellcolor[HTML]{000000}} \color[HTML]{F1F1F1} {\cellcolor[HTML]{FFFFFF}} \color{black} --- \\
 & WKND & {\cellcolor[HTML]{779AF7}} \color[HTML]{F1F1F1} \color{black} -0.54 & {\cellcolor[HTML]{D1DAE9}} \color[HTML]{000000} \color{black} -0.11 & {\cellcolor[HTML]{CC403A}} \color[HTML]{F1F1F1} \color{black} 0.65 & {\cellcolor[HTML]{000000}} \color[HTML]{F1F1F1} {\cellcolor[HTML]{FFFFFF}} \color{black} --- & {\cellcolor[HTML]{000000}} \color[HTML]{F1F1F1} {\cellcolor[HTML]{FFFFFF}} \color{black} --- \\
\cline{1-7}
\bottomrule
\end{tabular}
\end{table}

\begin{table}
\centering
\caption[Organised activity count coefficient estimates for local-level multinomial logistic regression predictive model]{Organised activity count coefficient estimates for local-level multinomial logistic regression predictive model for the week and weekend and by cluster and by city (Paris--Nancy shown in size order). The figure contains the strength of the association between each of the covariates. Coefficients are coloured according to the strength of the relationship. Data preprocessing was applied to the \texttt{NetMob23} dataset~\citep{martinez-duriveNetMob23DatasetHighresolution2023}. City names and Week/Weekend labels are abbreviated for brevity.}
\label{tab:organised-activity-count-coefficients}
\begin{tabular}{llrrrrr}
\toprule
 & Class & 1 & 2 & 3 & 4 & 5 \\
City & WK/WKND &  &  &  &  &  \\
\midrule
\multirow[c]{2}{*}{Pari} & WK & {\cellcolor[HTML]{D7DCE3}} \color[HTML]{000000} \color{black} -0.07 & {\cellcolor[HTML]{C5D6F2}} \color[HTML]{000000} \color{black} -0.17 & {\cellcolor[HTML]{F7B497}} \color[HTML]{000000} \color{black} 0.25 & {\cellcolor[HTML]{000000}} \color[HTML]{F1F1F1} {\cellcolor[HTML]{FFFFFF}} \color{black} --- & {\cellcolor[HTML]{000000}} \color[HTML]{F1F1F1} {\cellcolor[HTML]{FFFFFF}} \color{black} --- \\
 & WKND & {\cellcolor[HTML]{E2DAD5}} \color[HTML]{000000} \color{black} -0.01 & {\cellcolor[HTML]{D1DAE9}} \color[HTML]{000000} \color{black} -0.11 & {\cellcolor[HTML]{F2CBB7}} \color[HTML]{000000} \color{black} 0.12 & {\cellcolor[HTML]{000000}} \color[HTML]{F1F1F1} {\cellcolor[HTML]{FFFFFF}} \color{black} --- & {\cellcolor[HTML]{000000}} \color[HTML]{F1F1F1} {\cellcolor[HTML]{FFFFFF}} \color{black} --- \\
\cline{1-7}
\multirow[c]{2}{*}{Renn} & WK & {\cellcolor[HTML]{F7B599}} \color[HTML]{000000} \color{black} 0.23 & {\cellcolor[HTML]{F2CAB5}} \color[HTML]{000000} \color{black} 0.12 & {\cellcolor[HTML]{9FBFFF}} \color[HTML]{000000} \color{black} -0.36 & {\cellcolor[HTML]{000000}} \color[HTML]{F1F1F1} {\cellcolor[HTML]{FFFFFF}} \color{black} --- & {\cellcolor[HTML]{000000}} \color[HTML]{F1F1F1} {\cellcolor[HTML]{FFFFFF}} \color{black} --- \\
 & WKND & {\cellcolor[HTML]{F7AF91}} \color[HTML]{000000} \color{black} 0.27 & {\cellcolor[HTML]{F6BDA2}} \color[HTML]{000000} \color{black} 0.20 & {\cellcolor[HTML]{88ABFD}} \color[HTML]{000000} \color{black} -0.46 & {\cellcolor[HTML]{000000}} \color[HTML]{F1F1F1} {\cellcolor[HTML]{FFFFFF}} \color{black} --- & {\cellcolor[HTML]{000000}} \color[HTML]{F1F1F1} {\cellcolor[HTML]{FFFFFF}} \color{black} --- \\
\cline{1-7}
\multirow[c]{2}{*}{Lill} & WK & {\cellcolor[HTML]{F6A385}} \color[HTML]{000000} \color{black} 0.32 & {\cellcolor[HTML]{EAD5C9}} \color[HTML]{000000} \color{black} 0.05 & {\cellcolor[HTML]{9DBDFF}} \color[HTML]{000000} \color{black} -0.37 & {\cellcolor[HTML]{000000}} \color[HTML]{F1F1F1} {\cellcolor[HTML]{FFFFFF}} \color{black} --- & {\cellcolor[HTML]{000000}} \color[HTML]{F1F1F1} {\cellcolor[HTML]{FFFFFF}} \color{black} --- \\
 & WKND & {\cellcolor[HTML]{E8765C}} \color[HTML]{F1F1F1} \color{black} 0.49 & {\cellcolor[HTML]{F7B599}} \color[HTML]{000000} \color{black} 0.23 & {\cellcolor[HTML]{516DDB}} \color[HTML]{F1F1F1} \color{black} -0.72 & {\cellcolor[HTML]{000000}} \color[HTML]{F1F1F1} {\cellcolor[HTML]{FFFFFF}} \color{black} --- & {\cellcolor[HTML]{000000}} \color[HTML]{F1F1F1} {\cellcolor[HTML]{FFFFFF}} \color{black} --- \\
\cline{1-7}
\multirow[c]{2}{*}{Bord} & WK & {\cellcolor[HTML]{E5D8D1}} \color[HTML]{000000} \color{black} 0.01 & {\cellcolor[HTML]{E6D7CF}} \color[HTML]{000000} \color{black} 0.02 & {\cellcolor[HTML]{DEDCDB}} \color[HTML]{000000} \color{black} -0.03 & {\cellcolor[HTML]{000000}} \color[HTML]{F1F1F1} {\cellcolor[HTML]{FFFFFF}} \color{black} --- & {\cellcolor[HTML]{000000}} \color[HTML]{F1F1F1} {\cellcolor[HTML]{FFFFFF}} \color{black} --- \\
 & WKND & {\cellcolor[HTML]{DFDBD9}} \color[HTML]{000000} \color{black} -0.02 & {\cellcolor[HTML]{EBD3C6}} \color[HTML]{000000} \color{black} 0.05 & {\cellcolor[HTML]{DEDCDB}} \color[HTML]{000000} \color{black} -0.03 & {\cellcolor[HTML]{000000}} \color[HTML]{F1F1F1} {\cellcolor[HTML]{FFFFFF}} \color{black} --- & {\cellcolor[HTML]{000000}} \color[HTML]{F1F1F1} {\cellcolor[HTML]{FFFFFF}} \color{black} --- \\
\cline{1-7}
\multirow[c]{2}{*}{Gren} & WK & {\cellcolor[HTML]{EDD1C2}} \color[HTML]{000000} \color{black} 0.08 & {\cellcolor[HTML]{F7BCA1}} \color[HTML]{000000} \color{black} 0.21 & {\cellcolor[HTML]{AEC9FC}} \color[HTML]{000000} \color{black} -0.28 & {\cellcolor[HTML]{000000}} \color[HTML]{F1F1F1} {\cellcolor[HTML]{FFFFFF}} \color{black} --- & {\cellcolor[HTML]{000000}} \color[HTML]{F1F1F1} {\cellcolor[HTML]{FFFFFF}} \color{black} --- \\
 & WKND & {\cellcolor[HTML]{F6BFA6}} \color[HTML]{000000} \color{black} 0.19 & {\cellcolor[HTML]{F7B497}} \color[HTML]{000000} \color{black} 0.25 & {\cellcolor[HTML]{8FB1FE}} \color[HTML]{000000} \color{black} -0.43 & {\cellcolor[HTML]{000000}} \color[HTML]{F1F1F1} {\cellcolor[HTML]{FFFFFF}} \color{black} --- & {\cellcolor[HTML]{000000}} \color[HTML]{F1F1F1} {\cellcolor[HTML]{FFFFFF}} \color{black} --- \\
\cline{1-7}
\multirow[c]{2}{*}{Lyon} & WK & {\cellcolor[HTML]{EF886B}} \color[HTML]{F1F1F1} \color{black} 0.42 & {\cellcolor[HTML]{DBDCDE}} \color[HTML]{000000} \color{black} -0.05 & {\cellcolor[HTML]{9BBCFF}} \color[HTML]{000000} \color{black} -0.37 & {\cellcolor[HTML]{000000}} \color[HTML]{F1F1F1} {\cellcolor[HTML]{FFFFFF}} \color{black} --- & {\cellcolor[HTML]{000000}} \color[HTML]{F1F1F1} {\cellcolor[HTML]{FFFFFF}} \color{black} --- \\
 & WKND & {\cellcolor[HTML]{EF886B}} \color[HTML]{F1F1F1} \color{black} 0.43 & {\cellcolor[HTML]{DBDCDE}} \color[HTML]{000000} \color{black} -0.05 & {\cellcolor[HTML]{9BBCFF}} \color[HTML]{000000} \color{black} -0.37 & {\cellcolor[HTML]{000000}} \color[HTML]{F1F1F1} {\cellcolor[HTML]{FFFFFF}} \color{black} --- & {\cellcolor[HTML]{000000}} \color[HTML]{F1F1F1} {\cellcolor[HTML]{FFFFFF}} \color{black} --- \\
\cline{1-7}
\multirow[c]{2}{*}{Nant} & WK & {\cellcolor[HTML]{E1DAD6}} \color[HTML]{000000} \color{black} -0.01 & {\cellcolor[HTML]{C1D4F4}} \color[HTML]{000000} \color{black} -0.19 & {\cellcolor[HTML]{F7BCA1}} \color[HTML]{000000} \color{black} 0.21 & {\cellcolor[HTML]{000000}} \color[HTML]{F1F1F1} {\cellcolor[HTML]{FFFFFF}} \color{black} --- & {\cellcolor[HTML]{000000}} \color[HTML]{F1F1F1} {\cellcolor[HTML]{FFFFFF}} \color{black} --- \\
 & WKND & {\cellcolor[HTML]{F0CDBB}} \color[HTML]{000000} \color{black} 0.10 & {\cellcolor[HTML]{EFCEBD}} \color[HTML]{000000} \color{black} 0.10 & {\cellcolor[HTML]{C1D4F4}} \color[HTML]{000000} \color{black} -0.20 & {\cellcolor[HTML]{000000}} \color[HTML]{F1F1F1} {\cellcolor[HTML]{FFFFFF}} \color{black} --- & {\cellcolor[HTML]{000000}} \color[HTML]{F1F1F1} {\cellcolor[HTML]{FFFFFF}} \color{black} --- \\
\cline{1-7}
\multirow[c]{2}{*}{Toul} & WK & {\cellcolor[HTML]{F1CCB8}} \color[HTML]{000000} \color{black} 0.11 & {\cellcolor[HTML]{EAD4C8}} \color[HTML]{000000} \color{black} 0.05 & {\cellcolor[HTML]{C7D7F0}} \color[HTML]{000000} \color{black} -0.16 & {\cellcolor[HTML]{000000}} \color[HTML]{F1F1F1} {\cellcolor[HTML]{FFFFFF}} \color{black} --- & {\cellcolor[HTML]{000000}} \color[HTML]{F1F1F1} {\cellcolor[HTML]{FFFFFF}} \color{black} --- \\
 & WKND & {\cellcolor[HTML]{EDD2C3}} \color[HTML]{000000} \color{black} 0.07 & {\cellcolor[HTML]{EBD3C6}} \color[HTML]{000000} \color{black} 0.06 & {\cellcolor[HTML]{CEDAEB}} \color[HTML]{000000} \color{black} -0.12 & {\cellcolor[HTML]{000000}} \color[HTML]{F1F1F1} {\cellcolor[HTML]{FFFFFF}} \color{black} --- & {\cellcolor[HTML]{000000}} \color[HTML]{F1F1F1} {\cellcolor[HTML]{FFFFFF}} \color{black} --- \\
\cline{1-7}
\multirow[c]{2}{*}{Mont} & WK & {\cellcolor[HTML]{F6A283}} \color[HTML]{000000} \color{black} 0.32 & {\cellcolor[HTML]{F5C0A7}} \color[HTML]{000000} \color{black} 0.18 & {\cellcolor[HTML]{7EA1FA}} \color[HTML]{F1F1F1} \color{black} -0.50 & {\cellcolor[HTML]{000000}} \color[HTML]{F1F1F1} {\cellcolor[HTML]{FFFFFF}} \color{black} --- & {\cellcolor[HTML]{000000}} \color[HTML]{F1F1F1} {\cellcolor[HTML]{FFFFFF}} \color{black} --- \\
 & WKND & {\cellcolor[HTML]{F6A283}} \color[HTML]{000000} \color{black} 0.32 & {\cellcolor[HTML]{F6BEA4}} \color[HTML]{000000} \color{black} 0.19 & {\cellcolor[HTML]{7B9FF9}} \color[HTML]{F1F1F1} \color{black} -0.52 & {\cellcolor[HTML]{000000}} \color[HTML]{F1F1F1} {\cellcolor[HTML]{FFFFFF}} \color{black} --- & {\cellcolor[HTML]{000000}} \color[HTML]{F1F1F1} {\cellcolor[HTML]{FFFFFF}} \color{black} --- \\
\cline{1-7}
\multirow[c]{2}{*}{Tour} & WK & {\cellcolor[HTML]{F1CDBA}} \color[HTML]{000000} \color{black} 0.11 & {\cellcolor[HTML]{F5C2AA}} \color[HTML]{000000} \color{black} 0.17 & {\cellcolor[HTML]{98B9FF}} \color[HTML]{000000} \color{black} -0.39 & {\cellcolor[HTML]{F7BCA1}} \color[HTML]{000000} \color{black} 0.21 & {\cellcolor[HTML]{D4DBE6}} \color[HTML]{000000} \color{black} -0.09 \\
 & WKND & {\cellcolor[HTML]{F2CAB5}} \color[HTML]{000000} \color{black} 0.13 & {\cellcolor[HTML]{CEDAEB}} \color[HTML]{000000} \color{black} -0.12 & {\cellcolor[HTML]{E3D9D3}} \color[HTML]{000000} \color{black} -0.00 & {\cellcolor[HTML]{000000}} \color[HTML]{F1F1F1} {\cellcolor[HTML]{FFFFFF}} \color{black} --- & {\cellcolor[HTML]{000000}} \color[HTML]{F1F1F1} {\cellcolor[HTML]{FFFFFF}} \color{black} --- \\
\cline{1-7}
\multirow[c]{2}{*}{Stra} & WK & {\cellcolor[HTML]{F7A98B}} \color[HTML]{000000} \color{black} 0.29 & {\cellcolor[HTML]{F0CDBB}} \color[HTML]{000000} \color{black} 0.10 & {\cellcolor[HTML]{D5DBE5}} \color[HTML]{000000} \color{black} -0.09 & {\cellcolor[HTML]{AAC7FD}} \color[HTML]{000000} \color{black} -0.30 & {\cellcolor[HTML]{000000}} \color[HTML]{F1F1F1} {\cellcolor[HTML]{FFFFFF}} \color{black} --- \\
 & WKND & {\cellcolor[HTML]{F6BDA2}} \color[HTML]{000000} \color{black} 0.20 & {\cellcolor[HTML]{E3D9D3}} \color[HTML]{000000} \color{black} 0.00 & {\cellcolor[HTML]{BFD3F6}} \color[HTML]{000000} \color{black} -0.20 & {\cellcolor[HTML]{000000}} \color[HTML]{F1F1F1} {\cellcolor[HTML]{FFFFFF}} \color{black} --- & {\cellcolor[HTML]{000000}} \color[HTML]{F1F1F1} {\cellcolor[HTML]{FFFFFF}} \color{black} --- \\
\cline{1-7}
\multirow[c]{2}{*}{Orle} & WK & {\cellcolor[HTML]{F6BDA2}} \color[HTML]{000000} \color{black} 0.20 & {\cellcolor[HTML]{F3C7B1}} \color[HTML]{000000} \color{black} 0.14 & {\cellcolor[HTML]{A2C1FF}} \color[HTML]{000000} \color{black} -0.34 & {\cellcolor[HTML]{000000}} \color[HTML]{F1F1F1} {\cellcolor[HTML]{FFFFFF}} \color{black} --- & {\cellcolor[HTML]{000000}} \color[HTML]{F1F1F1} {\cellcolor[HTML]{FFFFFF}} \color{black} --- \\
 & WKND & {\cellcolor[HTML]{F5C2AA}} \color[HTML]{000000} \color{black} 0.17 & {\cellcolor[HTML]{F1CCB8}} \color[HTML]{000000} \color{black} 0.11 & {\cellcolor[HTML]{AFCAFC}} \color[HTML]{000000} \color{black} -0.28 & {\cellcolor[HTML]{000000}} \color[HTML]{F1F1F1} {\cellcolor[HTML]{FFFFFF}} \color{black} --- & {\cellcolor[HTML]{000000}} \color[HTML]{F1F1F1} {\cellcolor[HTML]{FFFFFF}} \color{black} --- \\
\cline{1-7}
\multirow[c]{2}{*}{Metz} & WK & {\cellcolor[HTML]{EED0C0}} \color[HTML]{000000} \color{black} 0.08 & {\cellcolor[HTML]{F2C9B4}} \color[HTML]{000000} \color{black} 0.13 & {\cellcolor[HTML]{BFD3F6}} \color[HTML]{000000} \color{black} -0.21 & {\cellcolor[HTML]{000000}} \color[HTML]{F1F1F1} {\cellcolor[HTML]{FFFFFF}} \color{black} --- & {\cellcolor[HTML]{000000}} \color[HTML]{F1F1F1} {\cellcolor[HTML]{FFFFFF}} \color{black} --- \\
 & WKND & {\cellcolor[HTML]{EDD2C3}} \color[HTML]{000000} \color{black} 0.07 & {\cellcolor[HTML]{F2C9B4}} \color[HTML]{000000} \color{black} 0.13 & {\cellcolor[HTML]{C1D4F4}} \color[HTML]{000000} \color{black} -0.19 & {\cellcolor[HTML]{000000}} \color[HTML]{F1F1F1} {\cellcolor[HTML]{FFFFFF}} \color{black} --- & {\cellcolor[HTML]{000000}} \color[HTML]{F1F1F1} {\cellcolor[HTML]{FFFFFF}} \color{black} --- \\
\cline{1-7}
\multirow[c]{2}{*}{Mans} & WK & {\cellcolor[HTML]{F7B093}} \color[HTML]{000000} \color{black} 0.26 & {\cellcolor[HTML]{C5D6F2}} \color[HTML]{000000} \color{black} -0.18 & {\cellcolor[HTML]{BAD0F8}} \color[HTML]{000000} \color{black} -0.23 & {\cellcolor[HTML]{F4C6AF}} \color[HTML]{000000} \color{black} 0.15 & {\cellcolor[HTML]{000000}} \color[HTML]{F1F1F1} {\cellcolor[HTML]{FFFFFF}} \color{black} --- \\
 & WKND & {\cellcolor[HTML]{ECD3C5}} \color[HTML]{000000} \color{black} 0.06 & {\cellcolor[HTML]{E9D5CB}} \color[HTML]{000000} \color{black} 0.04 & {\cellcolor[HTML]{BAD0F8}} \color[HTML]{000000} \color{black} -0.23 & {\cellcolor[HTML]{F2C9B4}} \color[HTML]{000000} \color{black} 0.13 & {\cellcolor[HTML]{000000}} \color[HTML]{F1F1F1} {\cellcolor[HTML]{FFFFFF}} \color{black} --- \\
\cline{1-7}
\multirow[c]{2}{*}{Dijo} & WK & {\cellcolor[HTML]{F6BFA6}} \color[HTML]{000000} \color{black} 0.19 & {\cellcolor[HTML]{F7AC8E}} \color[HTML]{000000} \color{black} 0.28 & {\cellcolor[HTML]{85A8FC}} \color[HTML]{F1F1F1} \color{black} -0.47 & {\cellcolor[HTML]{000000}} \color[HTML]{F1F1F1} {\cellcolor[HTML]{FFFFFF}} \color{black} --- & {\cellcolor[HTML]{000000}} \color[HTML]{F1F1F1} {\cellcolor[HTML]{FFFFFF}} \color{black} --- \\
 & WKND & {\cellcolor[HTML]{EAD5C9}} \color[HTML]{000000} \color{black} 0.04 & {\cellcolor[HTML]{F5C4AC}} \color[HTML]{000000} \color{black} 0.16 & {\cellcolor[HTML]{BFD3F6}} \color[HTML]{000000} \color{black} -0.20 & {\cellcolor[HTML]{000000}} \color[HTML]{F1F1F1} {\cellcolor[HTML]{FFFFFF}} \color{black} --- & {\cellcolor[HTML]{000000}} \color[HTML]{F1F1F1} {\cellcolor[HTML]{FFFFFF}} \color{black} --- \\
\cline{1-7}
\multirow[c]{2}{*}{Mars} & WK & {\cellcolor[HTML]{CFDAEA}} \color[HTML]{000000} \color{black} -0.12 & {\cellcolor[HTML]{D7DCE3}} \color[HTML]{000000} \color{black} -0.07 & {\cellcolor[HTML]{BAD0F8}} \color[HTML]{000000} \color{black} -0.23 & {\cellcolor[HTML]{F08A6C}} \color[HTML]{F1F1F1} \color{black} 0.42 & {\cellcolor[HTML]{000000}} \color[HTML]{F1F1F1} {\cellcolor[HTML]{FFFFFF}} \color{black} --- \\
 & WKND & {\cellcolor[HTML]{DFDBD9}} \color[HTML]{000000} \color{black} -0.02 & {\cellcolor[HTML]{EAD5C9}} \color[HTML]{000000} \color{black} 0.05 & {\cellcolor[HTML]{DFDBD9}} \color[HTML]{000000} \color{black} -0.02 & {\cellcolor[HTML]{000000}} \color[HTML]{F1F1F1} {\cellcolor[HTML]{FFFFFF}} \color{black} --- & {\cellcolor[HTML]{000000}} \color[HTML]{F1F1F1} {\cellcolor[HTML]{FFFFFF}} \color{black} --- \\
\cline{1-7}
\multirow[c]{2}{*}{Nice} & WK & {\cellcolor[HTML]{ECD3C5}} \color[HTML]{000000} \color{black} 0.06 & {\cellcolor[HTML]{F4C6AF}} \color[HTML]{000000} \color{black} 0.15 & {\cellcolor[HTML]{BFD3F6}} \color[HTML]{000000} \color{black} -0.21 & {\cellcolor[HTML]{000000}} \color[HTML]{F1F1F1} {\cellcolor[HTML]{FFFFFF}} \color{black} --- & {\cellcolor[HTML]{000000}} \color[HTML]{F1F1F1} {\cellcolor[HTML]{FFFFFF}} \color{black} --- \\
 & WKND & {\cellcolor[HTML]{DCDDDD}} \color[HTML]{000000} \color{black} -0.04 & {\cellcolor[HTML]{D3DBE7}} \color[HTML]{000000} \color{black} -0.10 & {\cellcolor[HTML]{F3C7B1}} \color[HTML]{000000} \color{black} 0.14 & {\cellcolor[HTML]{000000}} \color[HTML]{F1F1F1} {\cellcolor[HTML]{FFFFFF}} \color{black} --- & {\cellcolor[HTML]{000000}} \color[HTML]{F1F1F1} {\cellcolor[HTML]{FFFFFF}} \color{black} --- \\
\cline{1-7}
\multirow[c]{2}{*}{Nanc} & WK & {\cellcolor[HTML]{F0CDBB}} \color[HTML]{000000} \color{black} 0.10 & {\cellcolor[HTML]{F5C4AC}} \color[HTML]{000000} \color{black} 0.16 & {\cellcolor[HTML]{D8DCE2}} \color[HTML]{000000} \color{black} -0.07 & {\cellcolor[HTML]{C3D5F4}} \color[HTML]{000000} \color{black} -0.19 & {\cellcolor[HTML]{000000}} \color[HTML]{F1F1F1} {\cellcolor[HTML]{FFFFFF}} \color{black} --- \\
 & WKND & {\cellcolor[HTML]{EAD5C9}} \color[HTML]{000000} \color{black} 0.05 & {\cellcolor[HTML]{EFCEBD}} \color[HTML]{000000} \color{black} 0.10 & {\cellcolor[HTML]{CBD8EE}} \color[HTML]{000000} \color{black} -0.14 & {\cellcolor[HTML]{000000}} \color[HTML]{F1F1F1} {\cellcolor[HTML]{FFFFFF}} \color{black} --- & {\cellcolor[HTML]{000000}} \color[HTML]{F1F1F1} {\cellcolor[HTML]{FFFFFF}} \color{black} --- \\
\cline{1-7}
\bottomrule
\end{tabular}
\end{table}

\begin{table}
\centering
\caption[Outdoor count coefficient estimates for local-level multinomial logistic regression predictive model]{Outdoor count coefficient estimates for local-level multinomial logistic regression predictive model for the week and weekend and by cluster and by city (Paris--Nancy shown in size order). The figure contains the strength of the association between each of the covariates. Coefficients are coloured according to the strength of the relationship. Data preprocessing was applied to the \texttt{NetMob23} dataset~\citep{martinez-duriveNetMob23DatasetHighresolution2023}. City names and Week/Weekend labels are abbreviated for brevity.}
\label{tab:outdoor-count-coefficients}
\begin{tabular}{llrrrrr}
\toprule
 & Class & 1 & 2 & 3 & 4 & 5 \\
City & WK/WKND &  &  &  &  &  \\
\midrule
\multirow[c]{2}{*}{Pari} & WK & {\cellcolor[HTML]{DADCE0}} \color[HTML]{000000} \color{black} -0.05 & {\cellcolor[HTML]{E0DBD8}} \color[HTML]{000000} \color{black} -0.02 & {\cellcolor[HTML]{EDD1C2}} \color[HTML]{000000} \color{black} 0.07 & {\cellcolor[HTML]{000000}} \color[HTML]{F1F1F1} {\cellcolor[HTML]{FFFFFF}} \color{black} --- & {\cellcolor[HTML]{000000}} \color[HTML]{F1F1F1} {\cellcolor[HTML]{FFFFFF}} \color{black} --- \\
 & WKND & {\cellcolor[HTML]{DDDCDC}} \color[HTML]{000000} \color{black} -0.04 & {\cellcolor[HTML]{E5D8D1}} \color[HTML]{000000} \color{black} 0.01 & {\cellcolor[HTML]{E7D7CE}} \color[HTML]{000000} \color{black} 0.02 & {\cellcolor[HTML]{000000}} \color[HTML]{F1F1F1} {\cellcolor[HTML]{FFFFFF}} \color{black} --- & {\cellcolor[HTML]{000000}} \color[HTML]{F1F1F1} {\cellcolor[HTML]{FFFFFF}} \color{black} --- \\
\cline{1-7}
\multirow[c]{2}{*}{Renn} & WK & {\cellcolor[HTML]{E2DAD5}} \color[HTML]{000000} \color{black} -0.01 & {\cellcolor[HTML]{D9DCE1}} \color[HTML]{000000} \color{black} -0.06 & {\cellcolor[HTML]{EDD2C3}} \color[HTML]{000000} \color{black} 0.07 & {\cellcolor[HTML]{000000}} \color[HTML]{F1F1F1} {\cellcolor[HTML]{FFFFFF}} \color{black} --- & {\cellcolor[HTML]{000000}} \color[HTML]{F1F1F1} {\cellcolor[HTML]{FFFFFF}} \color{black} --- \\
 & WKND & {\cellcolor[HTML]{DEDCDB}} \color[HTML]{000000} \color{black} -0.03 & {\cellcolor[HTML]{DADCE0}} \color[HTML]{000000} \color{black} -0.06 & {\cellcolor[HTML]{EFCFBF}} \color[HTML]{000000} \color{black} 0.09 & {\cellcolor[HTML]{000000}} \color[HTML]{F1F1F1} {\cellcolor[HTML]{FFFFFF}} \color{black} --- & {\cellcolor[HTML]{000000}} \color[HTML]{F1F1F1} {\cellcolor[HTML]{FFFFFF}} \color{black} --- \\
\cline{1-7}
\multirow[c]{2}{*}{Lill} & WK & {\cellcolor[HTML]{C4D5F3}} \color[HTML]{000000} \color{black} -0.18 & {\cellcolor[HTML]{CEDAEB}} \color[HTML]{000000} \color{black} -0.12 & {\cellcolor[HTML]{F7A688}} \color[HTML]{000000} \color{black} 0.31 & {\cellcolor[HTML]{000000}} \color[HTML]{F1F1F1} {\cellcolor[HTML]{FFFFFF}} \color{black} --- & {\cellcolor[HTML]{000000}} \color[HTML]{F1F1F1} {\cellcolor[HTML]{FFFFFF}} \color{black} --- \\
 & WKND & {\cellcolor[HTML]{BCD2F7}} \color[HTML]{000000} \color{black} -0.22 & {\cellcolor[HTML]{C5D6F2}} \color[HTML]{000000} \color{black} -0.18 & {\cellcolor[HTML]{F29072}} \color[HTML]{F1F1F1} \color{black} 0.39 & {\cellcolor[HTML]{000000}} \color[HTML]{F1F1F1} {\cellcolor[HTML]{FFFFFF}} \color{black} --- & {\cellcolor[HTML]{000000}} \color[HTML]{F1F1F1} {\cellcolor[HTML]{FFFFFF}} \color{black} --- \\
\cline{1-7}
\multirow[c]{2}{*}{Bord} & WK & {\cellcolor[HTML]{DADCE0}} \color[HTML]{000000} \color{black} -0.06 & {\cellcolor[HTML]{EDD2C3}} \color[HTML]{000000} \color{black} 0.07 & {\cellcolor[HTML]{E1DAD6}} \color[HTML]{000000} \color{black} -0.02 & {\cellcolor[HTML]{000000}} \color[HTML]{F1F1F1} {\cellcolor[HTML]{FFFFFF}} \color{black} --- & {\cellcolor[HTML]{000000}} \color[HTML]{F1F1F1} {\cellcolor[HTML]{FFFFFF}} \color{black} --- \\
 & WKND & {\cellcolor[HTML]{E4D9D2}} \color[HTML]{000000} \color{black} 0.01 & {\cellcolor[HTML]{E5D8D1}} \color[HTML]{000000} \color{black} 0.01 & {\cellcolor[HTML]{E0DBD8}} \color[HTML]{000000} \color{black} -0.02 & {\cellcolor[HTML]{000000}} \color[HTML]{F1F1F1} {\cellcolor[HTML]{FFFFFF}} \color{black} --- & {\cellcolor[HTML]{000000}} \color[HTML]{F1F1F1} {\cellcolor[HTML]{FFFFFF}} \color{black} --- \\
\cline{1-7}
\multirow[c]{2}{*}{Gren} & WK & {\cellcolor[HTML]{E3D9D3}} \color[HTML]{000000} \color{black} 0.00 & {\cellcolor[HTML]{DDDCDC}} \color[HTML]{000000} \color{black} -0.04 & {\cellcolor[HTML]{E9D5CB}} \color[HTML]{000000} \color{black} 0.03 & {\cellcolor[HTML]{000000}} \color[HTML]{F1F1F1} {\cellcolor[HTML]{FFFFFF}} \color{black} --- & {\cellcolor[HTML]{000000}} \color[HTML]{F1F1F1} {\cellcolor[HTML]{FFFFFF}} \color{black} --- \\
 & WKND & {\cellcolor[HTML]{E0DBD8}} \color[HTML]{000000} \color{black} -0.02 & {\cellcolor[HTML]{DADCE0}} \color[HTML]{000000} \color{black} -0.05 & {\cellcolor[HTML]{EDD1C2}} \color[HTML]{000000} \color{black} 0.07 & {\cellcolor[HTML]{000000}} \color[HTML]{F1F1F1} {\cellcolor[HTML]{FFFFFF}} \color{black} --- & {\cellcolor[HTML]{000000}} \color[HTML]{F1F1F1} {\cellcolor[HTML]{FFFFFF}} \color{black} --- \\
\cline{1-7}
\multirow[c]{2}{*}{Lyon} & WK & {\cellcolor[HTML]{D4DBE6}} \color[HTML]{000000} \color{black} -0.10 & {\cellcolor[HTML]{E6D7CF}} \color[HTML]{000000} \color{black} 0.02 & {\cellcolor[HTML]{EED0C0}} \color[HTML]{000000} \color{black} 0.08 & {\cellcolor[HTML]{000000}} \color[HTML]{F1F1F1} {\cellcolor[HTML]{FFFFFF}} \color{black} --- & {\cellcolor[HTML]{000000}} \color[HTML]{F1F1F1} {\cellcolor[HTML]{FFFFFF}} \color{black} --- \\
 & WKND & {\cellcolor[HTML]{D3DBE7}} \color[HTML]{000000} \color{black} -0.10 & {\cellcolor[HTML]{E7D7CE}} \color[HTML]{000000} \color{black} 0.02 & {\cellcolor[HTML]{EDD1C2}} \color[HTML]{000000} \color{black} 0.08 & {\cellcolor[HTML]{000000}} \color[HTML]{F1F1F1} {\cellcolor[HTML]{FFFFFF}} \color{black} --- & {\cellcolor[HTML]{000000}} \color[HTML]{F1F1F1} {\cellcolor[HTML]{FFFFFF}} \color{black} --- \\
\cline{1-7}
\multirow[c]{2}{*}{Nant} & WK & {\cellcolor[HTML]{CDD9EC}} \color[HTML]{000000} \color{black} -0.13 & {\cellcolor[HTML]{DEDCDB}} \color[HTML]{000000} \color{black} -0.03 & {\cellcolor[HTML]{F5C4AC}} \color[HTML]{000000} \color{black} 0.16 & {\cellcolor[HTML]{000000}} \color[HTML]{F1F1F1} {\cellcolor[HTML]{FFFFFF}} \color{black} --- & {\cellcolor[HTML]{000000}} \color[HTML]{F1F1F1} {\cellcolor[HTML]{FFFFFF}} \color{black} --- \\
 & WKND & {\cellcolor[HTML]{CFDAEA}} \color[HTML]{000000} \color{black} -0.12 & {\cellcolor[HTML]{E9D5CB}} \color[HTML]{000000} \color{black} 0.04 & {\cellcolor[HTML]{EED0C0}} \color[HTML]{000000} \color{black} 0.08 & {\cellcolor[HTML]{000000}} \color[HTML]{F1F1F1} {\cellcolor[HTML]{FFFFFF}} \color{black} --- & {\cellcolor[HTML]{000000}} \color[HTML]{F1F1F1} {\cellcolor[HTML]{FFFFFF}} \color{black} --- \\
\cline{1-7}
\multirow[c]{2}{*}{Toul} & WK & {\cellcolor[HTML]{E5D8D1}} \color[HTML]{000000} \color{black} 0.01 & {\cellcolor[HTML]{DDDCDC}} \color[HTML]{000000} \color{black} -0.04 & {\cellcolor[HTML]{E7D7CE}} \color[HTML]{000000} \color{black} 0.02 & {\cellcolor[HTML]{000000}} \color[HTML]{F1F1F1} {\cellcolor[HTML]{FFFFFF}} \color{black} --- & {\cellcolor[HTML]{000000}} \color[HTML]{F1F1F1} {\cellcolor[HTML]{FFFFFF}} \color{black} --- \\
 & WKND & {\cellcolor[HTML]{E6D7CF}} \color[HTML]{000000} \color{black} 0.02 & {\cellcolor[HTML]{DDDCDC}} \color[HTML]{000000} \color{black} -0.04 & {\cellcolor[HTML]{E6D7CF}} \color[HTML]{000000} \color{black} 0.02 & {\cellcolor[HTML]{000000}} \color[HTML]{F1F1F1} {\cellcolor[HTML]{FFFFFF}} \color{black} --- & {\cellcolor[HTML]{000000}} \color[HTML]{F1F1F1} {\cellcolor[HTML]{FFFFFF}} \color{black} --- \\
\cline{1-7}
\multirow[c]{2}{*}{Mont} & WK & {\cellcolor[HTML]{CAD8EF}} \color[HTML]{000000} \color{black} -0.15 & {\cellcolor[HTML]{E0DBD8}} \color[HTML]{000000} \color{black} -0.02 & {\cellcolor[HTML]{F5C2AA}} \color[HTML]{000000} \color{black} 0.17 & {\cellcolor[HTML]{000000}} \color[HTML]{F1F1F1} {\cellcolor[HTML]{FFFFFF}} \color{black} --- & {\cellcolor[HTML]{000000}} \color[HTML]{F1F1F1} {\cellcolor[HTML]{FFFFFF}} \color{black} --- \\
 & WKND & {\cellcolor[HTML]{CAD8EF}} \color[HTML]{000000} \color{black} -0.15 & {\cellcolor[HTML]{DDDCDC}} \color[HTML]{000000} \color{black} -0.03 & {\cellcolor[HTML]{F6BFA6}} \color[HTML]{000000} \color{black} 0.19 & {\cellcolor[HTML]{000000}} \color[HTML]{F1F1F1} {\cellcolor[HTML]{FFFFFF}} \color{black} --- & {\cellcolor[HTML]{000000}} \color[HTML]{F1F1F1} {\cellcolor[HTML]{FFFFFF}} \color{black} --- \\
\cline{1-7}
\multirow[c]{2}{*}{Tour} & WK & {\cellcolor[HTML]{D6DCE4}} \color[HTML]{000000} \color{black} -0.08 & {\cellcolor[HTML]{DDDCDC}} \color[HTML]{000000} \color{black} -0.04 & {\cellcolor[HTML]{F4C5AD}} \color[HTML]{000000} \color{black} 0.16 & {\cellcolor[HTML]{E1DAD6}} \color[HTML]{000000} \color{black} -0.01 & {\cellcolor[HTML]{E0DBD8}} \color[HTML]{000000} \color{black} -0.02 \\
 & WKND & {\cellcolor[HTML]{EAD4C8}} \color[HTML]{000000} \color{black} 0.05 & {\cellcolor[HTML]{F4C6AF}} \color[HTML]{000000} \color{black} 0.15 & {\cellcolor[HTML]{C0D4F5}} \color[HTML]{000000} \color{black} -0.20 & {\cellcolor[HTML]{000000}} \color[HTML]{F1F1F1} {\cellcolor[HTML]{FFFFFF}} \color{black} --- & {\cellcolor[HTML]{000000}} \color[HTML]{F1F1F1} {\cellcolor[HTML]{FFFFFF}} \color{black} --- \\
\cline{1-7}
\multirow[c]{2}{*}{Stra} & WK & {\cellcolor[HTML]{AEC9FC}} \color[HTML]{000000} \color{black} -0.29 & {\cellcolor[HTML]{D7DCE3}} \color[HTML]{000000} \color{black} -0.07 & {\cellcolor[HTML]{E5D8D1}} \color[HTML]{000000} \color{black} 0.01 & {\cellcolor[HTML]{F59C7D}} \color[HTML]{000000} \color{black} 0.35 & {\cellcolor[HTML]{000000}} \color[HTML]{F1F1F1} {\cellcolor[HTML]{FFFFFF}} \color{black} --- \\
 & WKND & {\cellcolor[HTML]{C4D5F3}} \color[HTML]{000000} \color{black} -0.18 & {\cellcolor[HTML]{E8D6CC}} \color[HTML]{000000} \color{black} 0.03 & {\cellcolor[HTML]{F4C6AF}} \color[HTML]{000000} \color{black} 0.15 & {\cellcolor[HTML]{000000}} \color[HTML]{F1F1F1} {\cellcolor[HTML]{FFFFFF}} \color{black} --- & {\cellcolor[HTML]{000000}} \color[HTML]{F1F1F1} {\cellcolor[HTML]{FFFFFF}} \color{black} --- \\
\cline{1-7}
\multirow[c]{2}{*}{Orle} & WK & {\cellcolor[HTML]{DEDCDB}} \color[HTML]{000000} \color{black} -0.03 & {\cellcolor[HTML]{E2DAD5}} \color[HTML]{000000} \color{black} -0.01 & {\cellcolor[HTML]{E9D5CB}} \color[HTML]{000000} \color{black} 0.04 & {\cellcolor[HTML]{000000}} \color[HTML]{F1F1F1} {\cellcolor[HTML]{FFFFFF}} \color{black} --- & {\cellcolor[HTML]{000000}} \color[HTML]{F1F1F1} {\cellcolor[HTML]{FFFFFF}} \color{black} --- \\
 & WKND & {\cellcolor[HTML]{DEDCDB}} \color[HTML]{000000} \color{black} -0.03 & {\cellcolor[HTML]{E8D6CC}} \color[HTML]{000000} \color{black} 0.03 & {\cellcolor[HTML]{E4D9D2}} \color[HTML]{000000} \color{black} 0.01 & {\cellcolor[HTML]{000000}} \color[HTML]{F1F1F1} {\cellcolor[HTML]{FFFFFF}} \color{black} --- & {\cellcolor[HTML]{000000}} \color[HTML]{F1F1F1} {\cellcolor[HTML]{FFFFFF}} \color{black} --- \\
\cline{1-7}
\multirow[c]{2}{*}{Metz} & WK & {\cellcolor[HTML]{D1DAE9}} \color[HTML]{000000} \color{black} -0.11 & {\cellcolor[HTML]{E0DBD8}} \color[HTML]{000000} \color{black} -0.02 & {\cellcolor[HTML]{F2C9B4}} \color[HTML]{000000} \color{black} 0.13 & {\cellcolor[HTML]{000000}} \color[HTML]{F1F1F1} {\cellcolor[HTML]{FFFFFF}} \color{black} --- & {\cellcolor[HTML]{000000}} \color[HTML]{F1F1F1} {\cellcolor[HTML]{FFFFFF}} \color{black} --- \\
 & WKND & {\cellcolor[HTML]{D1DAE9}} \color[HTML]{000000} \color{black} -0.11 & {\cellcolor[HTML]{E0DBD8}} \color[HTML]{000000} \color{black} -0.02 & {\cellcolor[HTML]{F3C8B2}} \color[HTML]{000000} \color{black} 0.14 & {\cellcolor[HTML]{000000}} \color[HTML]{F1F1F1} {\cellcolor[HTML]{FFFFFF}} \color{black} --- & {\cellcolor[HTML]{000000}} \color[HTML]{F1F1F1} {\cellcolor[HTML]{FFFFFF}} \color{black} --- \\
\cline{1-7}
\multirow[c]{2}{*}{Mans} & WK & {\cellcolor[HTML]{A1C0FF}} \color[HTML]{000000} \color{black} -0.35 & {\cellcolor[HTML]{F2CAB5}} \color[HTML]{000000} \color{black} 0.12 & {\cellcolor[HTML]{F3C8B2}} \color[HTML]{000000} \color{black} 0.14 & {\cellcolor[HTML]{EFCFBF}} \color[HTML]{000000} \color{black} 0.09 & {\cellcolor[HTML]{000000}} \color[HTML]{F1F1F1} {\cellcolor[HTML]{FFFFFF}} \color{black} --- \\
 & WKND & {\cellcolor[HTML]{BED2F6}} \color[HTML]{000000} \color{black} -0.22 & {\cellcolor[HTML]{E9D5CB}} \color[HTML]{000000} \color{black} 0.04 & {\cellcolor[HTML]{F1CDBA}} \color[HTML]{000000} \color{black} 0.10 & {\cellcolor[HTML]{EDD1C2}} \color[HTML]{000000} \color{black} 0.07 & {\cellcolor[HTML]{000000}} \color[HTML]{F1F1F1} {\cellcolor[HTML]{FFFFFF}} \color{black} --- \\
\cline{1-7}
\multirow[c]{2}{*}{Dijo} & WK & {\cellcolor[HTML]{AFCAFC}} \color[HTML]{000000} \color{black} -0.28 & {\cellcolor[HTML]{E1DAD6}} \color[HTML]{000000} \color{black} -0.01 & {\cellcolor[HTML]{F7A98B}} \color[HTML]{000000} \color{black} 0.29 & {\cellcolor[HTML]{000000}} \color[HTML]{F1F1F1} {\cellcolor[HTML]{FFFFFF}} \color{black} --- & {\cellcolor[HTML]{000000}} \color[HTML]{F1F1F1} {\cellcolor[HTML]{FFFFFF}} \color{black} --- \\
 & WKND & {\cellcolor[HTML]{D2DBE8}} \color[HTML]{000000} \color{black} -0.10 & {\cellcolor[HTML]{F1CCB8}} \color[HTML]{000000} \color{black} 0.11 & {\cellcolor[HTML]{E1DAD6}} \color[HTML]{000000} \color{black} -0.01 & {\cellcolor[HTML]{000000}} \color[HTML]{F1F1F1} {\cellcolor[HTML]{FFFFFF}} \color{black} --- & {\cellcolor[HTML]{000000}} \color[HTML]{F1F1F1} {\cellcolor[HTML]{FFFFFF}} \color{black} --- \\
\cline{1-7}
\multirow[c]{2}{*}{Mars} & WK & {\cellcolor[HTML]{F7BCA1}} \color[HTML]{000000} \color{black} 0.21 & {\cellcolor[HTML]{E9D5CB}} \color[HTML]{000000} \color{black} 0.04 & {\cellcolor[HTML]{F2CBB7}} \color[HTML]{000000} \color{black} 0.12 & {\cellcolor[HTML]{9EBEFF}} \color[HTML]{000000} \color{black} -0.36 & {\cellcolor[HTML]{000000}} \color[HTML]{F1F1F1} {\cellcolor[HTML]{FFFFFF}} \color{black} --- \\
 & WKND & {\cellcolor[HTML]{EFCFBF}} \color[HTML]{000000} \color{black} 0.08 & {\cellcolor[HTML]{DADCE0}} \color[HTML]{000000} \color{black} -0.06 & {\cellcolor[HTML]{DFDBD9}} \color[HTML]{000000} \color{black} -0.03 & {\cellcolor[HTML]{000000}} \color[HTML]{F1F1F1} {\cellcolor[HTML]{FFFFFF}} \color{black} --- & {\cellcolor[HTML]{000000}} \color[HTML]{F1F1F1} {\cellcolor[HTML]{FFFFFF}} \color{black} --- \\
\cline{1-7}
\multirow[c]{2}{*}{Nice} & WK & {\cellcolor[HTML]{CEDAEB}} \color[HTML]{000000} \color{black} -0.13 & {\cellcolor[HTML]{F0CDBB}} \color[HTML]{000000} \color{black} 0.10 & {\cellcolor[HTML]{E7D7CE}} \color[HTML]{000000} \color{black} 0.03 & {\cellcolor[HTML]{000000}} \color[HTML]{F1F1F1} {\cellcolor[HTML]{FFFFFF}} \color{black} --- & {\cellcolor[HTML]{000000}} \color[HTML]{F1F1F1} {\cellcolor[HTML]{FFFFFF}} \color{black} --- \\
 & WKND & {\cellcolor[HTML]{9EBEFF}} \color[HTML]{000000} \color{black} -0.36 & {\cellcolor[HTML]{BFD3F6}} \color[HTML]{000000} \color{black} -0.20 & {\cellcolor[HTML]{DC5D4A}} \color[HTML]{F1F1F1} \color{black} 0.57 & {\cellcolor[HTML]{000000}} \color[HTML]{F1F1F1} {\cellcolor[HTML]{FFFFFF}} \color{black} --- & {\cellcolor[HTML]{000000}} \color[HTML]{F1F1F1} {\cellcolor[HTML]{FFFFFF}} \color{black} --- \\
\cline{1-7}
\multirow[c]{2}{*}{Nanc} & WK & {\cellcolor[HTML]{F7B99E}} \color[HTML]{000000} \color{black} 0.22 & {\cellcolor[HTML]{F7B194}} \color[HTML]{000000} \color{black} 0.26 & {\cellcolor[HTML]{F5C1A9}} \color[HTML]{000000} \color{black} 0.18 & {\cellcolor[HTML]{5E7DE7}} \color[HTML]{F1F1F1} \color{black} -0.65 & {\cellcolor[HTML]{000000}} \color[HTML]{F1F1F1} {\cellcolor[HTML]{FFFFFF}} \color{black} --- \\
 & WKND & {\cellcolor[HTML]{E9D5CB}} \color[HTML]{000000} \color{black} 0.03 & {\cellcolor[HTML]{E6D7CF}} \color[HTML]{000000} \color{black} 0.02 & {\cellcolor[HTML]{DBDCDE}} \color[HTML]{000000} \color{black} -0.05 & {\cellcolor[HTML]{000000}} \color[HTML]{F1F1F1} {\cellcolor[HTML]{FFFFFF}} \color{black} --- & {\cellcolor[HTML]{000000}} \color[HTML]{F1F1F1} {\cellcolor[HTML]{FFFFFF}} \color{black} --- \\
\cline{1-7}
\bottomrule
\end{tabular}
\end{table}

\end{document}